\documentclass[fleqn,usenatbib]{mnras}
\usepackage{cite}
\usepackage{graphicx}	
\usepackage{amsmath}	
\usepackage{amssymb}	
\usepackage{txfonts}
\usepackage{rotating}
\usepackage{accents}
\usepackage{float}
\usepackage{epsfig}
\usepackage{upgreek}
\usepackage{epstopdf}
\usepackage{graphicx}
\usepackage{color}
\usepackage{pdflscape}
\usepackage{appendix}
\usepackage{ulem}
\usepackage{soul}

\usepackage{newtxtext,newtxmath}
\usepackage[T1]{fontenc}

\begin{document}

\title[A new insight view of AGC\,198691 (Leoncino) galaxy with MEGARA at the GTC]{A new insight of AGC\,198691 (Leoncino) galaxy with MEGARA at the GTC}

\author[Carrasco et al. ]
{E. Carrasco$^{1}$\thanks{E-mail:}, M.L. Garc\'{i}a-Vargas$^{2}$, A. Gil de Paz$^{3, 4}$, M. Moll\'{a}$^{5}$, R. Izazaga-P\'erez$^{1}$,
\newauthor
A. Castillo-Morales$^{3, 4}$, P. G\'{o}mez-Alvarez$^{2}$,  J. Gallego$^{3, 4}$, J. Iglesias-P\'{a}ramo$^{6}$, 
\newauthor
 N. Cardiel$^{3, 4}$, S. Pascual$^{3, 4}$ and A. P\'{e}rez-Calpena$^{2}$
\\
$^{1}$ Instituto Nacional de Astrof{\'\i}sica, {\'O}ptica y Electr{\'o}nica (INAOE), Calle Luis Enrique Erro 1, C.P. 72840 Santa Mar{\'\i}a Tonantzintla, Puebla, M{\'e}xico\\
$^{2}$ FRACTAL S.L.N.E. Calle Tulip{\'a}n 2, portal 13, 1A, E-28231 Las Rozas de Madrid, Spain \\
$^{3}$ Dpto. de F{\'\i}sica de la Tierra y Astrof{\'\i}sica, Fac. CC. F{\'\i}sicas, Universidad Complutense de Madrid, Plaza de las Ciencias, 1, E-28040 Madrid, Spain \\
$^{4}$ Instituto de F{\'\i}sica de Part{\'\i}culas y del Cosmos, IPARCOS, Fac. CC. F{\'\i}sicas, Universidad Complutense de Madrid, Plaza de las Ciencias 1, E-28040 Madrid, Spain\\
$^{5}$ Dpto. de Investigaci\'{o}n B\'{a}sica, CIEMAT, Avda. Complutense 40, E-28040 Madrid, Spain\\
$^{6}$ Instituto de Astrof{\'\i}sica de Andaluc{\'\i}a, IAA-CSIC, Glorieta de la Astronom{\'\i}a s/n, E-18008  Granada, Spain\\
}

\date{Accepted Received ; in original form }
\pagerange{\pageref{firstpage}--\pageref{lastpage}} \pubyear{2019}

\maketitle
\label{firstpage}

\begin{abstract}
We describe the observations of the low-metallicity nearby galaxy AGC\,198691 (Leoncino Dwarf) obtained with the Integral Field Unit of the instrument MEGARA at the  Gran Telescopio Canarias.  The observations cover the wavelength ranges 4304 -- 5198\,\AA\ and 6098 -- 7306\,\AA\ with a resolving power  R\,$\approx$\,6000. We present 2D maps of the ionized gas, deriving the extension of the  H{\sc ii} region and gas kinematics from the observed emission lines.  We have not found any evidence of recent gas infall or loss of metals by means of outflows. This result is supported by the closed-box model predictions, consistent with the oxygen abundance found by other authors in this galaxy and points towards Leoncino being a genuine XMD galaxy.
We present for the first time spatially resolved spectroscopy allowing the detailed study of a star forming region. We use {\sc popstar}\,$+$\,{\sc cloudy} models to simulate the emission-line spectrum.   We find that the central emission line spectrum can be explained by a single young ionizing cluster with an age of $\approx$\,3.5\,$\pm$\,0.5\,Myr  and a  stellar mass of $\approx$\,2\,$\times\,$10$^{3}$\,M\textsubscript{\(\sun\)}. However, the radial profiles of [\ion{O}{iii}]\,$\lambda$\,5007\AA\ and the Balmer lines in emission demand photoionization by clusters of different ages between 3.5 and 6.5\,Myr that might respond either to the evolution of a single cluster evolving along the cooling time of the nebula ($\approx$\,3\,Myr at the metallicity of Leoncino, Z$\approx$\,0.0004) or to mass segregation of the cluster, being both scenarios consistent with the observed equivalent widths of the Balmer lines.

\end{abstract}


\section{Introduction}

The properties of eXtremely Metal-Deficient (XMD) galaxies, such as gas and stellar kinematics, and their relationship with the environment, along with the mechanisms in place that preserve their low abundance until today, have been a matter of scrutiny for the last few decades. Based on these properties, XMD galaxies have been proposed as the natural laboratories for testing the processes of chemical enrichment right after the primordial nucleosynthesis, in conditions similar to those of the early Universe.

Thus, nearby low metallicity galaxies are considered the local counterparts of the high-redshift primeval star-forming galaxies. This astrophysical subject has been boosted in the last five years due to new observational facilities and techniques pushing the limits of currently operating telescopes, both in the local universe and at intermediate redshift. The two main contributors to this progress have been recent large-scale surveys, such as the Sloan Digital Sky Survey, SDSS, \citep{Aba05} and new but still scarce detailed spectroscopic observations with medium-large aperture ($\geq$ 6.5m) telescopes.

 Local XMD galaxies have proven hard to find. Under the assumption that all normal galaxies follow the Luminosity (L) -- Metallicity (Z) correlation, where the luminosity is usually measured from the B-band absolute magnitude, over the whole abundance range, the L -- Z relationship would predict a low luminosity for the most metal-poor galaxies \citep{skillman89}, so XMD galaxies would be intrinsically weak, and therefore difficult to be detected,  unless a burst of star formation boosted the luminosity producing a strong emission line spectra. For this reason, most of the searches of these galaxies have been focused on finding objects with episodes of star formation, selecting the candidates for their blue excess or for the high values of [\ion{O}{iii}]\,$\lambda$\,5007/H$_{\beta}$, e.g., \citet{izotovandthuan07}, \citet{gusevaetal17}, \citet{hirs18}, \citet{izotovetal18a} and \citet{itg19}. Moreover,  [\ion{O}{iii}] lines are the result of the ionization by a relatively hard spectrum only present in very young star clusters of  age $\leq$ 5.5 Myr, as shown by \citet{gvbd95a, gvbd95b} and later  confirmed with  the evolutionary synthesis models {\sc popstar} \citep[hereafter M09]{mol09}.
 
 Galaxy luminosity is strongly linked to recent star formation activity, and all the galaxies cataloged as XMD so far have been observed while experiencing a strong burst of star formation.  The gas is ionized by a very young cluster, only a few Myr in age, making these XMD galaxies detectable in optical [OIII] lines.
However, {\sc popstar} models (M09) predict, for a single burst of star formation and a similar cluster's mass and H${\alpha}$ luminosity, that the lower the metallicity the larger the age range in which the emission line spectrum can be detected, so the low abundance by itself cannot justify the detection difficulty. The origin behind the bias of XMD in mid-aperture telescope surveys seems to be a lower probability of capturing the episodes of star formation, due to an intrinsic low luminosity or/and a slow and quiet evolution.

 A crucial point for XMD classification is the assignment of metallicity, usually done through the gas-phase oxygen abundance in the optical emission line spectrum of the ionized gas.  The detailed description of how the oxygen abundance is obtained from observations in the optical range can be found in \citet{osterbrockandferland} and, more recently, in \citet{pm14}. Briefly, the method consists in the determination of the average electron density,  $n_{\rm e}$,  of the gas from density-sensitive lines like [\ion{S}{ii}] or [\ion{O}{ii}], followed by the determination of the electronic temperature, T$_{\rm e}$, of the ionized nebula. In particular, T$_{\rm e}$ ([\ion{O}{iii}]) is obtained from the ratio $R_{\rm O3}$\,=\,(I$_{4959}$ + I$_{5007}$) / I$_{4363}$. The lines  [\ion{O}{iii}]\,$\lambda$\,4959, 5007\,\AA\ are very bright, especially in the low metallicity regions while the line [\ion{O}{iii}]\,$\lambda$\,4363\,\AA\ is very weak, of the order of 1-2 \% of the intensity of [\ion{O}{iii}]\,$\lambda$\,5007. Thus, the measurement of the [\ion{O}{iii}]\,$\lambda\,$4363 emission line is challenging, and imposes a bias in the low metallicity value depending on the telescope aperture, spatial resolution, spectroscopic field-of-view and affordable exposure times. When no detection of [\ion{O}{iii}]\,$\lambda$\,4363 line is available, the oxygen abundance could be inferred from other emission lines.  However, these methods have led to important discrepancies on the metallicity values \citep[and references therein]{andrews13, MA13} in the lowest range,  which is the case for XMD galaxies. 

 The metallicity could also be estimated from diagnostic diagrams when different emission line ratios are used simultaneously. \citet[] [hereafter M10]{mman10} 
used Spectral Energy Distributions (SEDs) synthesis\,$+$\,photoionization code {\sc cloudy} to predict the emission line spectrum produced by an ionizing Single Stellar Population, SSP,  under certain hypotheses on the Initial Mass Function, IMF, the chemical composition of the ionizing star clusters, Z, surrounding gas abundance and electron density, $n_{\rm e}$, and geometry (ionization bounded). They reached metallicities as low as Z\,$=$\,0.0004 (0.023 Z\textsubscript{\(\sun\)} equivalent to 12\,+\,log (O/H)\,=\,7.05) and Z\,$=$\,0.0001 (0.0058 Z\textsubscript{\(\sun\)}, equivalent to 12\,+\,log (O/H)\,=\, 6.45). M10 presented four  diagnostic diagrams (see their figures 9, 10, 11 and 12) at different metallicities for  the ionizing cluster and the surrounding gas. In particular, for values as low as those needed for reaching the lowest metallicity HII regions when plotting models, for a wide range of ionizing cluster ages and masses. This low-metallicity HII region locus in the diagnostic diagrams might be representative of XMD observations. The lowest metallicity models have intrinsic limitations, such as the uncertainty in both evolutionary tracks and atmosphere models of low metallicity stars and the cooling time in the ionized nebula. This cooling time is comparable to the age of the ionizing cluster itself, as computed and discuss in M10. Nevertheless, there are still very few observations for XMD with reliable metallicity and good values of all these lines to test the goodness of these diagrams as XMD abundance estimators.

 A few years ago, a new technique based on selecting galaxies with low \ion{H}{i} mass and bright optical emission line counterpart, has allowed the discoveries of few more XMD as Leoncino \citep[hereafter H16]{hirs16}. However, these XMD galaxies studies are still based on emission line spectra, biased to the time when these galaxies are undergoing  a burst of star formation. This in turn,  biases the samples as most quiet low-metallicity old galaxies without recent star formation are missing, pointing  towards  these galaxies being not less numerous but less detectable as we need to go deeper in luminosity. Moreover, these galaxies would have low gas mass content and/or a low Star Formation Rate, SFR, at least for genuine XMDs, making the detection less likely.

The maximum oxygen abundance value to classify a galaxy as XMD differs among different authors and works \citep{kunth00, izotovandthuan07, izotovetal12, gusevaetal15, sanchezalmeida16, kojima20} and, for years, was modulated by the value of the solar abundance itself. Just 20 years ago, the value of 12\,+\,log(O/H)$\leq$ 7.65 was setup as the limit to be an XMD galaxy by \citet{kunth00}.  
However, the number of the discoveries of XMD galaxies has increased so much in the last few years that the abundance of 12\,+\,log(O/H)\,$\leq$\,7.35, equivalent to 0.046\,Z\textsubscript{\(\sun\)}, roughly a twentieth of the current value of the solar metallicity, is now considered to be the upper limit, see e.g.  \citet{gusevaetal17} and   \citet [hereafter MQ20]{mcquinn20}. Some authors even point to a value of 12\,+\,log(O/H)\,$\leq$\,7.15, as minimum oxygen abundance to be XMD (H16). The competition for finding these objects is open and, until now, the lowest metallicity galaxies are J1631+4426 \citep{kojima20}, J0811+4730 \citep{izotovetal18a}, Leoncino galaxy (H16) and J1234+3901 \citep{itg19}, with 12\,+\,log(O/H) equal to 6.90\,$\pm$\,0.03,  6.98\,$\pm$\,0.02, 7.02\,$\pm$\,0.03,  recently reported  as   7.12\,$\pm$\,0.04   (unpublished) by MQ20  and 7.035\,$\pm$\,0.026, respectively.

There are  other two fundamental aspects related to these galaxies. The first one is how the gas abundance is linked to the stellar metallicity of the galaxy while the second one is what is the origin of the low metallicity, being both associated to the history of star formation in the galaxy and the interaction with the surrounding gas.  Assuming a sole old stellar population in the galaxy, its metallicity would map the gas metal content when stars formed, while the gas abundance would fit the current metallicity. However, the underlying stellar population is the combination of the different generations of stars created at different epochs tuned by the SFR of the galaxy and enriched according to its chemical evolution. For example, if the SFR was high in the past, we would expect to observe an enriched gas, so an  explanation of 
the observed low metallicity would be the dilution by  infall gas or the loss of metals by means of  outflows. In contrast, if the star formation history was slow, due to an inefficient star formation or a small amount of gas available to fuel the star formation, the gas and old star population might have similar (low) metallicity. The latter scenario reinforces the scarcity  of these XMD galaxies due to the low probability of capturing the star 
formation burst episode.

There have been many papers focused in disentangling the star and gas abundances in the XMD galaxies. One of the most used tools have been the relationship between the mass in stars, M\textsubscript{\(\ast\)}, and the metallicity, Z \citep{bergetal12, gusevaetal17}. The stellar mass has to be indirectly inferred by comparison of an observable, e.g. spectra or colours, with galaxy evolution models, each one with its own hypotheses and limitations, leading to different conclusions in terms of the dominant stellar populations in the galaxy and its total stellar mass. In the first published works on this topic, a universal M\textsubscript{\(\ast\)} -- Z relation was put forward. However, with the advent of more observations and models, this relation became a matter of intense debate. On the one hand, for massive spiral galaxies (M\textsubscript{\(\ast\)} $\geq$ 10$^{10.5}$ M\textsubscript{\(\sun\)}) the {\mbox M\textsubscript{\(\ast\)} -- Z} relation starts to flatten \citep{tremontietal04, hirs18}, while for XMD galaxies (M\textsubscript{\(\ast\)} $\leq$ 10$^{9}$ M\textsubscript{\(\sun\)}) large discrepancies were found resulting in significant disagreements among different works \citep[and references therein]{hirs18}. \citet{bergetal12} obtained a strong correlation between M\textsubscript{\(\ast\)} and Z, with measured oxygen abundance higher for higher galaxy mass. They used observed values of the luminosity at 4.5\,$\mu$m and the (B -- K) colour and compared them with the values predicted by the models of \citet{bc03}, assuming a Salpeter IMF and under the hypothesis that the 4.5\,$\mu$m luminosity is mostly dominated by low-mass main-sequence and Red Giant Branch (RGB) stars. However, \citet{gusevaetal17} presented a different M\textsubscript{\(\ast\)} -- Z relation for XMD  galaxies, finding a flatter relation of abundance when the stellar mass is derived from SEDs including nebular continuum emission, not present in the \citet{bc03} models \citet{gusevaetal17} worked under the hypothesis that the star formation history of XMD galaxies can be represented by a short burst with age $\leq$ 10\,Myr plus an old population formed in a rather continuous way. By fitting the model to their observations, they find values of the SFR between 0.1 and 1.0 M\textsubscript{\(\odot\)} yr$^{-1}$ and a specific SFR, defined as SFR/M\textsubscript{\(\ast\)}, of about 50\,Gyr$^{-1}$, similar to the one found by these same authors in previous works for star forming galaxies at redshifts 2\,$<$\,z\,$<\,$4.

Finally, the answer to the question on the origin of the lack of metals in XMD galaxies is also a matter of debate. \citet{ekta10} summarized the following three possibilities: 1) a genuine low metallicity, 2) a dilution of the metals by pristine gas falling and 3) a preferential loss of metals.

The first scenario assumes that the low metal content is genuine, so that there is and always was a small amount of metals, implying that the galaxy has held a very low rate of star formation over a long period of time. This might occur if the galaxy is isolated enough, although there is some controversial in the interpretation of data supporting this argument \citep{KD02}. Alternatively, the cause can be a slow and different chemical evolution, lower than usual SFR over the galaxy age, inferring an inefficient triggering of star formation \citep{gavilan13}. Under this assumption, the low metal content of these galaxies has been studied to put an upper limit to the primordial helium abundance \citep[and references therein]{cyburt16}, to constrain simulations of the formation of very low-mass galaxies and to test evolutionary models of massive stars out of pristine gas \citep[and references therein]{szecsi16}. 

The second scenario assumes a dilution of the metals by pristine gas falling into the galaxy from the outer disk, if existent, or from the local environment, like a consequence of an interaction with a neighbour galaxy. In the latter case,  the infalling gas would trigger a quick mix of the galaxy material, producing a lower abundance \citep[MQ20]{ekta10}. However, \citet{dalcanton07} showed that for such galaxies, the effective yield tends asymptotically to a constant value independent of the gas mass, implying that the inflow of metal-poor gas cannot substantially lower the effective yield of extremely gas-rich galaxies. This mechanism, therefore, does not seem to be the one responsible of the low metallicity in very gas rich galaxies, as the XMD found in HI surveys. 

The two scenarios described above are very different as they  pointed to distinct causes of the low abundance. In the first one, the metals are not created in the XMD galaxy because the SFR is very low and, consequently, the mass created in new stars will be very low too. In the second scenario, metals are created and they are present in the gas, but an infall of primordial (or less enriched) gas would dilute the proportion of them in the total gas mass, with the corresponding abundance decrease. \citet{gavilan13} developed self-consistent chemical evolution and spectrophotometric models of the formation and evolution of  gas-rich dwarf galaxies, with recent star formation but not bursting (i.e. dIrr), considering infall of primordial external gas but excluding outflow or galactic winds. They compare their predictions with observational data and  conclude that such  galaxies with moderate to low SFRs may be able to preserve the vast majority of their gas and metals.  

The third scenario assumes that there was a more metallic content, but a significant part of these metals have been lost by violent phenomena, e.g. supernova (SN)  events or outflows like enriched galactic winds. In the last years, this metal loss scenario has been the preferred one to explain the low abundance, as  in the case of the LeoP galaxy, where, with an abundance of 12\,+\,log(O/H)\,=\,7.17\,$\pm$\, 0.04 \citep[MQ20]{skillman13},  it seems that  95\,$\pm$\,2 per cent of its oxygen would have been lost through galactic winds \citep{mcquinn15a, mcquinn15b}.

The XMD galaxies study demands to move to the next step: 2D spectroscopy in the optical range to derive, on the one hand, the age and metallicity of the stellar populations and, on the other hand, the gas metallicity and kinematics along the extension of the galaxy. To accomplish this goal, we need a large aperture telescope, to reach the low luminosity objects, with a spectral resolution high enough to decouple the emission line fluxes and kinematics of different physical components, and good spatial resolution to minimize the aperture effects and to place the different burst of star formation. These observations in the visible are more useful if the galaxies have HST high spatial resolution images or UV spectroscopy that unveil the young stellar clusters, hidden by the ionized gas in the ground-based optical spectra. Last but not least, the use of evolutionary synthesis and chemical evolution models is crucial to infer the physical properties of the stellar populations (mass, age and metallicity) and the star formation history.

MEGARA  (Multi Espectr{\'o}grafo en GTC de Alta Resoluci{\'o}n para Astronom{\'\i}a)  offers all the required capabilities in the world largest optical telescope, the 10.4\,m  Gran Telescopio Canarias (GTC), at the Observatorio del Roque de los Muchachos located in La Palma, Spain.  MEGARA observations allow us to derive maps of the  gas properties at different galactocentric distances that can  provide clues on the origin of the low metal content and the confinement (or not) of the ionized gas around the young clusters.

This paper is dedicated to present and discuss the results taken with MEGARA Integral Field Unit, IFU, on Leoncino galaxy. The spectra have resolving power of R\,$\approx$\,6000, in two different configurations corresponding to the spectral regions around H${\beta}$ and [\ion{O}{iii}]$\lambda$\,5007, and H${\alpha}$, respectively.
Section~\ref{Sec:Leoncino} summarizes the published results on Leoncino galaxy. Section~\ref{Sec:Observationsandanalyses} presents our observations, data reduction and analysis. The results are detailed in section~\ref{Sec:Results} with the discussion in section~\ref{Sec:Discussion}, finishing with the conclusions in section~\ref{Sec:Conclusions}.

\section {Leoncino galaxy data from literature}
\label{Sec:Leoncino}

The galaxy AGC\,198691 (also known as the {\sl little lion} and nicknamed Leoncino) was discovered and catalogued as the most metal-poor gas-rich galaxy known at the time (H16). This galaxy formed part of the the Arecibo Legacy Fast ALFA blind [$\ion{H}{i}$] survey, ALFALFA \citep{giovanelli05, haynes11, giovanelli13,  haynes18}, and has an estimated gas mass M$_\ion{H}{i}\,\leq 10^{7}$\,M$_{\sun}$.  It was included for following up in the Survey of HI in Extremely Low-mass Dwarfs, SHIELD  \citep{cannon11}  for ALFALFA galaxies with  10$^{6}\,$M$_{\odot}\,\leq$\, M$_\ion{H}{i}\,\leq\, 10^{7.2}$ M$_{\odot}$ and optical counterparts in the SDSS \citep{Aba05}. H16 obtained R-band and H${\alpha}$ images  with the WIYN 0.9\,m telescope  and  optical long-slit (1.2\,$\times$\,10\, arcsec$^{2}$) spectroscopic data on the 4\,m Mayall  telescope at the Kitt Peak National Observatory (KPNO),  under  seeing conditions  of 1.4\,arcsec FWHM using the Kitt Peak Ohio State Multi-Object Spectrograph, KOSMOS, in both its blue and red arms with reciprocal dispersion of  0.66 and 0.99~\AA\,pixel$^{-1}$, respectively. They also presented 0.8\,arcsec seeing observations using the Blue Channel Spectrograph at the  6.5\,m Multi Mirror Telescope (MMT) with slits of 1 and 1.5 arcsec in width and a linear dispersion of 1.19~\AA\,pixel$^{-1}$. All observations were flux calibrated and taken along the parallactic angle. The extraction slit lengths were of 3.21 and 2.70 arcsec for the KPNO and  the MMT observations,  respectively. These authors reported a complete emission line analysis for that central spectrum with MMT composite data obtaining  values of $T_{\rm e}$\,=\,19130\,$\pm$\,800\,K,   $n_{\rm e}$\,=\,270\,$\pm$\,200\,cm$^{-3}$ and 12\,+\,log(O/H)\,=\,7.02\,$\pm$\,0.03. They  estimated  a distance to the galaxy, obtained using the velocity flow model employed by the ALFALFA team, of 7.7 Mpc. 

Recently, MQ20  published a comprehensive work on Leoncino imaging data obtained with HST. The  galaxy was observed by the HST/WFC3-UVIS instrument (date: 2018-04-24, PID: 15243, PI: McQuinn) using the F606W and F814W UVIS2 filters for a total of 15018 and 18618\,s, respectively. 
MQ20 presented a detailed photometric analysis using the {\sc dolphot} package \citep{dolphin00} and built a final single-star Colour -- Magnitude Diagram, CMD, including 147 stars, with a mix of young and old populations composed by Red Super Giants (RSG), red HeB and Asympthotic Giant Branch (AGB) stars to which they assigned ages of 25, 50 and 100\,Myr, respectively, using the  {\sc parsec} isochrones \citep{bressan12}. MQ20  derived a minimum mass for the cluster hosting RSG stars of  3~$\times$~10$^{5}$~M\textsubscript{\(\sun\)}.  On the other hand,  based on the integrated flux at the 3.6\,$\mu m$ IRAC band of ($1.50\pm 0.07$)\,$\times$\,10$^{-5}$\,Jy (J. M. Cannon et al., in preparation)  and assuming a mass-to-luminosity ratio of 0.47 \citep{mcgaugh14}, they report a  total stellar mass, M\textsubscript{\(\ast\)}, of 7.3\,$^{+2.2}_{-4.3}$\,$\times$\,10$^{5}$\,M\textsubscript{\(\sun\)}. Using the brightness of the Tip of the Red Giant Branch (TRGB) in their CMD as a distance indicator,  MQ20 obtained a distance modulus of 30.40~$^{+0.31}_{-0.60}$\,mag, which corresponds to a distance of 12.1~$^{+1.7}_{-3.4}$\,Mpc. This value  places Leoncino in an under-dense galaxy environment at the position of void number 12 of the Pustilnik catalogue \citep{pustilnik19}. Likewise,  using the TRGB method on single-star photometry with HST data \citet[]{tg19} calculated  the  distances to 18 dwarf galaxies from the Arecibo survey, including Leoncino, which was found to be located at 8.8\,Mpc, marginally consistent, with the MQ20 determinations. 

\begin{figure}
\includegraphics[width=0.45\textwidth,angle=0]{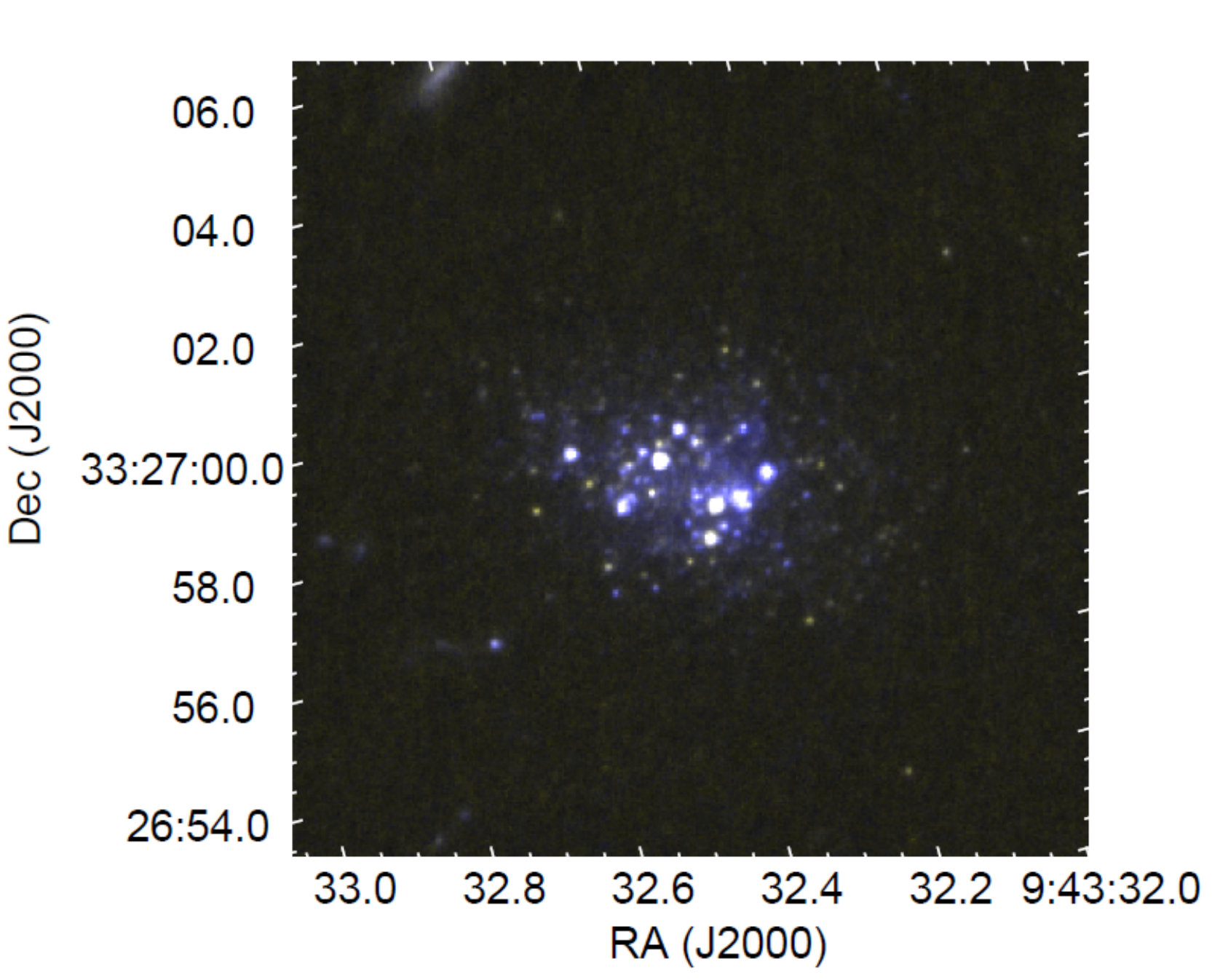}
\caption{Image of the galaxy Leoncino based on HST/WFC3-UVIS data. The F606W image is used as blue channel while the F814W image is used for both the green and red channels. The plate scale of these HST images is 0.04\,arcsec\,pixel$^{-1}$. This image shows a region of $14\, \times 14$\, arcsec$^2$ around the target coordinates,  slightly larger than the field-of-view covered by our MEGARA observations, $12.5\, \times\, 11.3$\, arcsec$^2$.}
\label{HST-falsecolor}
\end{figure}

\begin{figure}
\includegraphics[width=0.45\textwidth,angle=0]{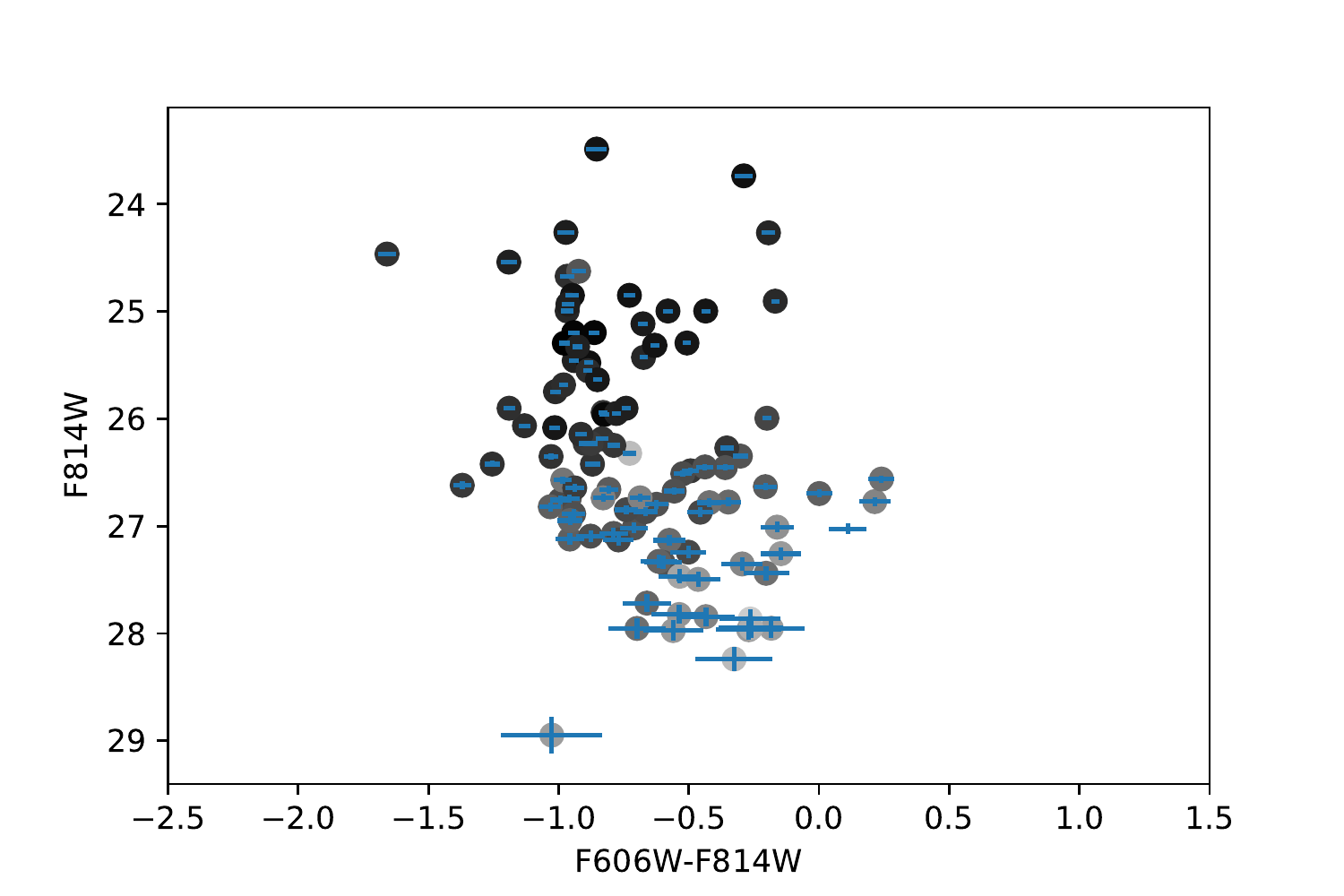}
\caption{CMD of a region of 6\,arcsec (150 pixels) in radius around the central coordinates of the galaxy. The figure shows, color-coded, the sources detected as a function of the distance to the galaxy center (as derived by MQ20 from HST images), darker shades indicate smaller galactocentric distances. The stars with values of F814W lower than 26.5 mag are basically outside the central 2\,arcsec (50 pixels). The magnitudes are given in STMAG system.}
\label{HST-CMD}
\end{figure}

Fig.\,\ref{HST-falsecolor} shows an image of the galaxy that  we obtained from HST archive  using the F606W image as blue channel and the F814W image for both the green and red channels. This \-image shows a region of $14\times14$\,arcsec$^2$ around the target center coordinates, 
$\alpha~$(J2000.0 FK5)\,=\, 09$^{\mathrm{h}}$\,43$^{\mathrm{m}}$\,32.40$^{\mathrm{s}}$ and $\delta~$(J2000.0 FK5) = $+33^{\circ} 26^{\prime} 57.9^{\prime\prime}$, with a plate scale of 0.04\,arcsec\,pixel$^{-1}$.
We had re-done the single-star photometry analysis using the drizzle images available through MAST and running the {\sc daofind} and {\sc phot iraf } tasks to obtain V (F606W) and I (F814W) magnitudes for all sources detected above 5$\sigma$ (4$\sigma$) in the combined F606W (F814W) image. Fig.\,\ref{HST-CMD} shows the resulting CMD of a 6\,arcsec (150 pixels) radius region around the center of the galaxy  (we have taken the one derived by MQ20 from HST images). The diagram shows in gray-scale the sources detected as a function of the distance to the galaxy center, where points of darker shade correspond to stars placed closer to the galaxy center. The stars with values of F814W lower than 26.5\,mag are basically outside the central 2\,arcsec (50 pixels). The magnitudes are given in STMAG system. A bright and blue main-sequence is clearly visible along with some RSG stars. Outside the central 2\,arcsec, equivalent to 110\,pc at the distance derived by MQ20, less massive red stars are detected. This could be indicative of the presence of an intermediate-aged stellar population in the outskirts of this system. These results are fully in agreement with the ones derived by MQ20.

Leoncino is located at a projected distance of 46\,kpc from UGC\,5186 galaxy, and with an offset of only 35\,km\,s$^{-1}$, which led MQ20 to suggest a possible interaction between both galaxies, further supported by the detection of \ion{H}{i}  between them. Based on this fact and the positioning of Leoncino in the M -- Z relation, they proposed that Leoncino has been experimenting an inefficient star formation history overall, dominated by individual episodic bursts whose associated galactic winds  would be the cause of its low metal abundance. The most recent star formation bursts would have been triggered as the result of a minor interaction with its neighbor UGC\,5186. 
Regarding its gas mass content, at the  distance determined by MQ20, M$_\ion{H}{i}$ results in $1.83\times 10^{7}$\,M\textsubscript{\(\sun\)}.  As mentioned, these authors report a   value  of 12\,+\,log(O/H)\,=\,7.12\,$\pm\,$0.04 from an observation with the Large Binocular Telescope (LBT).  Recently,  \citet{aver21}  revised this  value to 12\,+\,log(O/H)\,=\,7.06\,$\pm\,$0.03, obtained from the same LBT observation.


We used the results from  MQ20 for the gas mass of $1.83\times 10^{7}$\,M\textsubscript{\(\sun\)} and stellar mass ranging between 6.5 and 7.3\,$\times\, 10^{5}$\,M\textsubscript{\(\sun\)} to calculate the predicted oxygen mass abundance, Z$_{\rm O}$, in a closed-box chemical evolutionary model. Z$_{\rm O}$ is determined through the equation Z$_{\rm O}$ = p$_{\rm O}\times \ln{\mu}^{-1}$, where p$_{\rm O}$ is the oxygen integrated yield and $\mu$ is the ratio between the gas and the total (stellar and gas) mass.  By using the integrated stellar yields from \citet{mol15}, we obtain  a value of Z$_{\rm O}$ between 2.3 $\times 10^{-5}$ and 8.4 $\times 10^{-4}$, which would correspond to  12\, $+$\, log(O/H) between 6.28 and 7.85 depending on the combination IMF + stellar yields used for the integrated yield, p$_{\rm O}$.  The average predicted oxygen abundance using the whole set of 144 integrated yields from \citet{mol15} is 12\,+\,log(O/H)\,$=$\,7.17, which corresponds to the mode of the distribution obtained for all of them, being 6.28 and 7.85 the minimum and the maximum, respectively. Therefore, any oxygen abundance between the two is possible, with a higher probability in the range between 6.80 and 7.40, and being the most likely value 7.17 (7.15  when using the IMF of Kroupa \citep{Kroupa2001} and Z\,=\,0.0004, for which the integrated yield varies between 0.001 and 0.007), close to the reported abundances. These  closed-box model results  are consistent with a genuine low metallicity in Leoncino.

\section{MEGARA observations and data analysis}
\label{Sec:Observationsandanalyses}

We present and discuss IFU observations of Leoncino galaxy obtained with MEGARA at  GTC. MEGARA is an optical (3650 -- 9750~\AA) fiber-fed spectrograph. The instrument offers two modes: integral-field and multi-object spectroscopy. The  IFU provides a field of view (FoV) of 12.5\,$\times$\,11.3\,arcsec$^2$, plus eight additional mini-bundles, located at the edge of the FoV, for simultaneous sky subtraction. The spatial sampling is 0.62\,arcsec per fiber\footnote{This size corresponds to the diameter of the circle on which the hexagonal spaxel is inscribed}, thanks to the combination of a 100~$\upmu$m core fibre coupled to a microlens that converts the f/17 entrance telescope beam to f/3, to   minimize  the fibers focal ratio degration and  to  maximize the  efficiency. The spaxel projection is over-sized relative to the fiber core for a precise fiber-to-fiber flux uniformity. A fiber link 44.5\,m in length drives the light, coming from the folded Cassegrain focal plane into the spectrograph, which is placed at the GTC Nasmyth A platform. The spectrograph includes a set of 18 Volume Phase Holographic (VPH) gratings, offering three spectral modes with  resolving power of R$_{\mathrm{FWHM}}$\,$\approx$\,6\,000, 12\,000 and 20\,000, for low, medium   and high resolution, respectively. The scientific data are recorded by a deep-depleted Teledyne-e2V 4096\,pix\,$\times$\,4096\,pix detector with 15\,$\upmu$m-pixel pitch. The instrument design and final performance on the GTC based on commissioning results can be found in \citet{carspie18}, \citet{gilspie18}, and Gil de Paz et al. (2021, submitted). 

\subsection{Observations and Data Reduction}
\label{Sec:Observations}

We observed Leoncino using as target coordinates $\alpha~$(J2000.0 FK5)\,=\, 09$^{\mathrm{h}}$\,43$^{\mathrm{m}}$\,32.36$^{\mathrm{s}}$ and $\delta~$(J2000.0 FK5) = $+33^{\circ} 26^{\prime} 57.6^{\prime\prime}$. The data were taken in the dark, clear night of February 2$^{\mathrm{nd}}$ 2019.  We used the spectrograph configurations given by the VPH gratings labelled as LR-B and LR-R. LR-B ranges from 4304 to 5198\,\AA\, with a reciprocal linear dispersion $\delta\lambda=0.23$\,\AA\,pixel$^{-1}$ and a FWHM spectral resolution element  $\Delta\lambda=0.92$\,\AA, while the LR-R setup covers the interval from 6098 to 7306\,\AA\, with $\delta\lambda=0.32$\,\AA\, pixel$^{-1}$ and $\Delta\lambda=1.24$\,\AA. The  average seeing was 0.8\,arcsec and 0.6\,arcsec for the LR-B and LR-R observations, respectively. Six 1200\,s images (7200\,s) and three 900\,s exposures (2700\,s) were taken with the LR-B and LR-R setups, respectively. 

The observations had their associated calibrations: bias, halogen, ThAr (for LR-B) and ThNe (for LR-R) lamps, obtained for bias subtraction and modelling, tracing and flatfielding, and wavelength calibration, respectively. These auxiliary images were obtained in daytime and in each case three identical exposures were done to facilitate the cosmic ray removal. Observations of the spectrophotometric standard star HR3454 were made to generate the response function for absolute flux calibration. The data were reduced with the MEGARA Data Reduction Pipeline, DRP, a {\sc python} based software tool operating in command-line \citep{carandpas18, pasetal18}. 

The resulting product of the DRP is a Row Stacked Spectra (RSS) FITS file, which contains the individual flux-calibrated spectra for all 623 fibers in the IFU. The MEGARA DRP also produces a final RSS image in which the sky has been subtracted by combining the signal of the seven (or a subsample of them) sky mini-bundles placed along the IFU pseudo-slit. We applied this standard sky-subtraction procedure for the LR-B data. The Quick Look Analysis ({\sc qla} tool), developed by \citet[]{gomezalv18}, allows the user to choose the sky bundles, optimizing the   spectra extraction and visualization. We followed this sky-subtraction method to obtain the LR-R final RSS combined image. Additionally, we used the {\sc qla} package to obtain reconstructed 2D images in specific spectral windows from the RSS spectra. 

From the comparison of the HST and MEGARA coordinates of the clusters, we derived the shift between the two images and updated the headers of the MEGARA synthetic images to match the reference coordinates and orientation of the LR-R-based and HST reference images. A final minor astrometric correction  was performed. 

To ensure a correct absolute astrometry, the availability of HST imaging was crucial as there were no stars within the field of MEGARA IFU. The main sources of information come from the LR-R continuum image, where two knots can be detected matching the position of the brightest continuum sources in the HST F606W image, and the H$\alpha$ and H$\beta$ flux images.
First, we aligned our LR-R continuum image with a precision that we estimate to be below half a spaxel ($\sim$0.3\,arcsec). This automatically corrected the H$\alpha$ flux data as they come from the same spectral setup as our 6590\,--\,7100\,\AA\ continuum image. We then aligned the H$\beta$ flux image to the H$\alpha$ one (the low dust reddening ensures that potential offsets due to differential extinction should be negligible at these scales). This relative alignment is needed since the noise in the LR-B continuum reconstructed image is too high and no HST image bluer than F606W is available.
The two setups turned out to be offset by 1.35\,arcsec in right ascension  and 0.35\,arcsec in declination. So that the coordinates (J2000.0 FK5) for the center of MEGARA images are: $\alpha~$\,=\, 09$^{\mathrm{h}}$\,43$^{\mathrm{m}}$\,32.40$^{\mathrm{s}}$; $\delta~$\,=\,+33$^{\circ} 26^{\prime} 57.56^{\prime\prime}$. The centers of the brightest spaxels in the Balmer lines in LR-B and LR-R are $\alpha~$\,=\, 09$^{\mathrm{h}}$\,43$^{\mathrm{m}}$\,32.59$^{\mathrm{s}}$;  $\delta~$ = $+33^{\circ} 26^{\prime} 55.89^{\prime\prime}$  and $\alpha~$\,=\, 09$^{\mathrm{h}}$\,43$^{\mathrm{m}}$\,32.62$^{\mathrm{s}}$;  $\delta~$ = $+33^{\circ} 26^{\prime} 55.49^{\prime\prime}$  respectively, coincident at the level of half a spaxel and coherent within the seeing values.

We remark that the continuum emission level derived from the analysis of the LR-R data is of the order of 10$^{-6}$\, Jy\,spaxel$^{-1}$, 
which matches the $\sim$\,4$\times10^{-6}$\,Jy\,arcsec$^{-2}$ surface brightness level in the HST image (0.05\,electrons\,s$^{-1}$\,pixel$^{-1}$) for that same region, given that the MEGARA IFU spaxel size is 0.25\, arcsec$^{2}$. Here, we have obtained the conversion from  electrons\,s$^{-1}$ to Jy in the HST data using the PHOTFNU keyword from the F606W image.

The corresponding astrometry correction was  applied to all the results in this paper involving absolute positions (e.g. maps or comparison of the MEGARA-IFU and HST morphologies). 

\section{RESULTS}
\label{Sec:Results}

We produced several 2D maps throughout the galaxy. The  description  of the maps and the  gas kinematics are presented  in subsections \ref{maps} and \ref{kinematics}, respectively. We  performed extractions at  different apertures, i.e. in an accumulative way,  to show how the emission line fluxes and lines ratios change  due aperture effects, as described in subsection \ref{aperture-effect}. To analyze the emission line spectrum  in different ionization rings we carried out a spaxel-to-spaxel extraction, using the {\sc qla} package and  the results are presented in   subsection~\ref{ionization-rings}

In all the spectra we measured both, the line fluxes and their errors, using three different analysis packages: {\sc figaro}, {\sc  qla tool} and the utility {\sc analyzed-spectrum} of  {\sc megaratools} \citep {gilspie18b}. The three utilities produced the same values of the emission lines within the given errors. In case of discrepancy, we always used the largest error bar. The lines detected and measured wherever possible were in the LR-B spectra: H${\gamma}$, [\ion{O}{iii}]$\lambda$\,4363\AA\ (hereafter [\ion{O}{iii}]4363), \ion{He}{i}\,$\lambda$\,4471\AA\ (hereafter \ion{He}{i}\,4471), H${\beta}$, [\ion{O}{iii}]$\lambda$\,4959\AA\ (hereafter [\ion{O}{iii}]\,4959),
[\ion{O}{iii}]$\lambda$\,5007\AA\ (hereafter [\ion{O}{iii}]\,5007)
and \ion{He}{i}\,$\lambda$\,5015\AA\ (hereafter  \ion{He}{i}\,\,5015). In the LR-R spectra:  H${\alpha}$, [\ion{N}{ii}]$\lambda$\,6584\AA\ (hereafter [\ion{N}{ii}]\,6584),
\ion{He}{i}$\,\lambda$\,6678\AA\ (hereafter \ion{He}{i}\,6678), 
[\ion{S}{ii}]$\lambda$\,6717\AA\ (hereafter [\ion{S}{ii}\,6717),
[\ion{S}{ii}]$\lambda$\,6731\AA\ (hereafter [\ion{S}{ii}]$\lambda$\,6731),
\ion{He}{i}\,$\lambda$\,7065\AA\ (hereafter \ion{He}{i}\,7065),
[\ion{Ar}{iii}]$\lambda$\,7135\AA\ (hereafter [\ion{Ar}{iii}]\,7135) and
\ion{He}{i}\,$\lambda$\,7281\AA\ (hereafter \ion{He}{i}\,7281). 

For the brightest lines in each setup, i.e.  H${\beta}$, [\ion{O}{iii}]\,4959 and [\ion{O}{iii}]\,5007 in LR-B and H${\alpha}$, [\ion{N}{ii}]\,6584, [\ion{S}{ii}]\,6717 and [\ion{S}{ii}]\,6731 in LR-R,  we  systematically measured the emission-line properties including only spaxels with signal-to-noise per pixel (at the peak of the line) S/N\,$\geq$\,5. The emission-line properties, flux, recession velocity and velocity dispersion  correspond to the  zeroth, first, and second momenta of the best-fitting Gauss-Hermite line profile decomposition used by the  {\sc analyzed-rss} utility of  {\sc megaratools} \citep {gilspie18b}.

 For our analysis, we have considered  a distance to Leoncino  of 7.7\,Mpc (H16).  This gives  a scale of 22\,pc spaxel$^{-1}$ and 37\,pc\,arcsec$^{-1}$,  consistent with the radial velocity of our own emission line spectrum and similar to other published results. The implications of considering other values of the distance are explained in \ref{Sec:constrainingparameters}. In particular, the spatial resolution given by the spaxel size would change from the value we assume of 22\,pc for a distance of 7.7\,Mpc to 42\,pc for a 13.8\,Mpc distance. In the tables and figures of this section, the observed  projected distances are indicated in arcsec  to facilitate a reference for scale conversion if using a different distance to Leoncino.

 

\begin{figure*}
\includegraphics[width=0.48\textwidth]{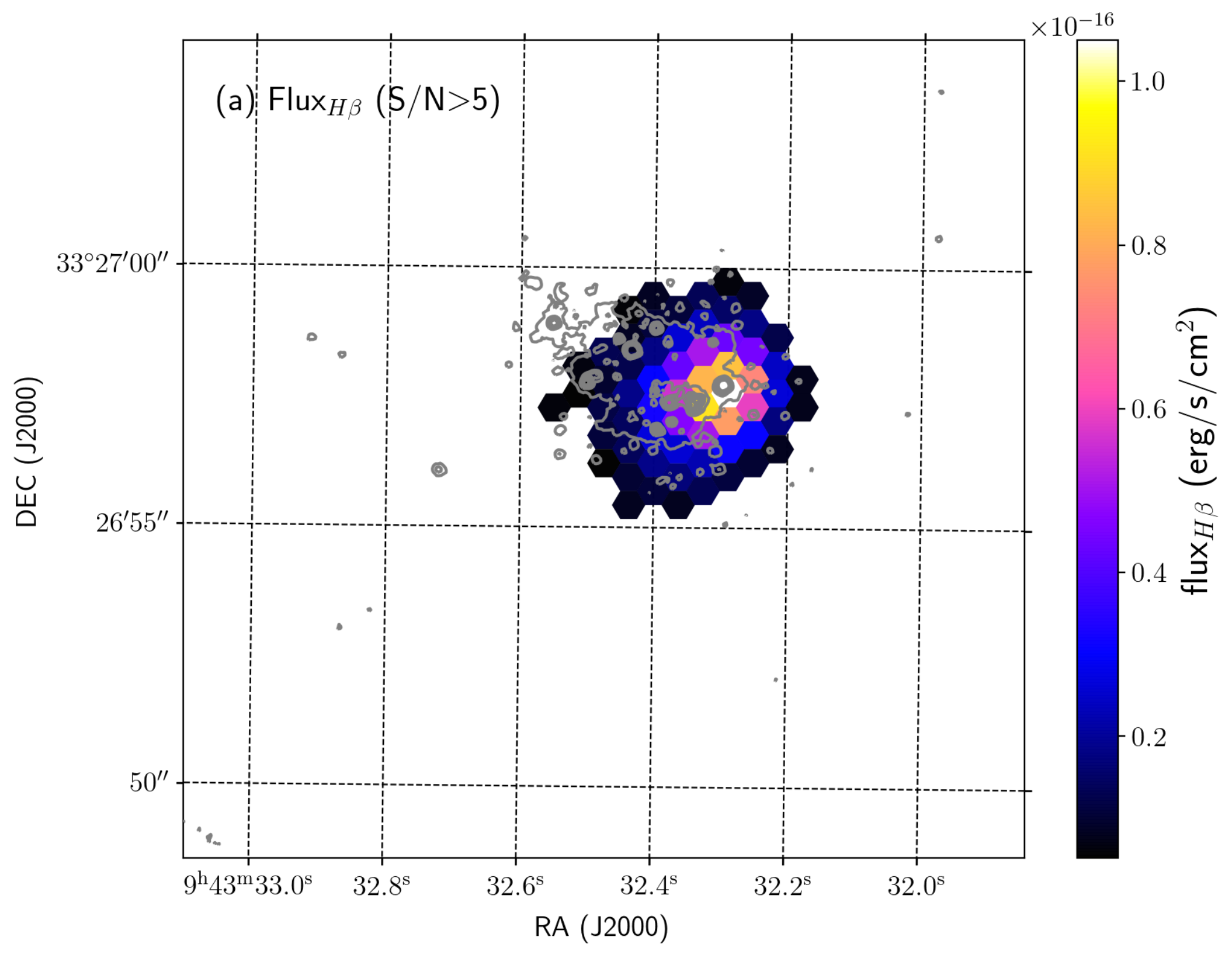}
\includegraphics[width=0.48\textwidth]{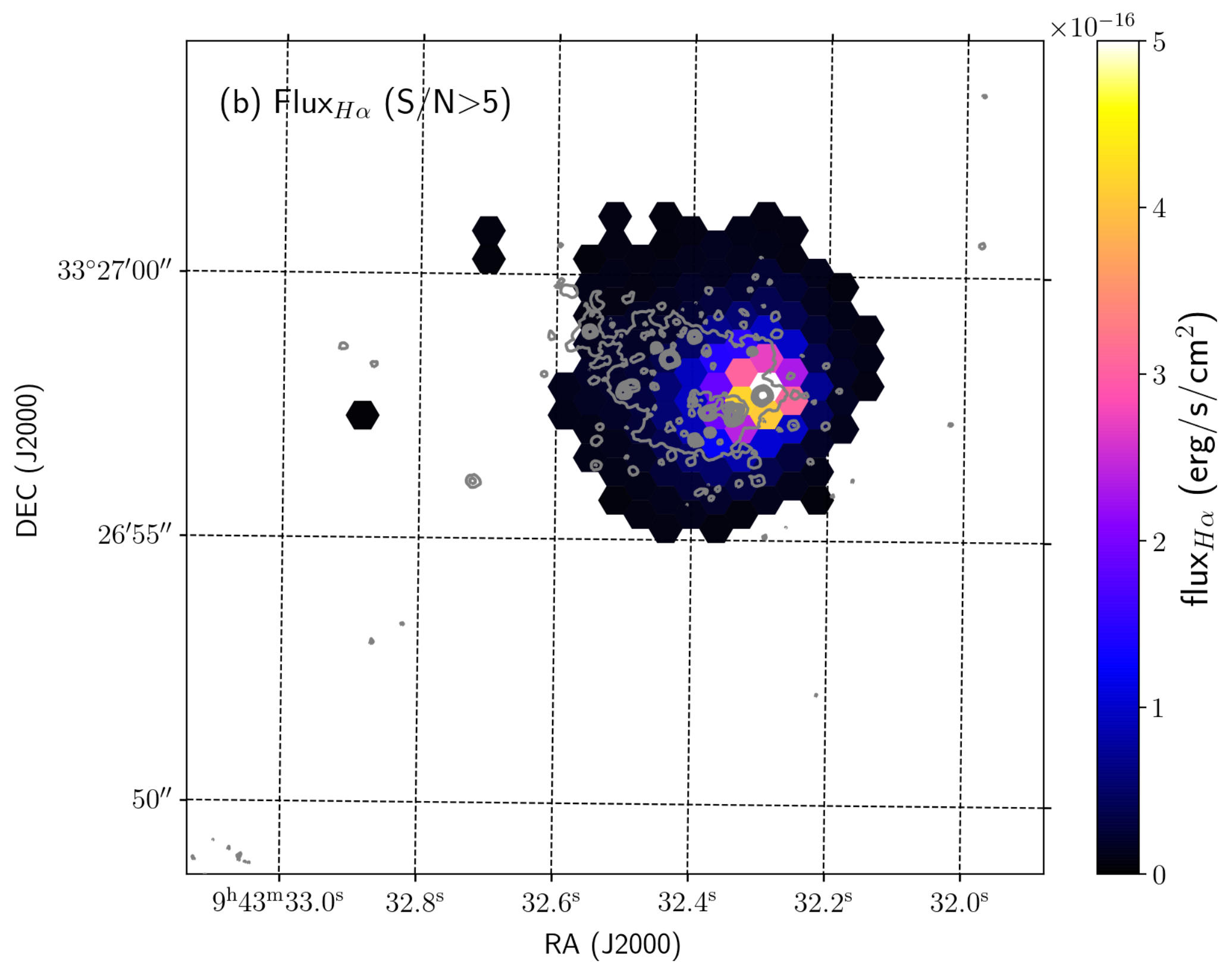}
\includegraphics[width=0.48\textwidth]{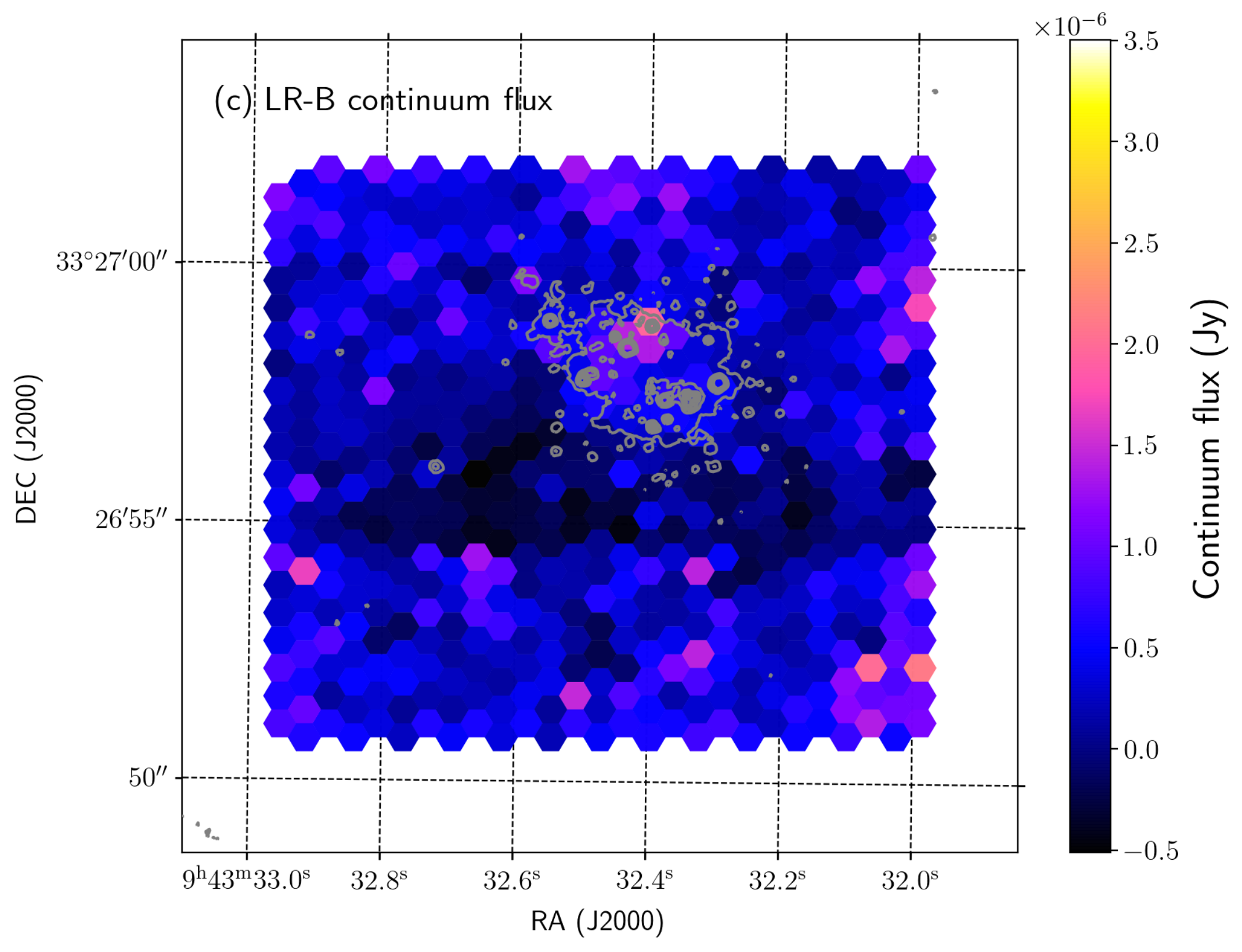}
\includegraphics[width=0.48\textwidth]{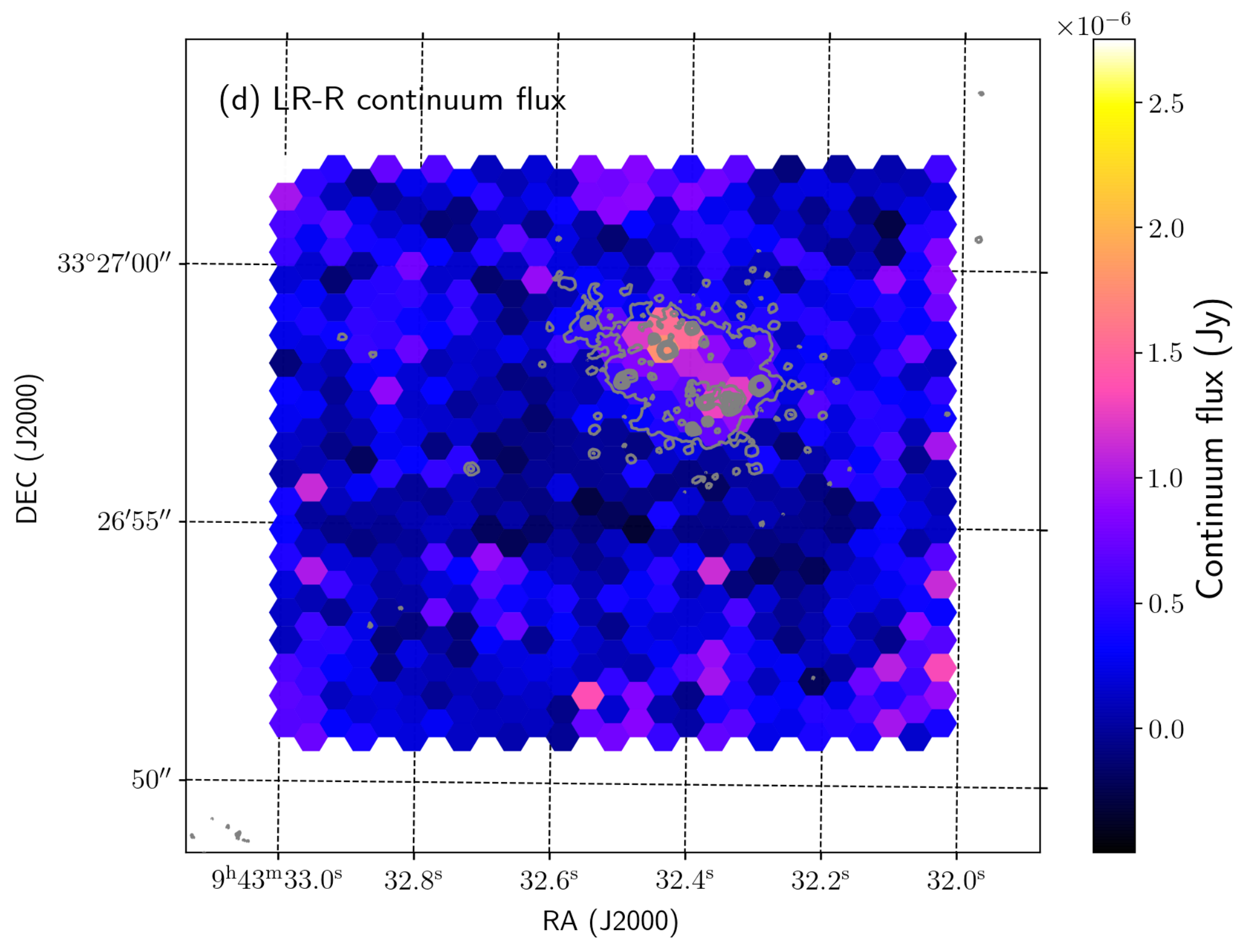}
\includegraphics[width=0.48\textwidth]{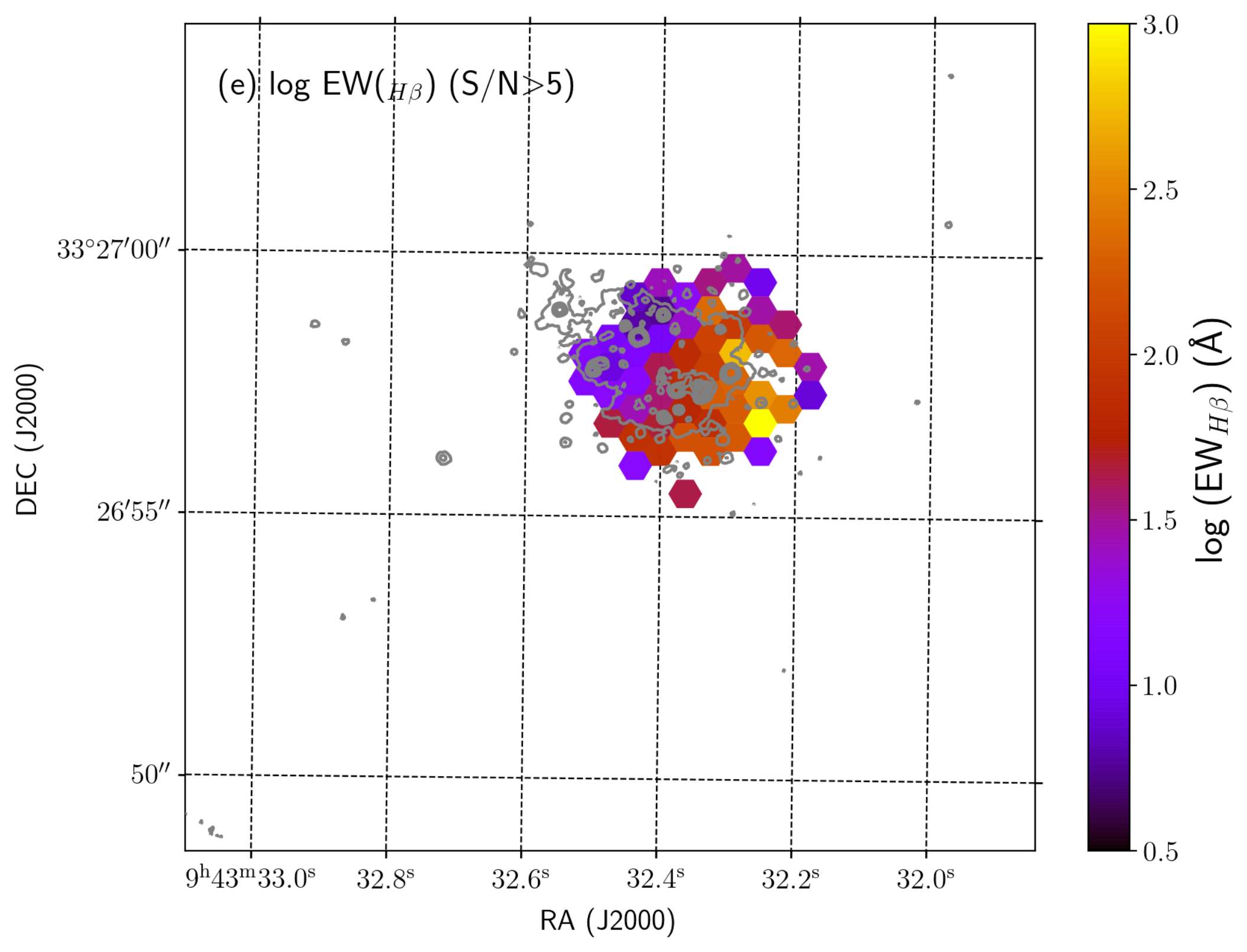}
\includegraphics[width=0.48\textwidth]{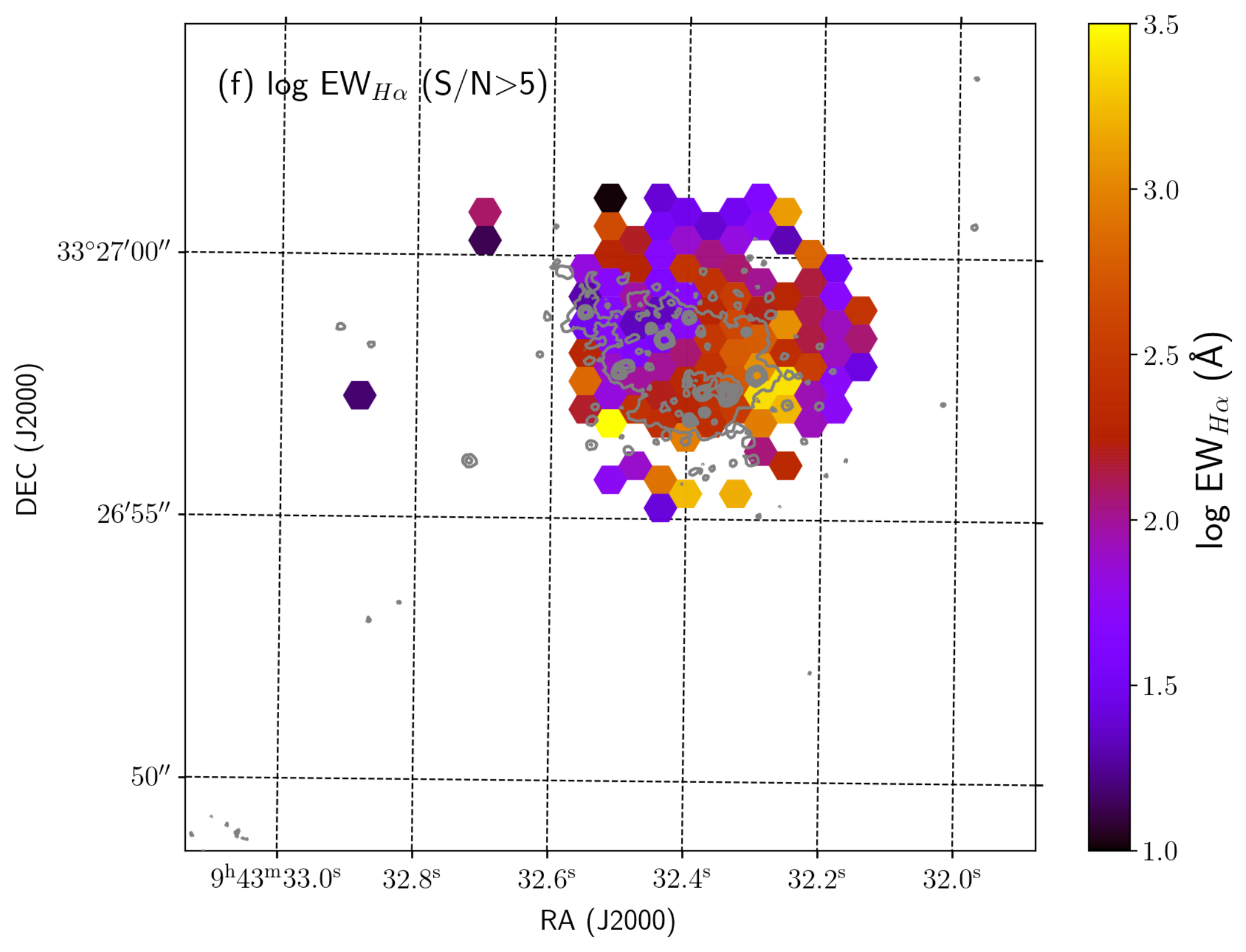}
\caption{
Top panels: emission line flux maps. Middle panels: maps (in Jy) for the H${\beta}$ continuum averaged between 4400 and 4850\,\AA\ (left) and for the H${\alpha}$ continuum averaged between 6590 and 7100\,\AA\ (right).  Bottom panels: logarithm of the EW (in \AA) of H${\beta}$ (left) and  H${\alpha}$ (right). The actual FoV of the entire MEGARA IFU can be seen outlined in the continuum plots. In all cases the contours come from the HST archival image obtained with the WFC3-UVIS/HST instrument in the F606W band and correspond to 0.02, 0.07, 0.12, 0.17, 0.22, 0.27, 0.32 and 0.37 electrons\,s$^{-1}$ or 0.23, 0.8, 1.4, 1.9, 2.5, 3.1, 3.6 and 4.2 $\times$ 10$^{-20}$ erg\,s$^{-1}$\,cm$^{-2}$\,\AA$^{-1}$.}
\label{HalphaHbetaEW}
\end{figure*}

\begin{figure*}
\includegraphics[width=0.48\textwidth]{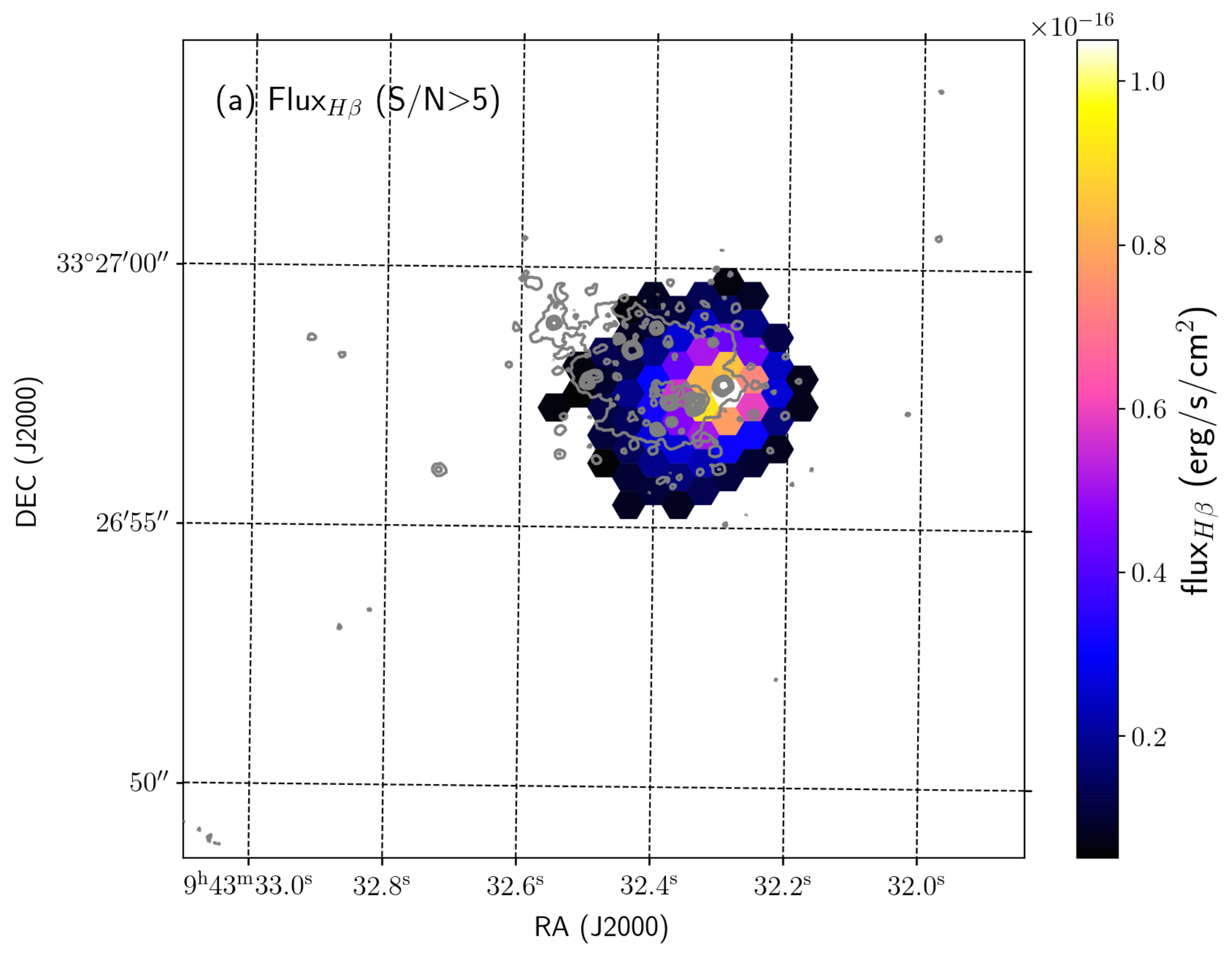}
\includegraphics[width=0.48\textwidth]{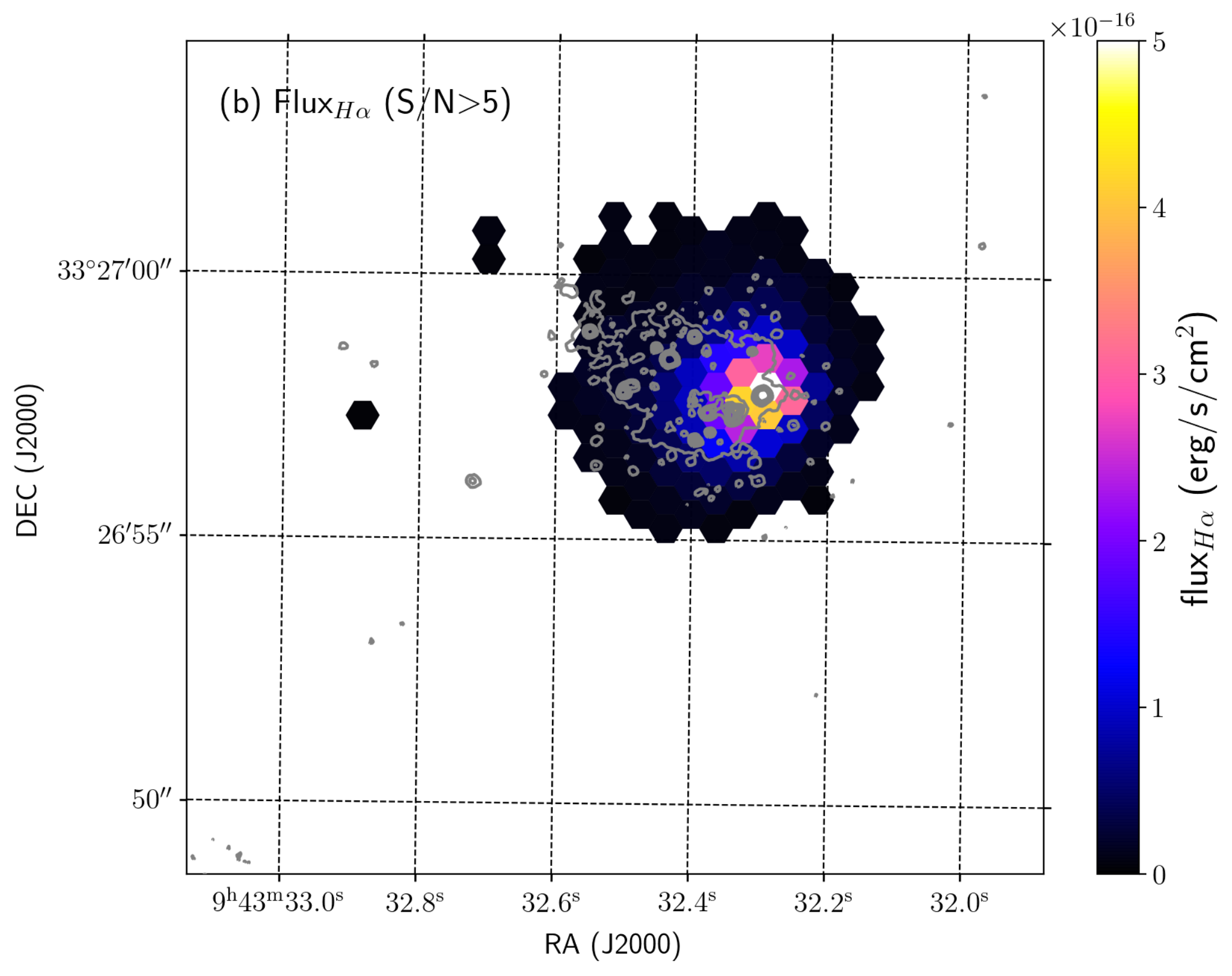}
\includegraphics[width=0.48\textwidth]{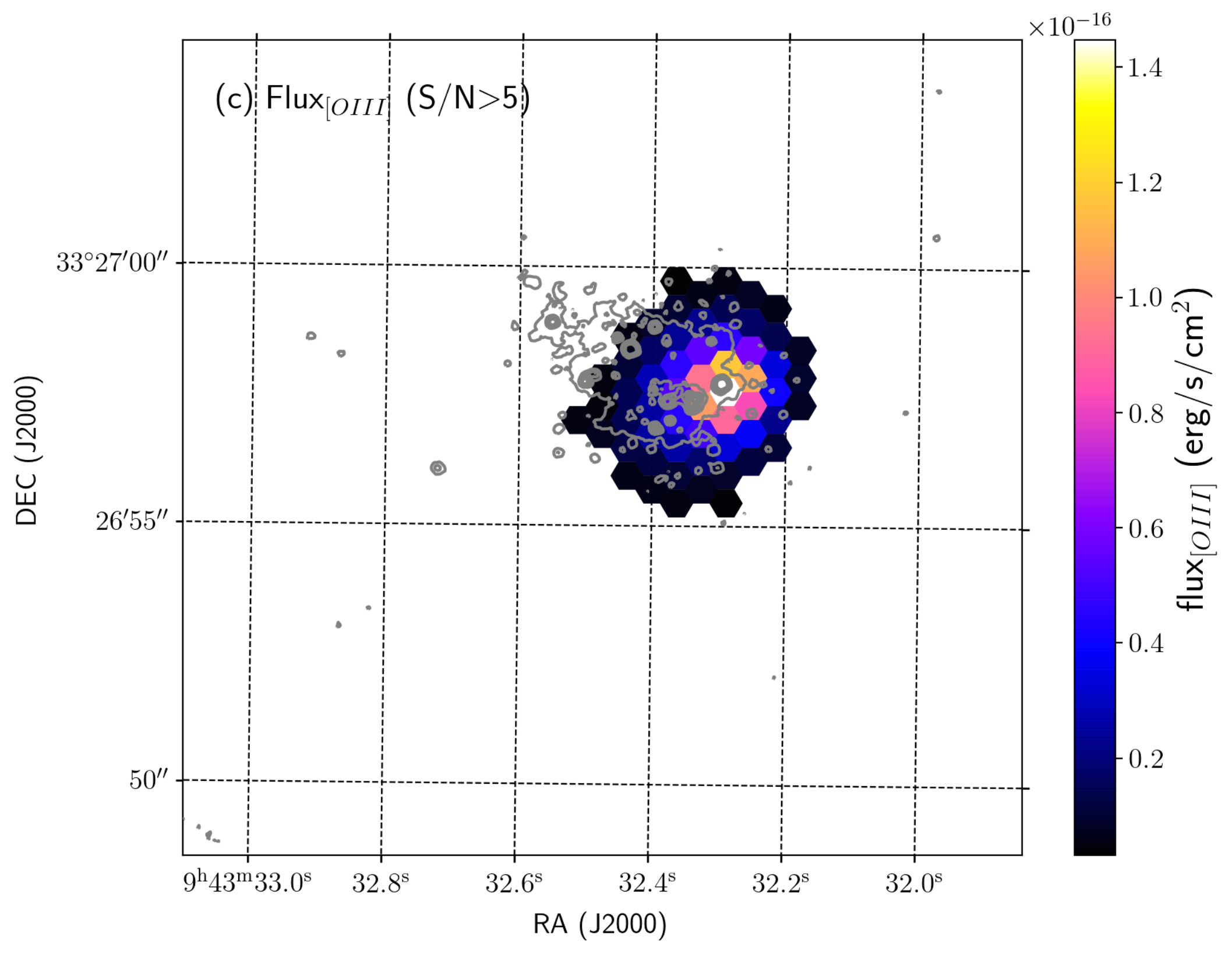}
\includegraphics[width=0.48\textwidth]{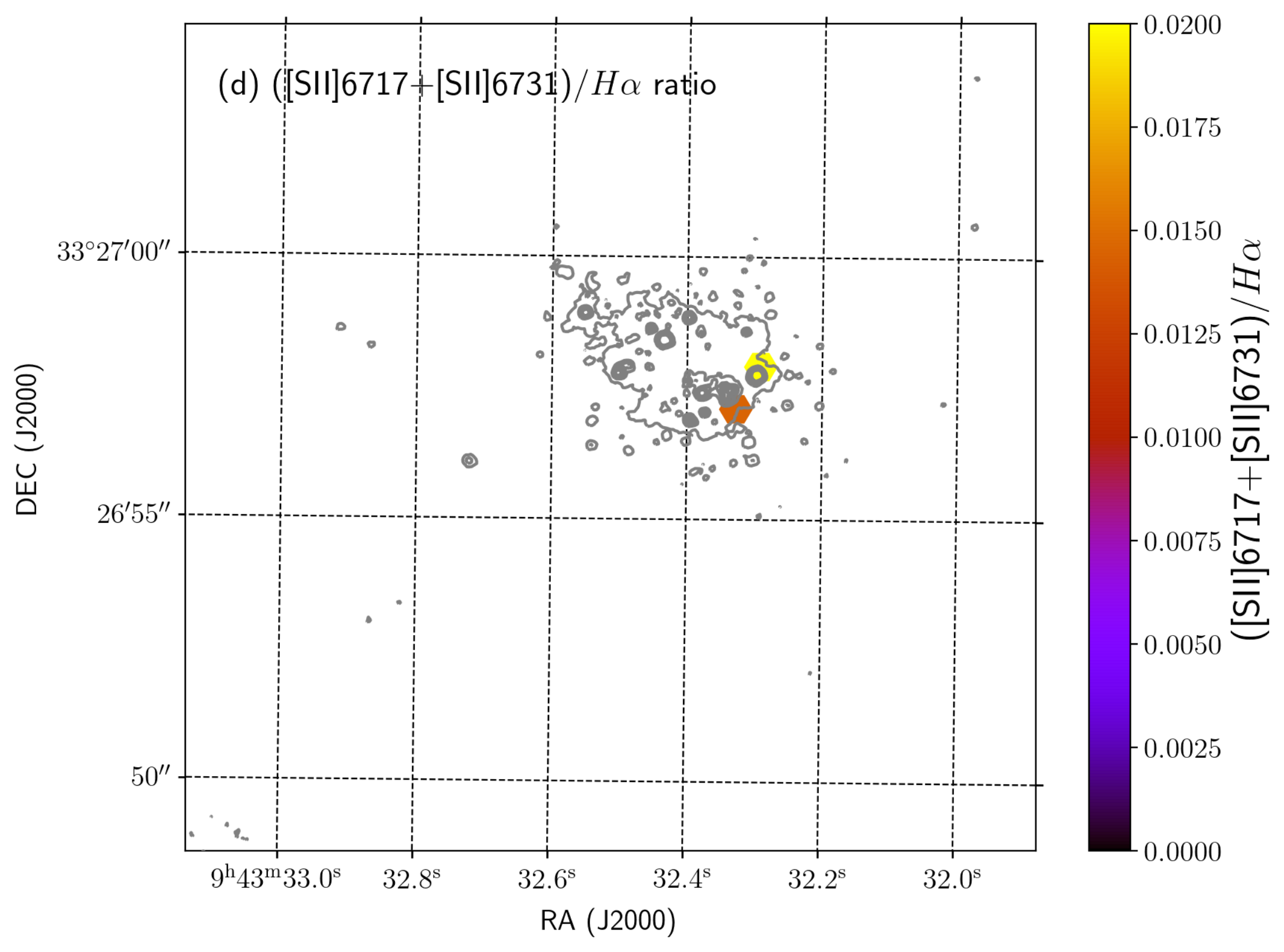}
\includegraphics[width=0.48\textwidth]{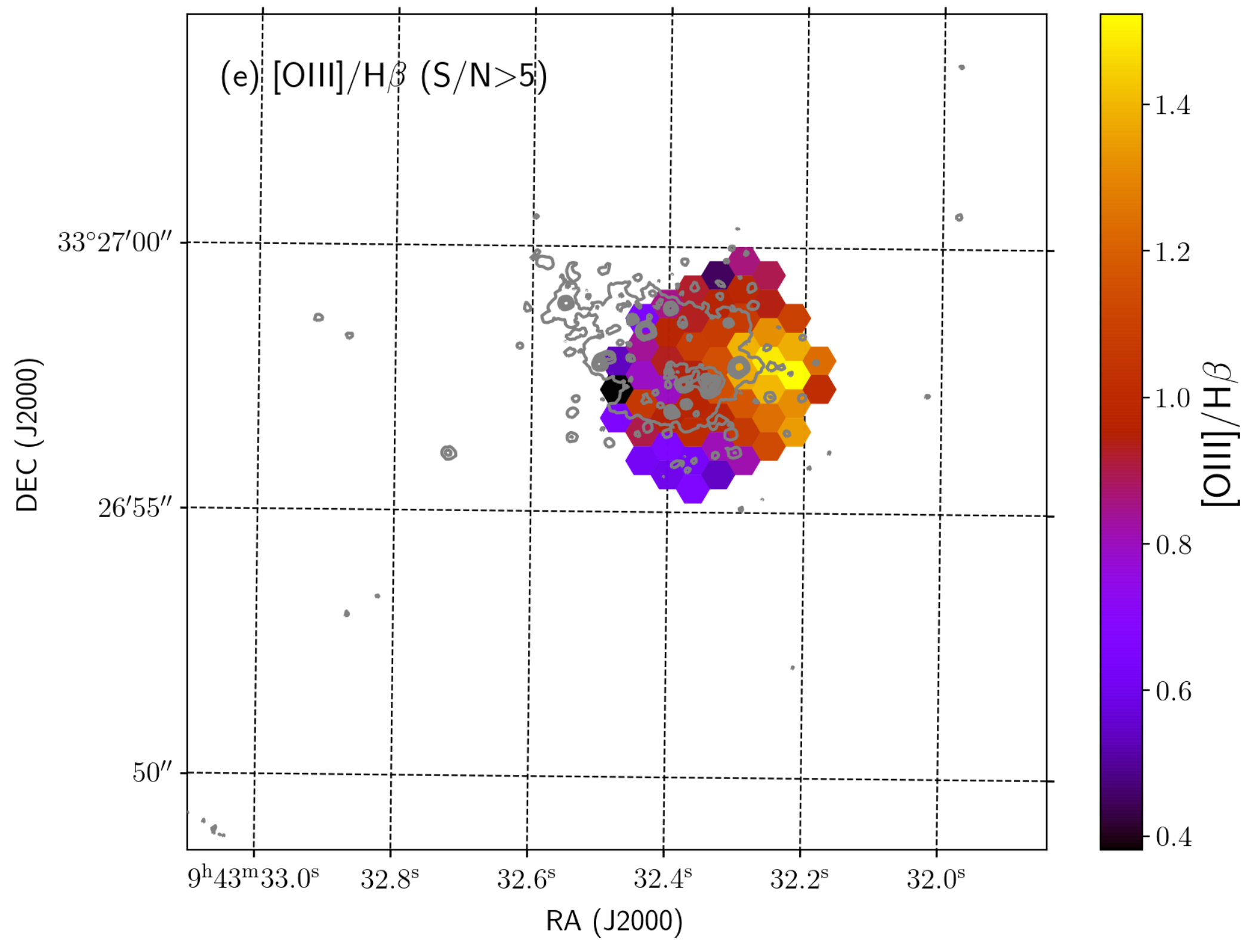}
\includegraphics[width=0.48\textwidth]{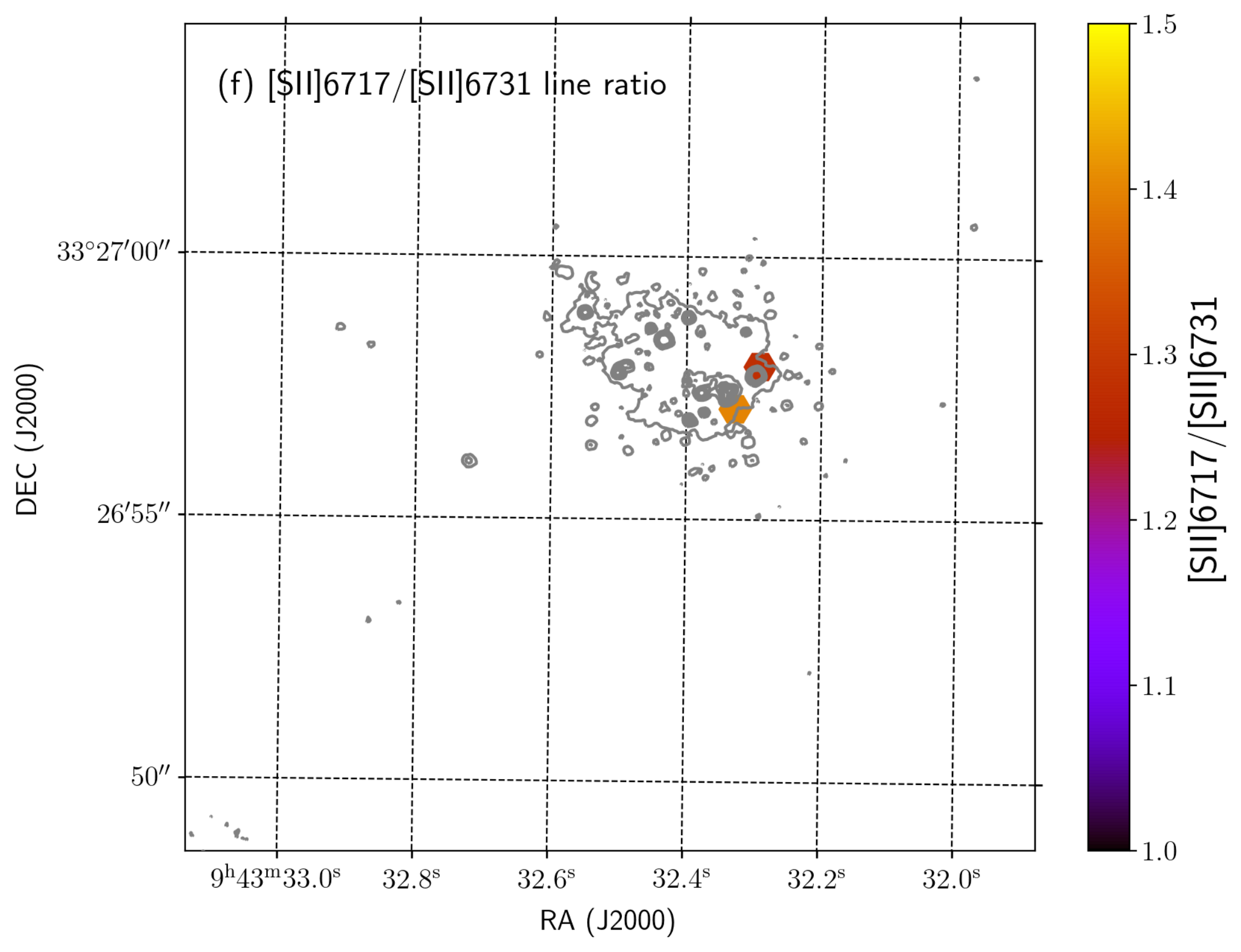}
\caption{
Left:  maps of  H${\beta}$ flux (top) and  [\ion{O}{iii}]\,5007 (middle) flux (in erg\,s$^{-1}$\,cm$^{-2}$) and [\ion{O}{iii}]\,5007/H${\beta}$ (bottom). Right: map of  H${\alpha}$ flux (in erg\,s$^{-1}$\,cm$^{-2}$) (top) and [\ion{S}{ii}]~6717+6731/H${\alpha}$ (middle) and [\ion{S}{ii}]\,6717/[\ion{S}{ii}]\,6731 (bottom).  In all cases the contours come from the HST archival image obtained with the WFC3-UVIS/HST instrument in the F606W band and correspond to 0.02, 0.07, 0.12, 0.17, 0.22, 0.27, 0.32 and 0.37 electrons\,s$^{-1}$ or 0.23, 0.8, 1.4, 1.9, 2.5, 3.1, 3.6 and 4.2 $\times$ 10$^{-20}$ erg\,s$^{-1}$\,cm$^{-2}$\,\AA$^{-1}$.}
\label{Emissionlines}
\end{figure*}

\begin{figure*}
\includegraphics[width=0.42\textwidth]{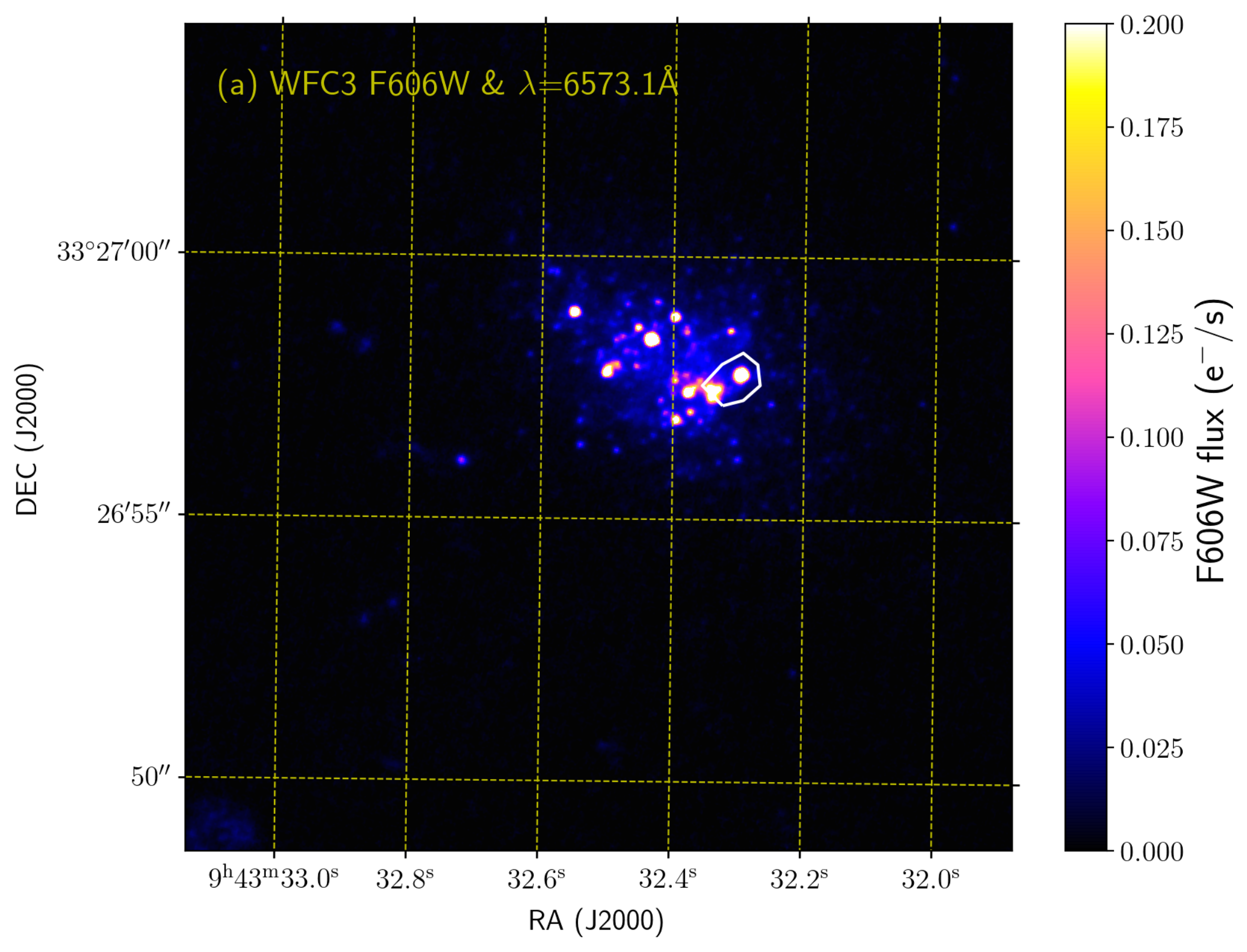}
\includegraphics[width=0.42\textwidth]{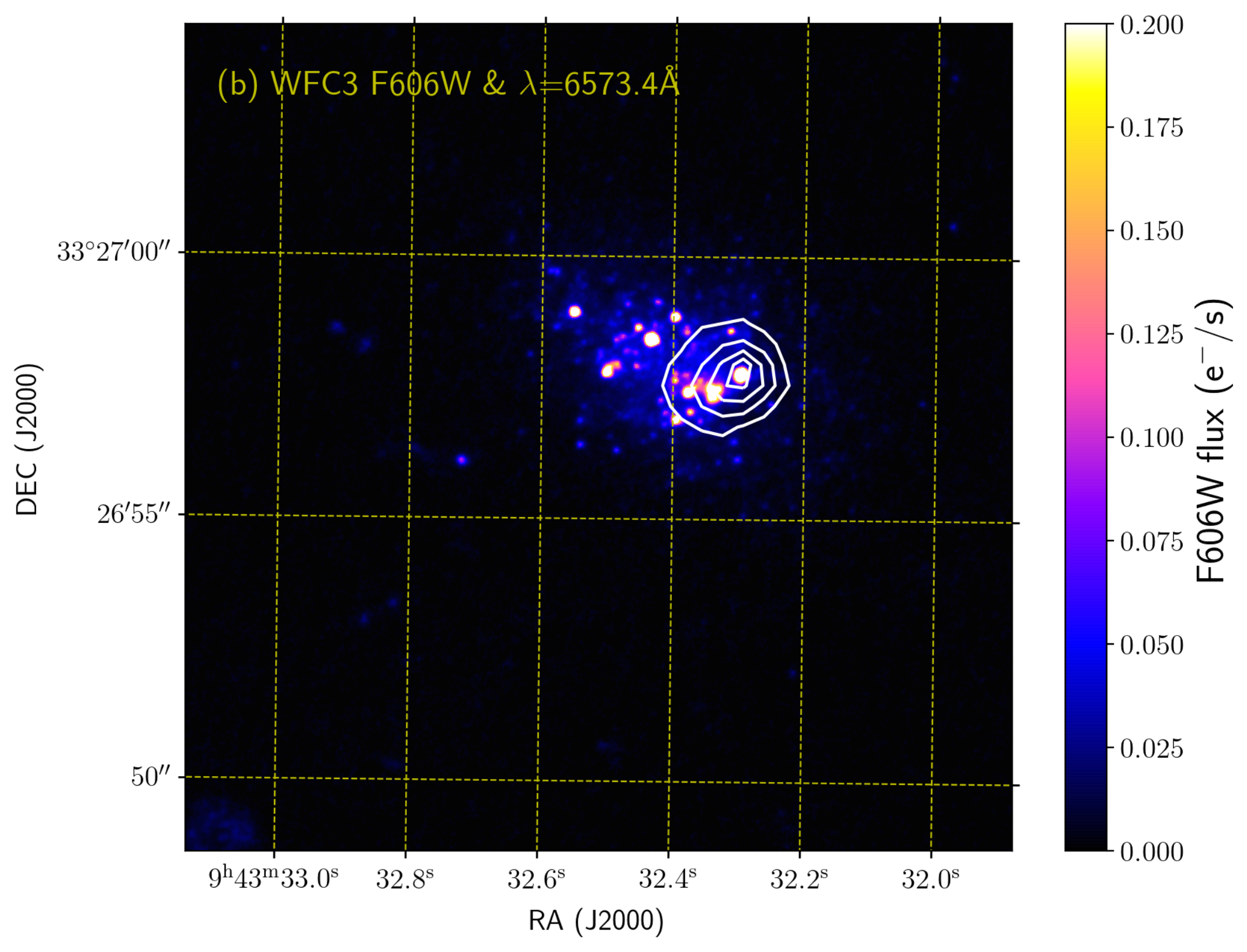}
\includegraphics[width=0.42\textwidth]{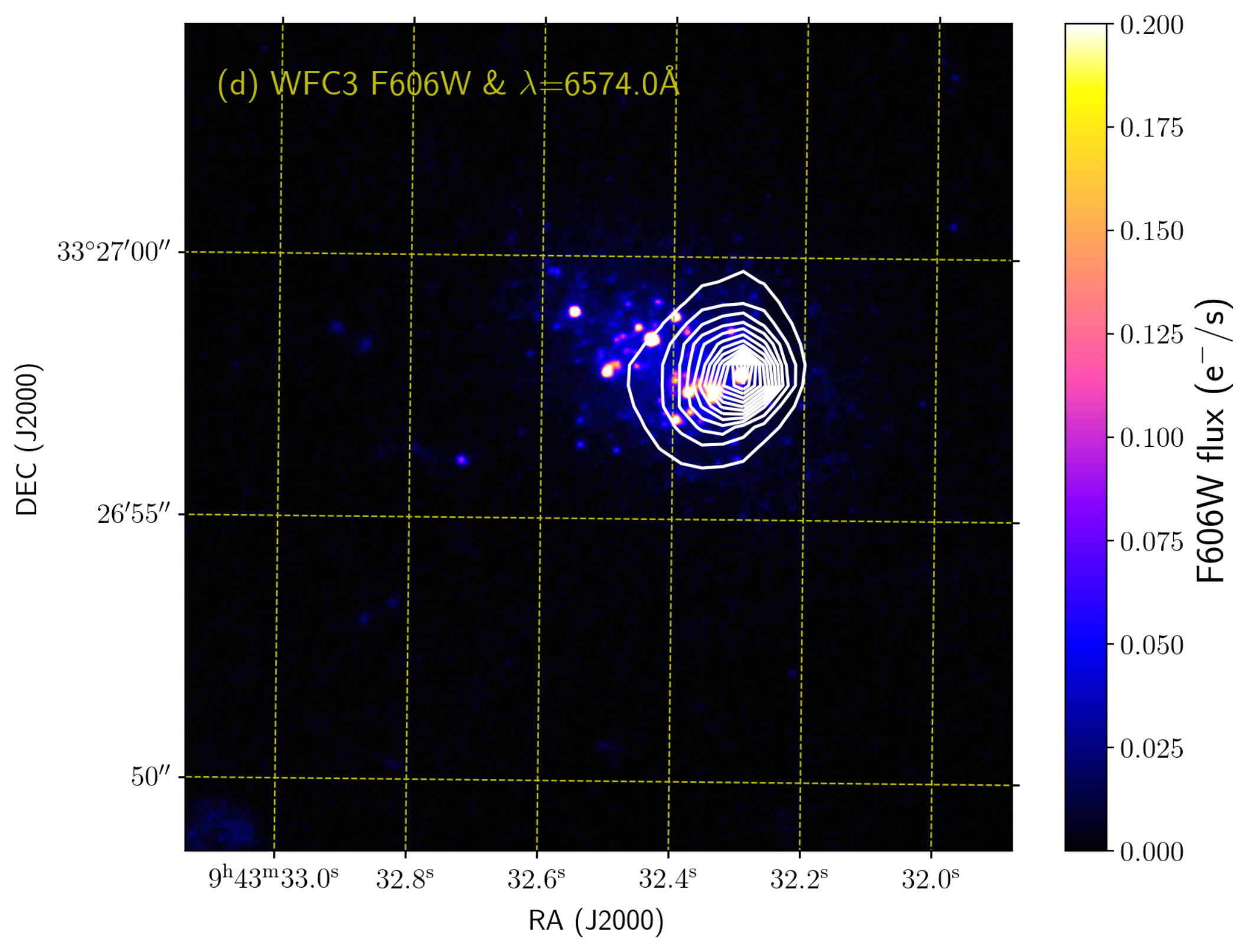}
\includegraphics[width=0.42\textwidth]{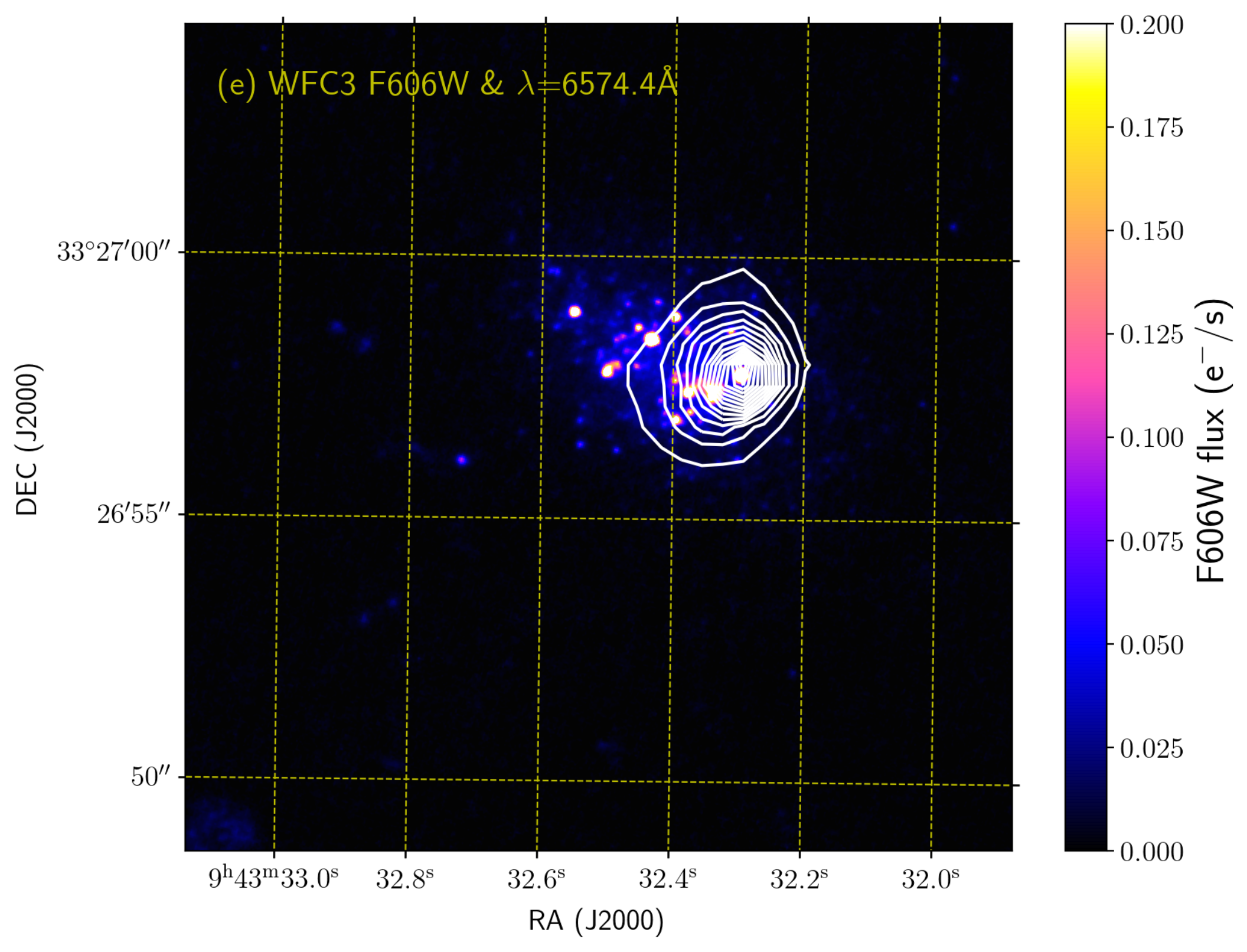}
\includegraphics[width=0.42\textwidth]{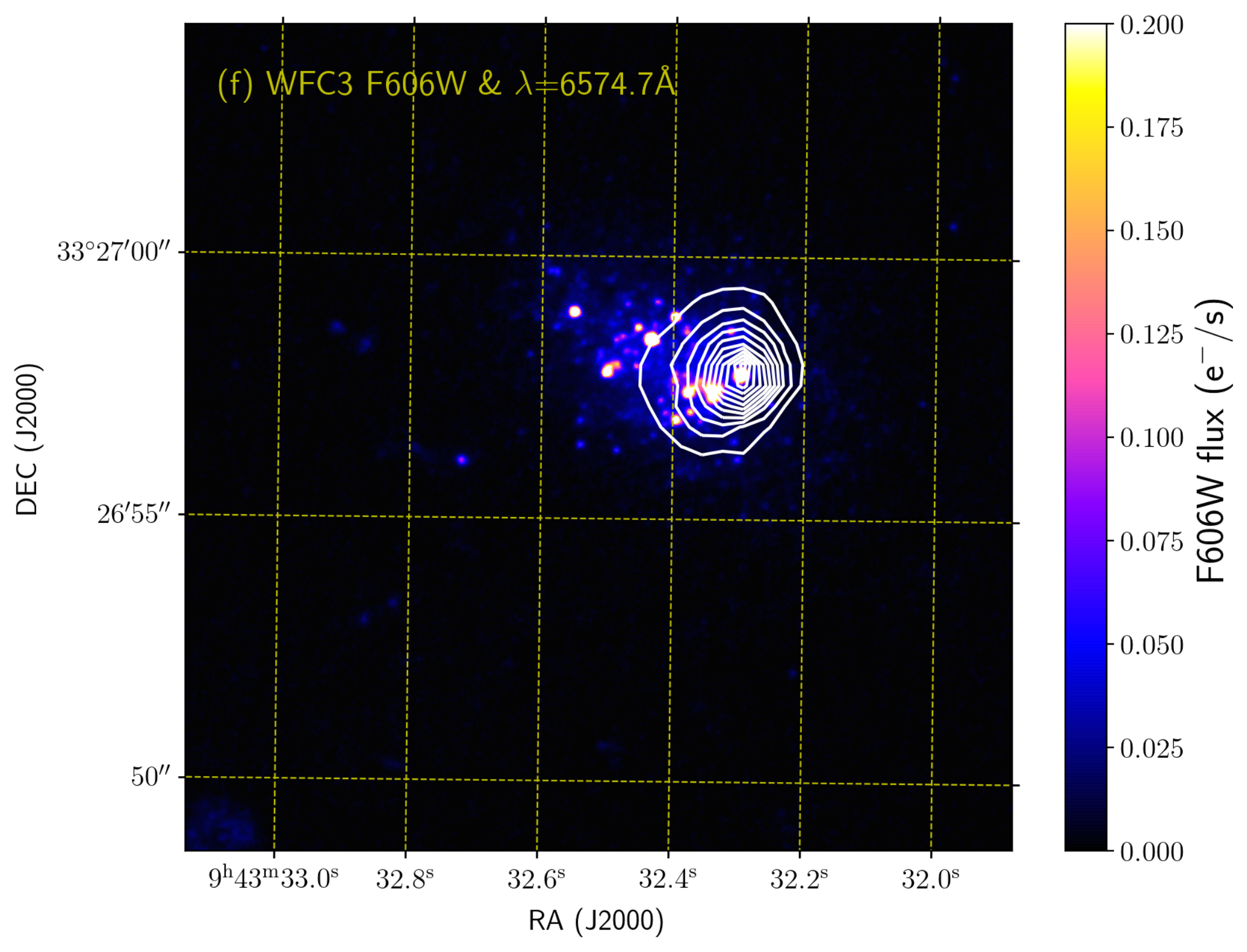}
\includegraphics[width=0.42\textwidth]{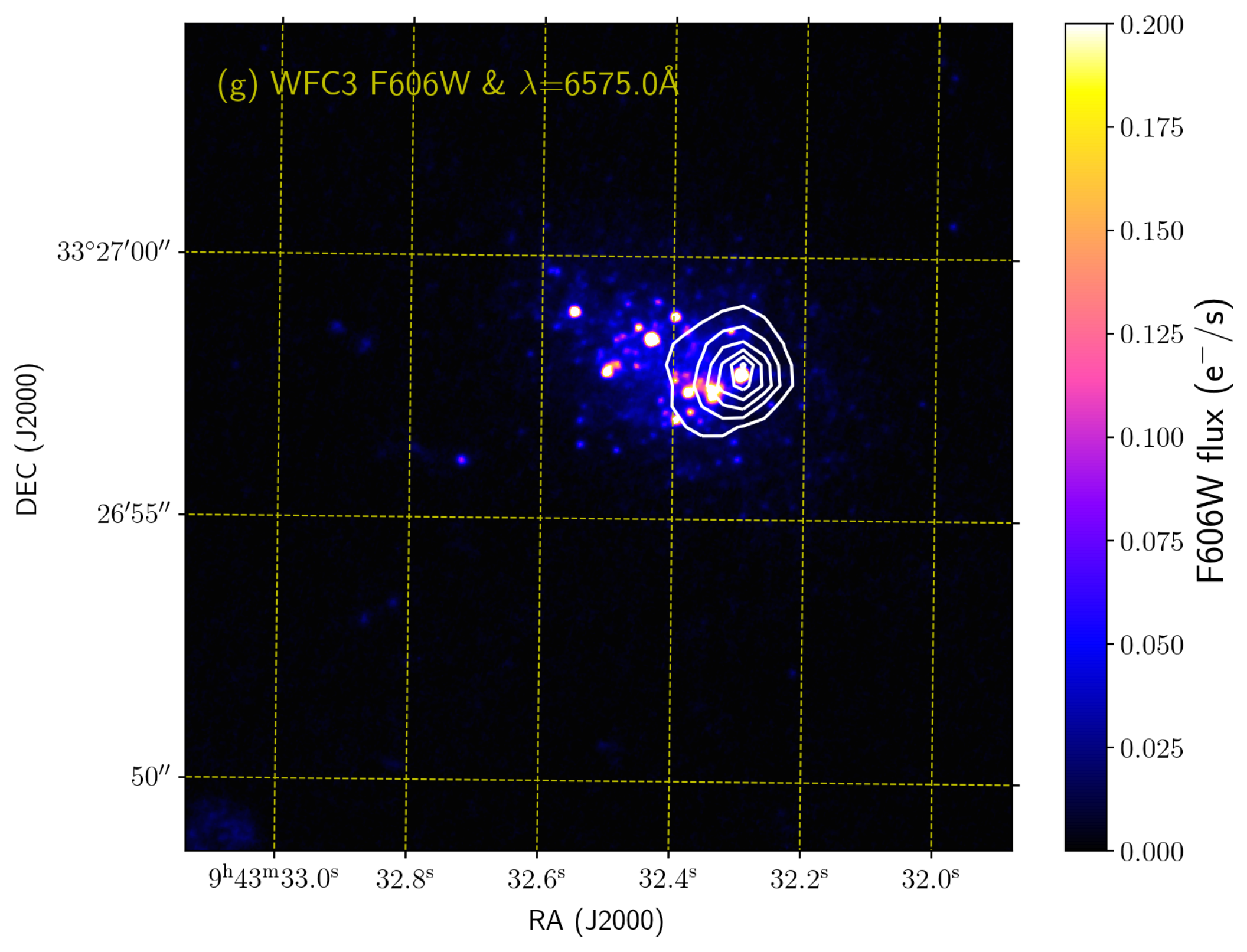}
\includegraphics[width=0.42\textwidth]{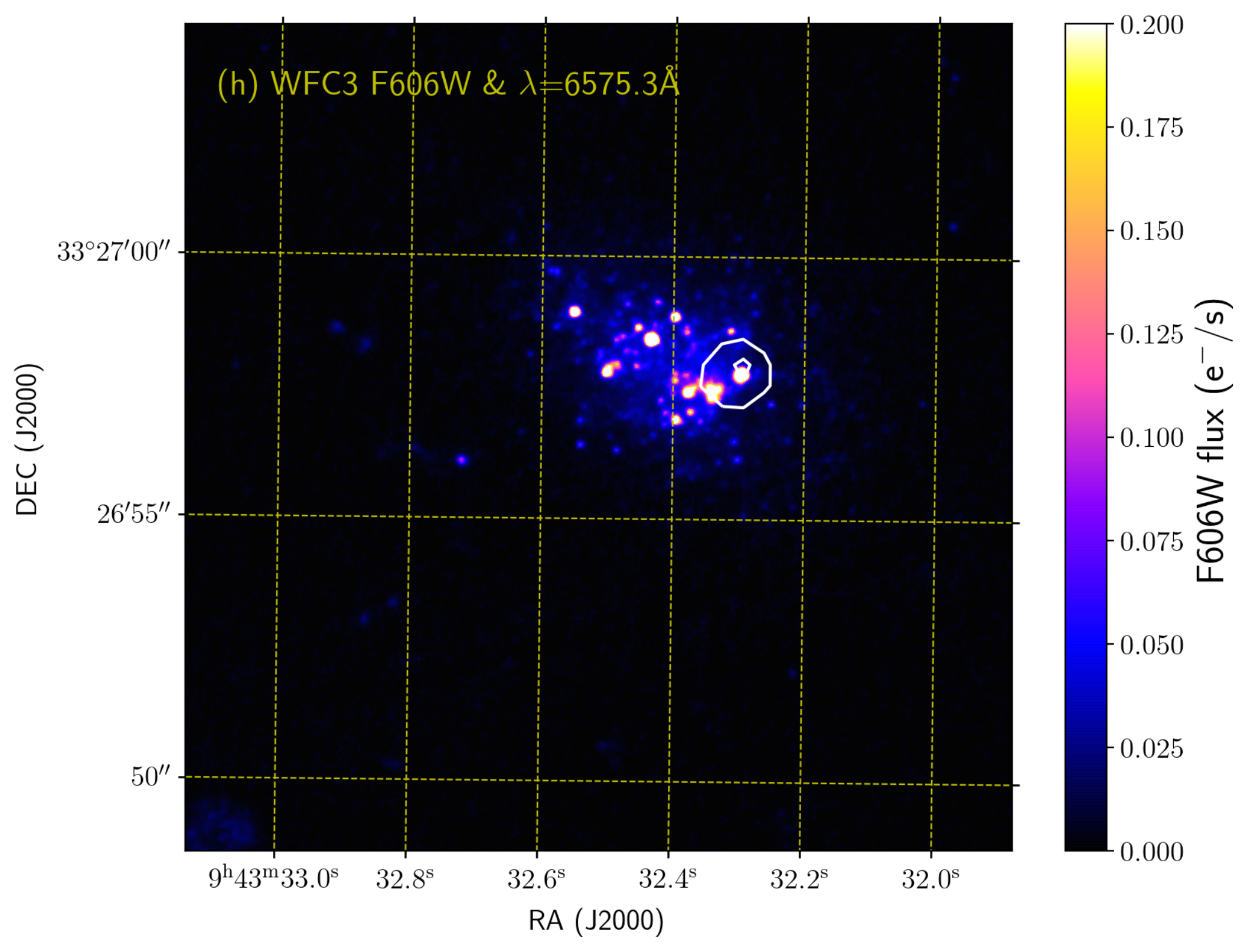}
\includegraphics[width=0.42\textwidth]{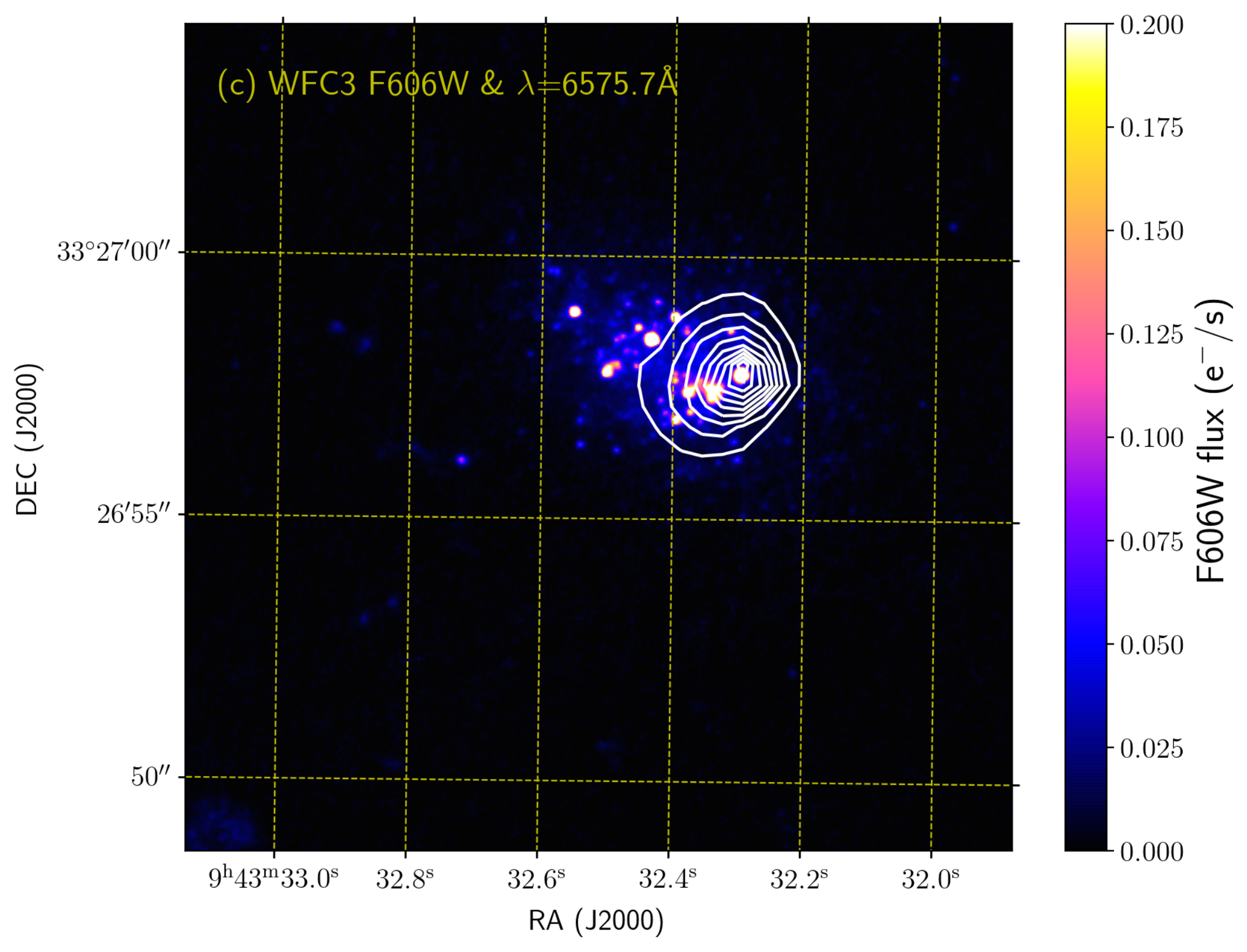}
\caption{WFC3-UVIS/HST F606W image on which we have overlapped the velocity channels around H${\alpha}$ line, as described in subsection~\ref{kinematics}.}
\label{Halphavelocitychannels}
\end{figure*}

\begin{figure*}
\includegraphics[width=0.42\textwidth]{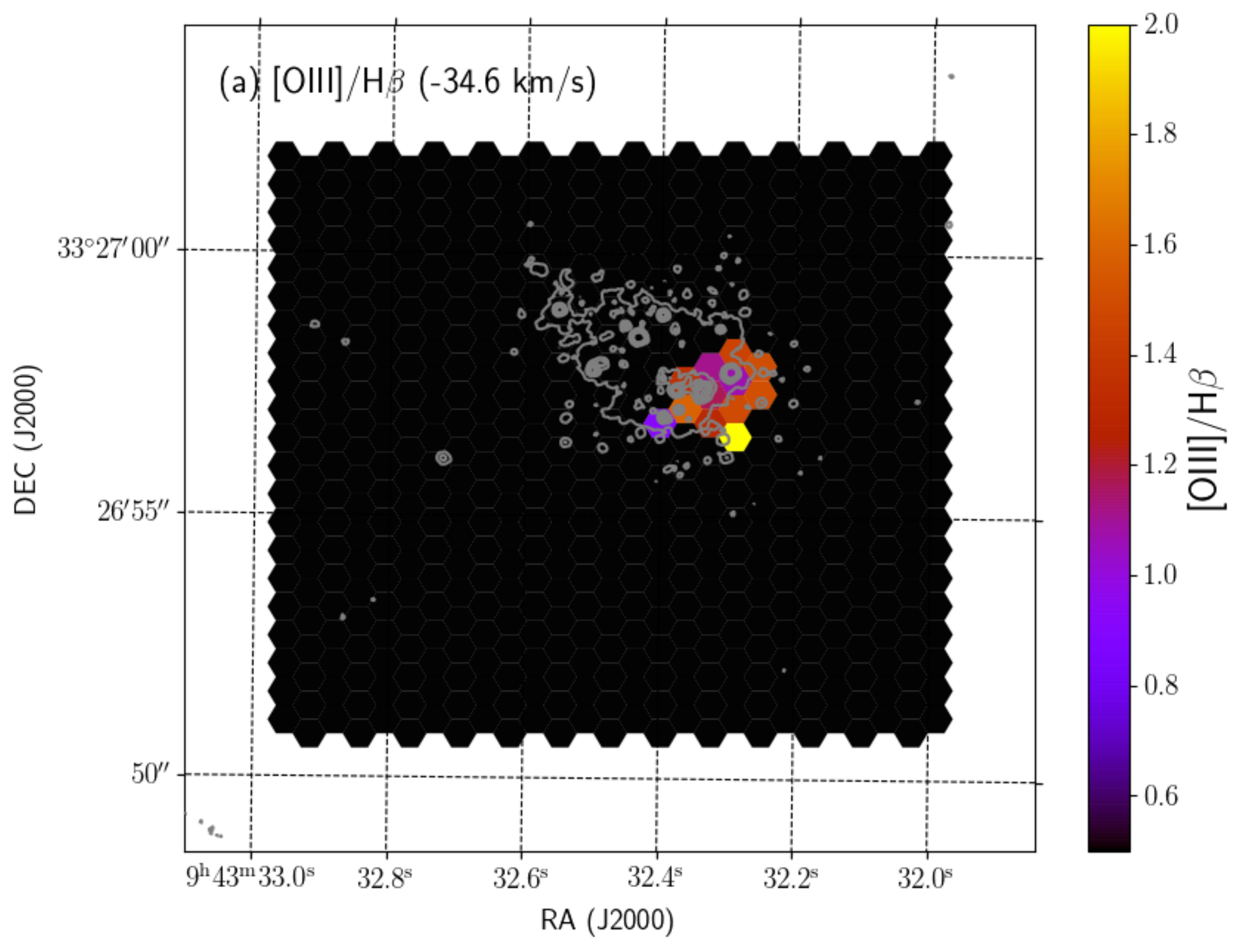}
\includegraphics[width=0.42\textwidth]{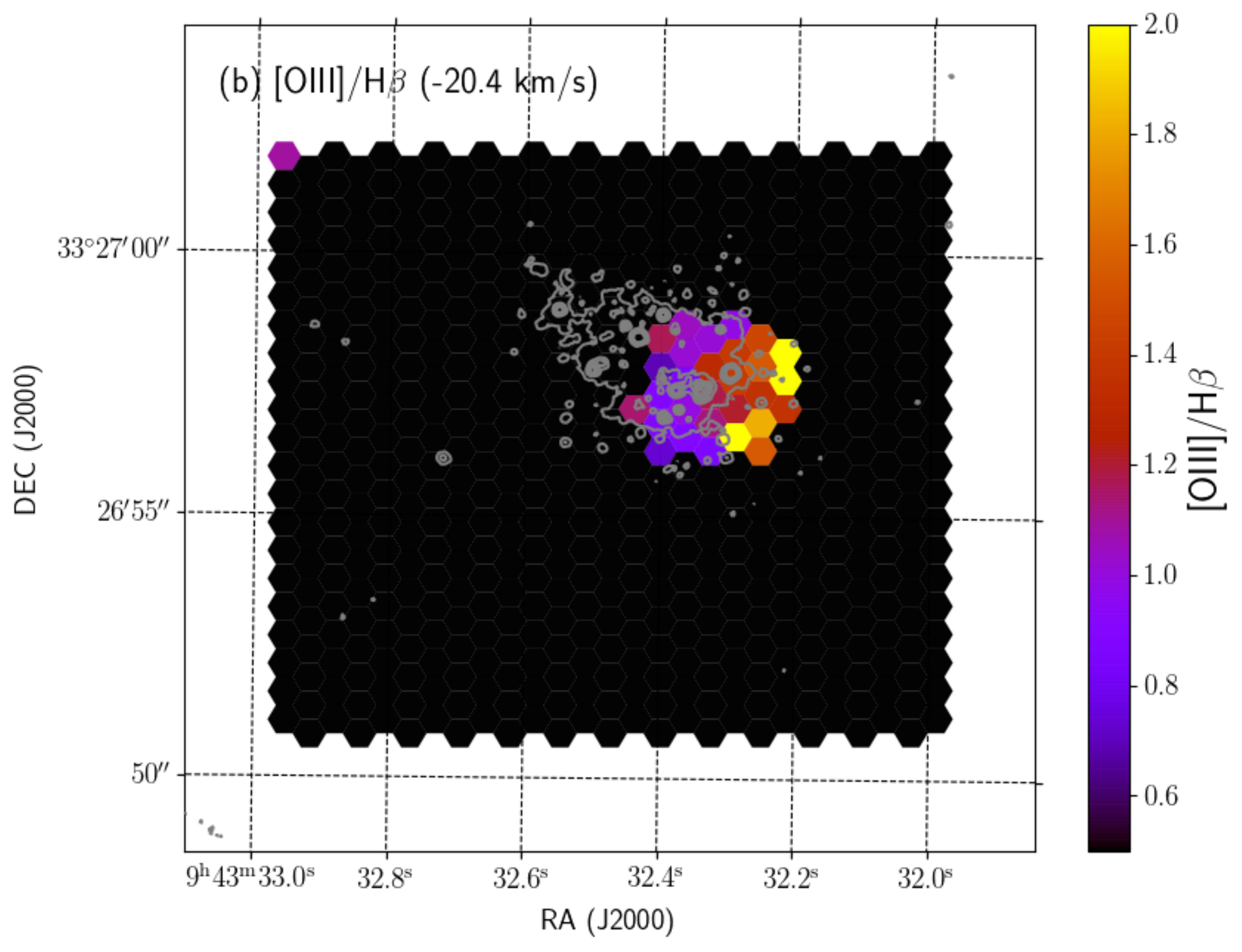}
\includegraphics[width=0.42\textwidth]{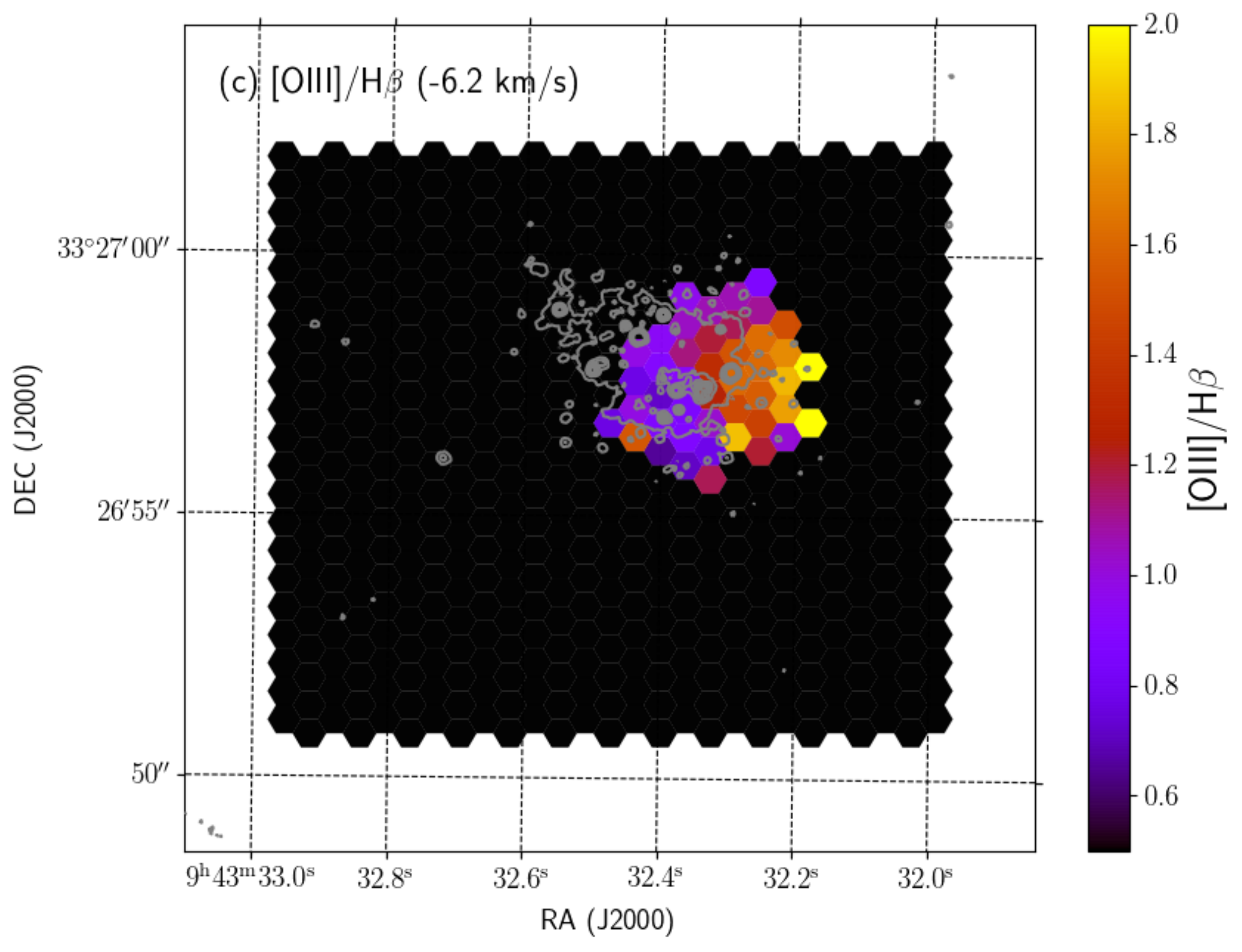}
\includegraphics[width=0.42\textwidth]{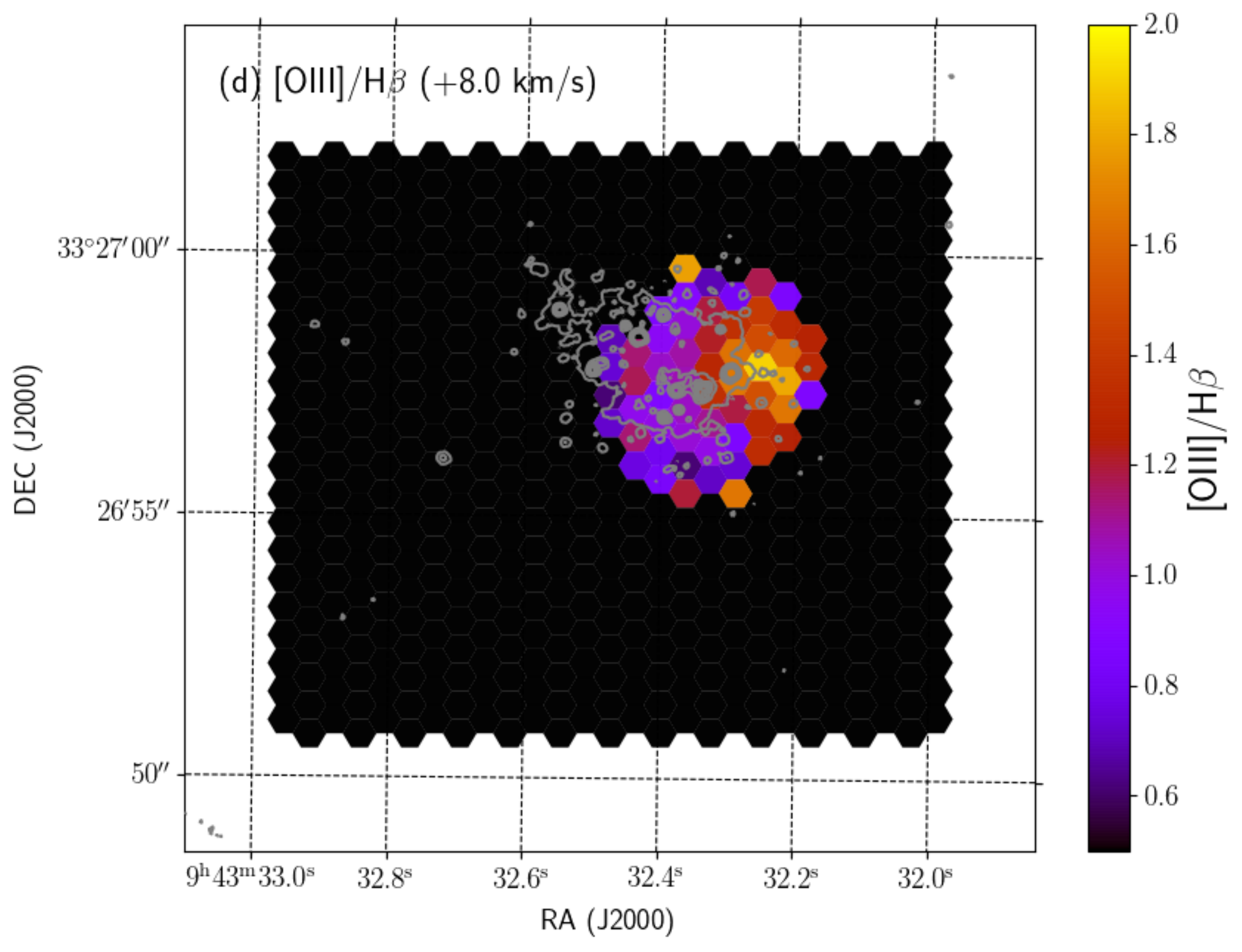}
\includegraphics[width=0.42\textwidth]{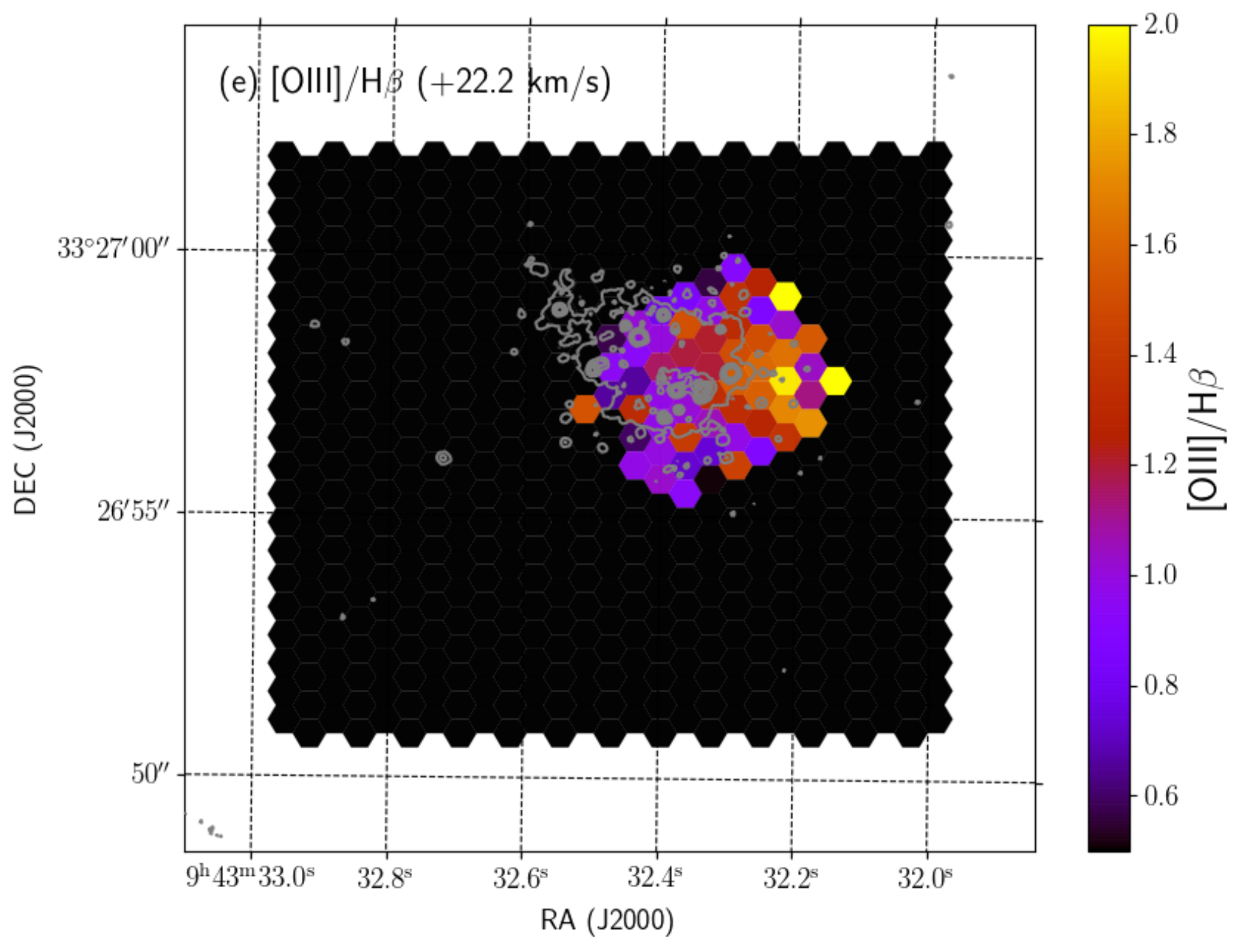}
\includegraphics[width=0.42\textwidth]{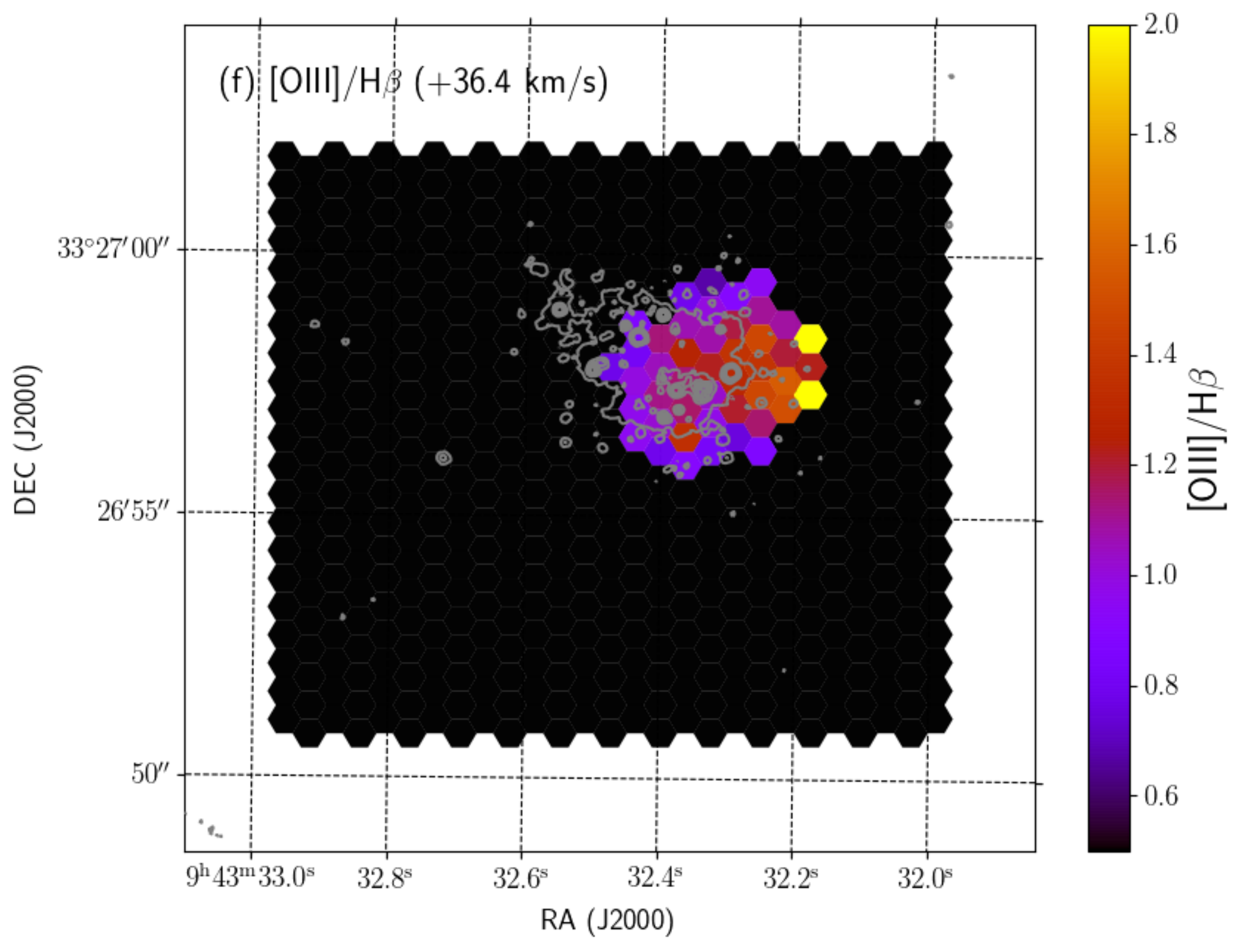}
\includegraphics[width=0.42\textwidth]{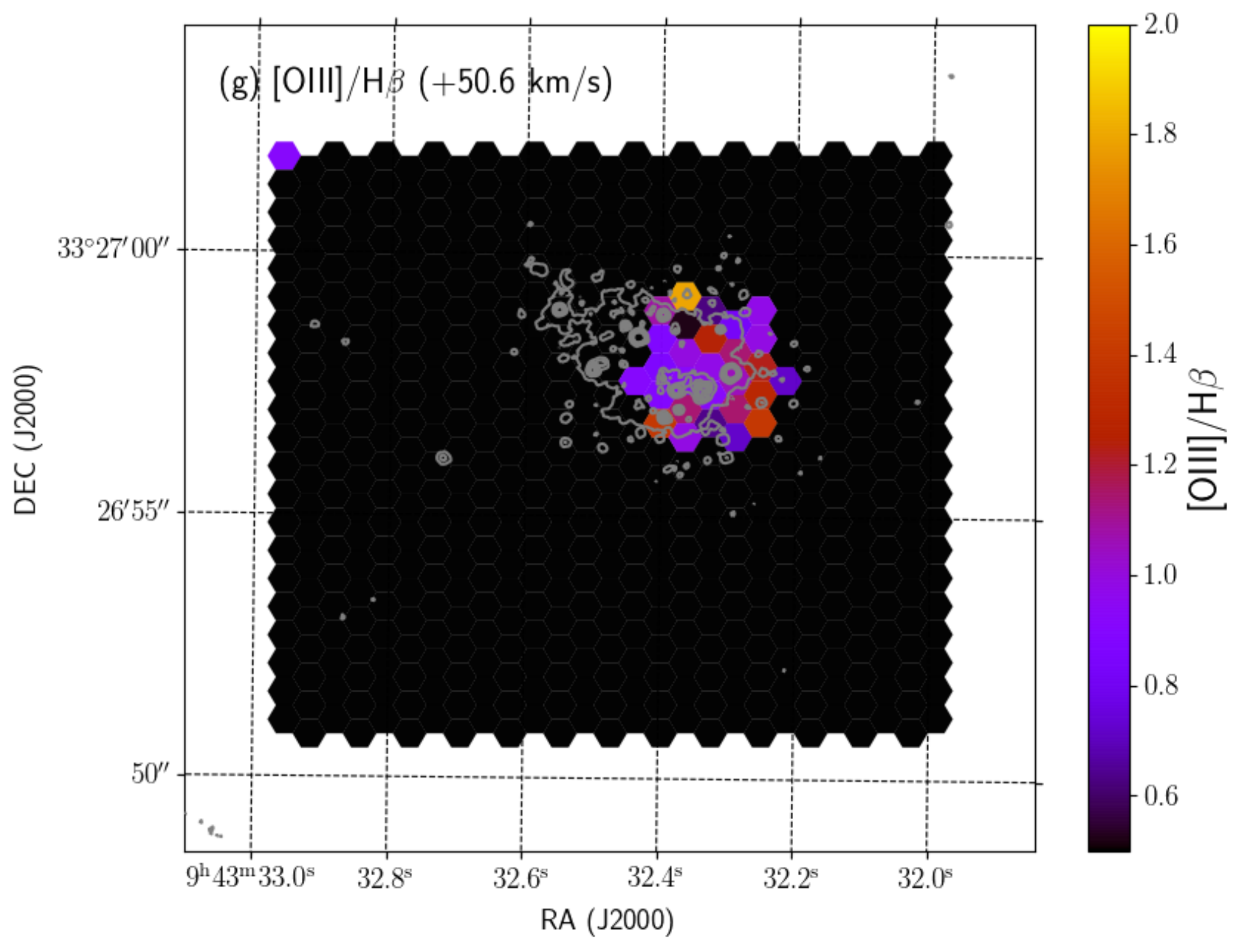}
\includegraphics[width=0.42\textwidth]{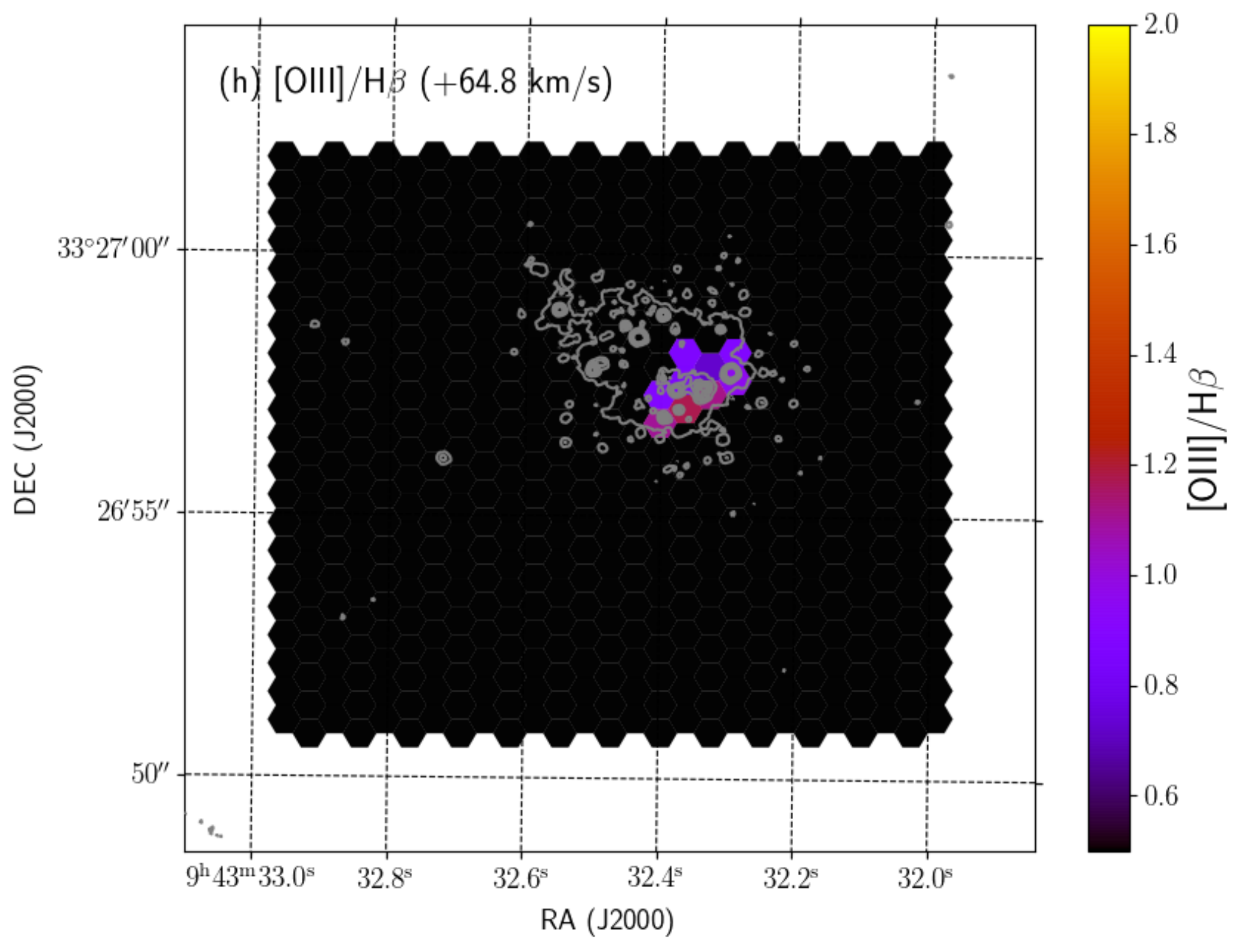}
\caption{[OIII]/H$\beta$ ratio for different velocity channels. See description in subsection~\ref{kinematics}.}
\label{OIIIHbvelocitychannels}
\end{figure*}

\subsection{Emission line 2D maps}
\label{maps}

In Fig.\,\ref{HalphaHbetaEW} we show the results for the case of the H$\beta$ (left) and H$\alpha$ (right) lines from LR-B and LR-R data, respectively.  The top panels show the emission line flux maps.  The central panels,  where proper size of the MEGARA IFU can be seen, display the continuum maps (in Jy) of H${\beta}$ (left), averaged between 4400 and 4850\,\AA\ and H${\alpha}$ (right), averaged between 6590 and 7100\,\AA. The bottom panels show the map of the logarithm of the  equivalent width, EW, (in \AA) for H${\beta}$ (left) and H${\alpha}$ (right), respectively. 
 The contours come from the WFC3-UVIS F606W HST image and correspond to 0.02, 0.07, 0.12, 0.17, 0.22, 0.27, 0.32 and 0.37 electrons\,s$^{-1}$ or 0.23, 0.8, 1.4, 1.9, 2.5, 3.1, 3.6 and 4.2 $\times$$\times$ 10$^{-20}$~erg\,s$^{-1}$\,cm$^{-2}$\,\AA$^{-1}$. 

The H${\beta}$ and H${\alpha}$  line-flux maps in Fig.\,\ref{HalphaHbetaEW} show that the peak of line emission is associated to the westernmost blue stellar cluster detected in the HST data, see also the HST image shown in Fig.~\ref{HST-falsecolor}. In terms of physical extension we detect H$\alpha$ emission with S/N\,$>$\,5 reaching as far as $\sim$\,4\,arcsec from this peak emission ($\sim$150\,pc away from it). Besides, it is also worth noting that  the region  is slightly asymmetric, being more extended towards the NE than to the SW of the line-emission peak. This is discussed in greater depth in section\,\ref{Sec:Discussion} but given the distribution of the continuum in the F606W HST image this could be due either to  photoionization by photons coming from other clusters located NE of the brightest cluster or to a change in the bounding regime in both directions: radiation bounded to the NE and density bounded instead in the SW direction. The same behavior is observed in the H$\beta$ flux map. Regarding the distribution in EW showed in the bottom panels of Fig.\,\ref{HalphaHbetaEW}, we note that a negative gradient  from the SW, where the line-emission peak is located, towards the NE of the galaxy. In addition, the close inspection of the region around this peak shows that the highest EW ($>$1000\,\AA) is reached not at the position of the spaxel with the brightest line emission but in the two spaxels located immediately SW from it, which suggests a spatial segregation between the ionized gas and the ionizing stars in spatial scales of a few tens of parsecs. 

\begin{figure*}
\includegraphics[width=0.45\textwidth]{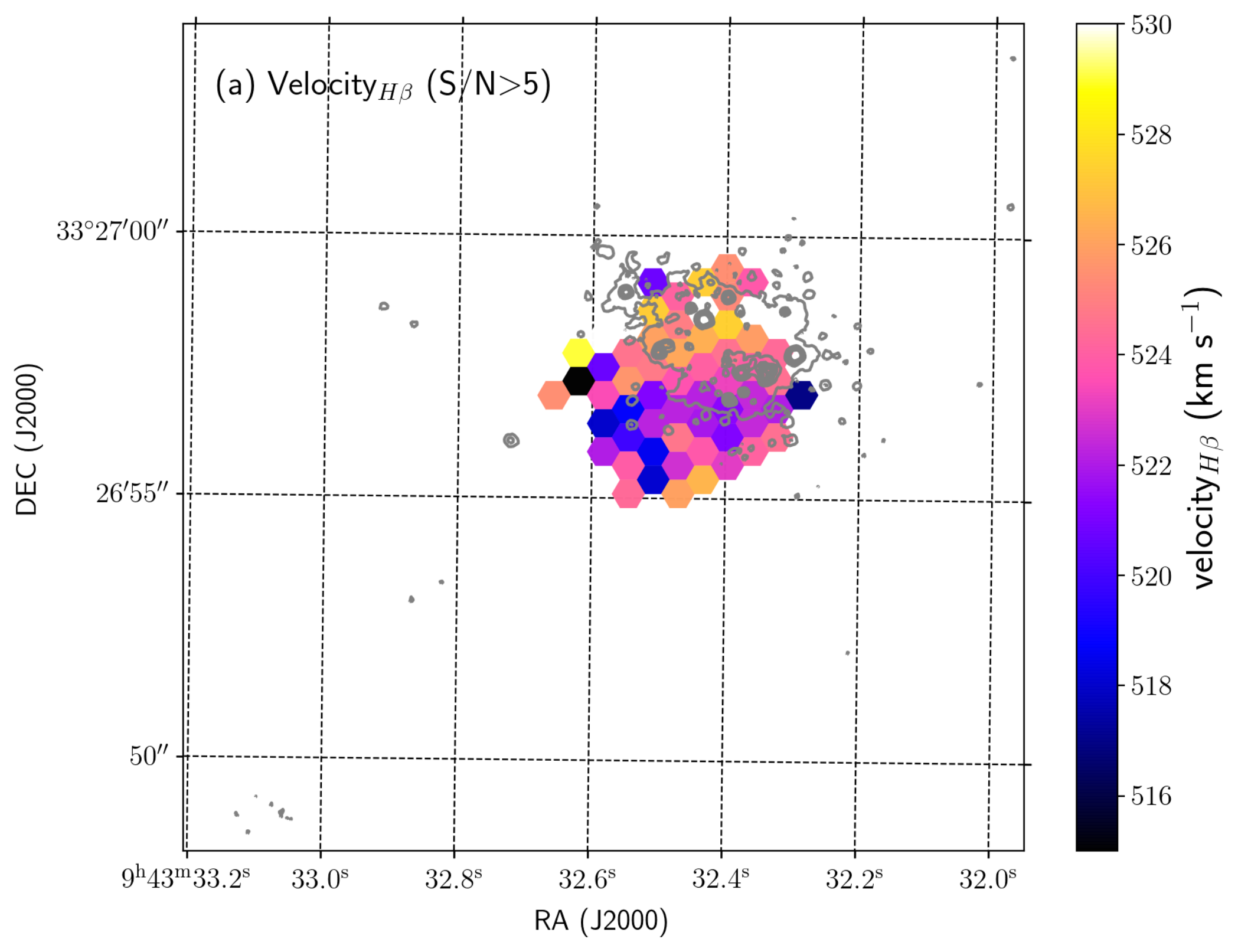}
\includegraphics[width=0.45\textwidth]{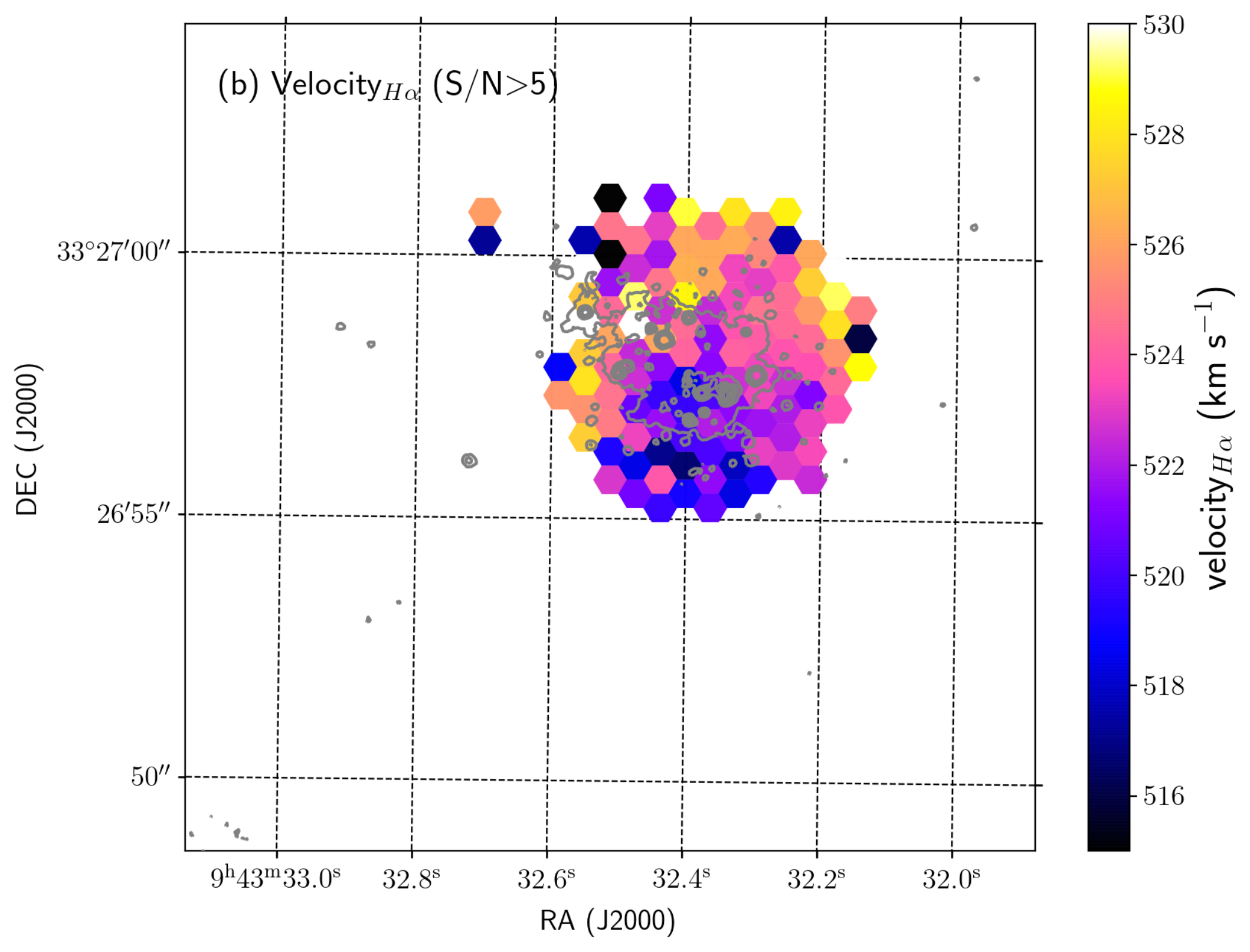}
\includegraphics[width=0.45\textwidth]{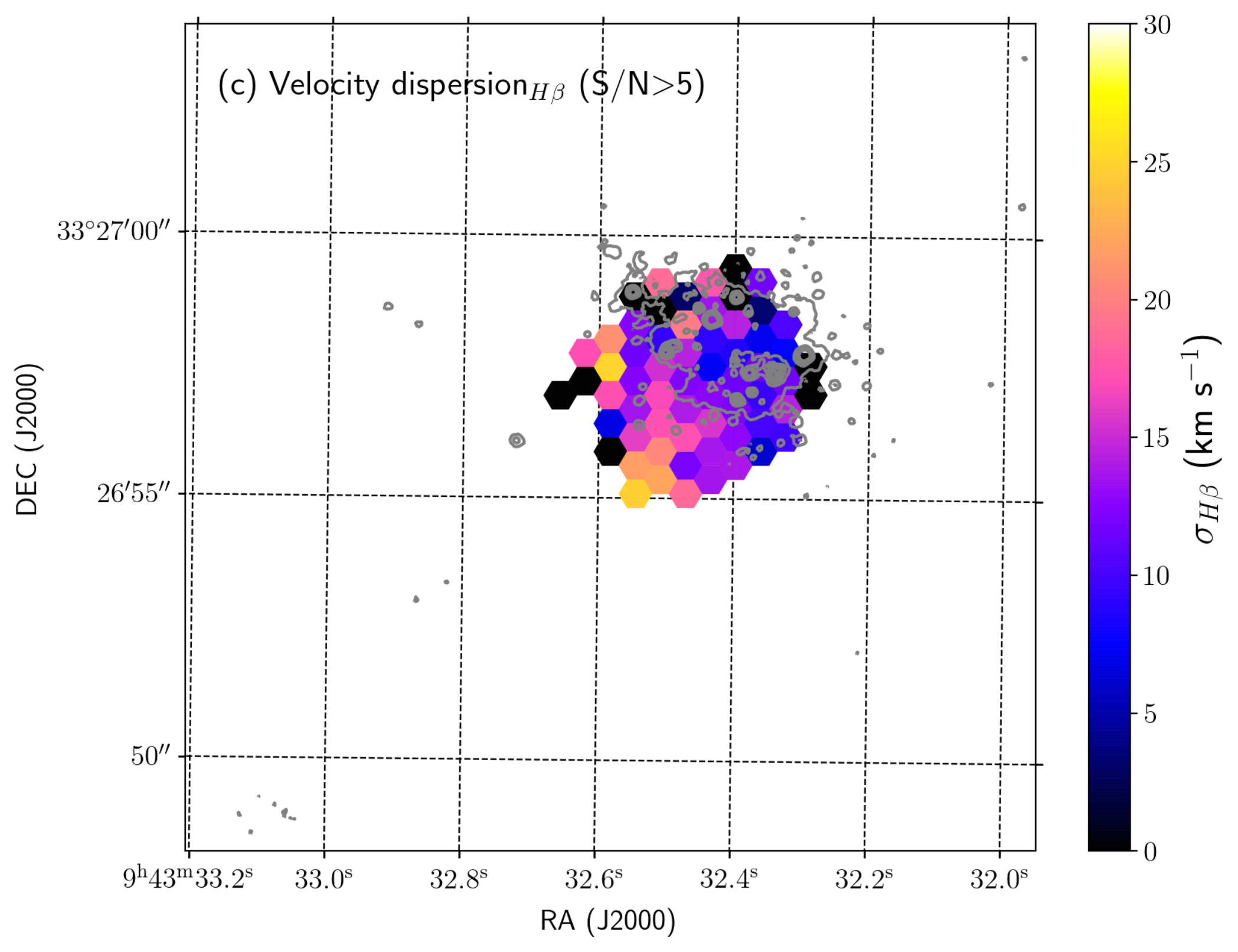}
\includegraphics[width=0.45\textwidth]{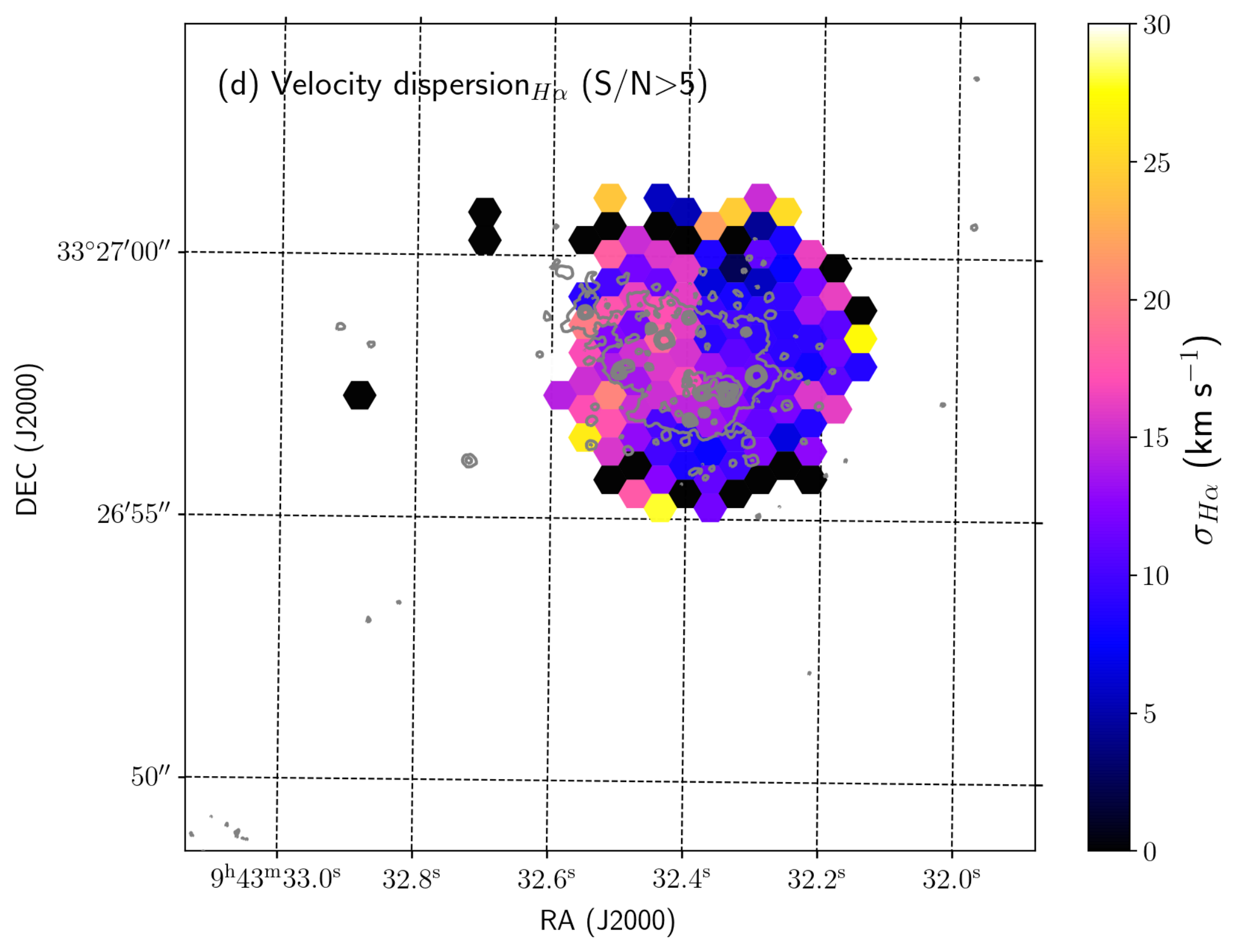}
\caption{Maps of radial topocentric velocity (top) and velocity dispersion ${\sigma}$ (bottom), corrected from instrumental contribution, derived from the positions and widths of the H${\beta}$ (left) and H${\alpha}$ (right) lines. All values are  in km\,s$^{-1}$.}
\label{HalphaOIIIV}
\end{figure*}

We present in Fig.\,\ref{Emissionlines} some emission lines fluxes and flux ratio maps for the spaxels in which we have measured S/N\,$>$\,5 in every line involved. The left panels present the maps obtained from the LR-B image: flux (in erg\,s$^{-1}$cm$^{-2}$) of H${\beta}$ (top) and  [\ion{O}{iii}]\,5007 (middle) and the ratio [\ion{O}{iii}]\,5007/H$\beta$ (bottom). The right panels display the maps from LR-R: flux (in erg\,s$^{-1}\,$cm$^{-2}$) in  H$\alpha$ (top), [\ion{S}{ii}]\,6717+6731/H${\alpha}$  (middle)  and the ratio [\ion{S}{ii}]\,6717/[\ion{S}{ii}]6731 (bottom). Fluxes with S/N\,$>$\,5 were measured in the [\ion{S}{ii}]\,6717, 6731 doublet for only two bright central spaxels. The contours are the same as in Fig. \ref{HalphaHbetaEW}.  Note that in this figure we include the same H$\beta$ and H$\alpha$ line-flux maps already shown in Fig.\,\ref{HalphaHbetaEW} for convenience when discussing our results below. 

While the spatial distribution of the [\ion{O}{iii}]\,5007 line emission resembles closely that seen in H$\beta$, the line ratio map (bottom-left panel) shows significant radial and azimuthal variations, with the maximum being found W of the line-flux peak. We further discuss on the radial variation of the [\ion{O}{iii}]\,5007/H$\beta$ line ratio in section\,\ref{possiblescenarios}. In the middle  and bottom-right panels we show the properties of the [\ion{S}{ii}]~6717, 6731 doublet emission. Unfortunately, its emission is only detected in two of the spaxels, although they yield [\ion{S}{ii}]/H${\alpha}$ line ratios  of the order of 0.015-0.020. 

\subsection{Gas kinematics}
\label{kinematics}

\begin{table*}
\caption{LR-B emission lines measurements in apertures around the central brightest spaxel. Fluxes are given in units of $\times$ 10$^{-17}$ erg cm$^{-2}$ s$^{-1}$ and EW in \AA.}
\label{TabLRBaper}
\begin{tabular}{rcccccccccccc}
\hline
(1) & (2) & (3) & (4) & (5) & (6) & (7)  & (8) & (9) & (10) & (11) & (12) & (13)\\
d & d & nr & Ns & H${\beta}$ & H${\beta}$ & H${\gamma}$  & H${\gamma}$ & [\ion{O}{iii}]4363 & [\ion{O}{iii}]4959 & [\ion{O}{iii}]5007 & \ion{He}{i}4471 & \ion{He}{i}5015 \\
 pc & ´´ &  &  & Flux & EW & Flux & EW & Flux & Flux & Flux & Flux & Flux \\
11	&	0.31	&	0	&	1	&	10.50$\pm$1.03	&	213$\pm$168	& 5.09$\pm$1.45	&	44$\pm$43	&	0.38$\pm$0.44	&	5.18$\pm$0.73	&	14.42$\pm$1.06	&	0.26$\pm$0.26	&	0.20$\pm$0.17	  \\
33	&	0.93	&	1	&	7	&	57.71$\pm$2.64	&	306$\pm$136	&	28.53$\pm$3.86	&	24$\pm$8	&	1.93$\pm$1.10	&	26.51$\pm$2.22	&	74.19$\pm$3.19	&	2.78$\pm$0.80	&	1.07$\pm$0.47	  \\
55	&	1.55	&	2	&	19	&	104.60$\pm$4.19	&	177$\pm$52	&	55.05$\pm$6.78	&	18$\pm$5	&	4.08$\pm$2.23	&	44.92$\pm$2.74	&	126.30$\pm$4.05	&	4.26$\pm$1.43	&	1.41$\pm$0.70	  \\
77	&	2.17	&	3	&	37	&	134.70$\pm$5.15	&	101$\pm$21	&	77.44$\pm$8.71	&	13$\pm$3	&	5.44$\pm$3.14	&	56.40$\pm$3.33	&	155.70$\pm$4.88	&	4.71$\pm$1.82	&	2.53$\pm$0.98	  \\
121	&	3.41	&	All	&	63	&	160.70$\pm$6.40	&	76$\pm$15	& 96.27$\pm$11.44	&	9$\pm$2	&	7.71$\pm$4.74	&	62.86$\pm$3.99	&	171.90$\pm$6.01	&	6.32$\pm$2.75	&	2.73$\pm$1.43	  \\
\hline
\end{tabular}
\end{table*}

\begin{table*}
\caption{LR-R emission lines measurements in apertures around the central brightest spaxel. Fluxes are given in units of $\times$ 10$^{-17}$ erg cm$^{-2}$ s$^{-1}$ and EW in \AA.}
\label{TabLRRaper}
\begin{tabular}{rcccccccccccc}
\hline
(1) & (2) & (3) & (4) & (5) & (6) & (7)  & (8) & (9) & (10) & (11) & (12) & (13)\\
d & d & nr & Ns & H${\alpha}$ & H${\alpha}$ & [\ion{N}{ii}]6584 & \ion{He}{i}6678 & [\ion{S}{ii}]6717 &  [\ion{S}{ii}]6731 & \ion{He}{i}7065 & [\ion{Ar}{iii}]7135 &  \ion{He}{i}7281 \\
 pc & ´´ &  &  & Flux & EW  & Flux & Flux & Flux & Flux & Flux  & Flux  & Flux  \\
11	&	0.31	&	0	&	1	&	58.67$\pm$1.89	&	1407$\pm$847	&	0.40$\pm$0.21	&	0.43$\pm$0.18	&	0.73$\pm$0.27	&	0.53$\pm$0.20	&	0.50$\pm$0.18	&	0.27$\pm$0.21	&	0.07$\pm$0.10	  \\
33	&	0.93	&	1	&	7	&	252.60$\pm$3.36	&	543$\pm$84	&	1.92$\pm$0.56	&	1.84$\pm$0.48	&	2.84$\pm$0.53	&	1.79$\pm$0.44	&	1.97$\pm$0.56	&	1.46$\pm$0.58	&	0.36$\pm$0.38	  \\
55	&	1.55	&	2	&	19	&	398.70$\pm$5.03	&	346$\pm$40	&	2.59$\pm$0.86	&	2.80$\pm$0.75	&	5.02$\pm$0.84	&	2.44$\pm$0.74	&	2.93$\pm$0.85	&	2.39$\pm$1.05	&	1.05$\pm$0.74	  \\
77	&	2.17	&	3	&	37	&	489.00$\pm$8.30	& 259$\pm$35	&	3.17$\pm$1.24	&	3.59$\pm$0.94	&	6.42$\pm$1.24	&	3.36$\pm$1.12	&	2.84$\pm$1.13	&	3.82$\pm$1.44	&	1.01$\pm$1.05	  \\
 121	&	3.41	&	All	&	61	&	545.10$\pm$11.85	&	202	$\pm$31	&	4.09$\pm$1.78	&	4.17$\pm$1.45	&	7.72$\pm$1.91	&	2.41$\pm$1.44	&	2.68$\pm$1.41	&	3.59$\pm$2.01	&	1.66$\pm$1.42  \\
\hline
\end{tabular}
\end{table*}

In Fig.\,\ref{Halphavelocitychannels} we show the WFC3-UVIS/HST F606W image and  overlapped are the velocity channels around the H${\alpha}$ line: the intensity contours correspond to 0.25, 0.50, 0.75, 1.00, 1.25, 1.50, 1.75, 2.00, 2.25, 2.50, 2.75, 3.00, 3.25, 3.50, 3.75, 4.00, 4.25, 4.50 and 4.75 $\times$ 10$^{-4}$\,Jy. Note that the brightest contours are missing in some channels as these are tracing the wings of the H$\alpha$ line. In fact, only the 0.25\,$\times$\, 10$^{-4}$\,Jy contour is depicted in all 8 panels. This figure shows that there is little velocity structure within the region showing ionized-gas emission in the galaxy, since the peak of emission at different wavelengths is centered in the same region for all velocity channels. This  indicates that either all the emission comes from a region with a common bulk motion or any rotational motion is mainly taking place in the plane of the sky. Besides, the mild asymmetry in the H$\alpha$ flux is also seen in the velocity channel maps for a wide range of velocities ($\pm$\,40\,km\,s$^{-1}$ around its systemic velocity). 

As mentioned in subsection\,\ref{maps} we find radial and azimuthal anisotropies in the [\ion{O}{iii}]\,5007/H$\beta$ ratio. One interpretation  is the presence of a hot gas component flowing out through an opening in the giant H{\sc ii} region \citep {menacho2019}. Interestingly, this region of high excitation is located in the region where the H$\alpha$ surface brightness drops fastest. Such a drop could be due to the presence of density-bounded conditions in the SW of the galaxy, which would indeed favor the escape of hot gas from inside the nebula. To provide further clues to this possible scenario we have followed the methodology described by \citet {menacho2019} for MUSE data but using our higher spectral resolution MEGARA-IFU data. In Fig.\,\ref{OIIIHbvelocitychannels} we show the [\ion{O}{iii}]\,5007/H$\beta$  maps measured for each velocity channel in the range $-$34.6 to $+$64.8\,km\,s$^{-1}$ around each of the lines with the F606W HST contours overlapped (0.02, 0.07, 0.12, 0.17, 0.22, 0.27, 0.32 and 0.37 electrons\,s$^{-1}$ or 0.23, 0.8, 1.4, 1.9, 2.5, 3.1, 3.6 and 4.2 $\times$ 10$^{-20}$ erg\,s$^{-1}$\,cm$^{-2}$\,\AA$^{-1}$). Again, the high excitation gas is present at all velocities, i.e$.$ both at the core and the wings of the emission lines, and the overall spatial variation of the [\ion{O}{iii}]~5007/H$\beta$ ratio is also rather similar. This result indicate that either there is no high excitation gas flowing out of the region or  its motion is taking place in the plane of the sky. 

Additionally, in Fig.\,\ref{HalphaOIIIV} we show the maps of topocentric radial velocity and velocity dispersion (in km\,s$^{-1}$) corrected from instrumental contribution, obtained from the analysis of the H${\beta}$ (left) and H$\alpha$ (right).  While the velocity channels (Fig.~\ref{Halphavelocitychannels}) do not show a clear rotation pattern, when we map the velocities of the best-fitting Gauss-Hermite line profile in the case of the brightest H$\alpha$ line, we infer a mild velocity gradient with a projected maximum velocity of $\pm$\,5\,km\,s$^{-1}$ assuming that the dynamical center is near the center of the region where radial velocities can be measured,  as shown in the top panels.

Adopting this rotational velocity and a spherically symmetric mass distribution, the total enclosed mass at a galactocentric distance of 110\,pc (3\,arcsec) would be $6.4\times 10^{5}$\,M\textsubscript{\(\sun\)}. Note that although the assumption of spherical symmetry could yield this value as an upper limit (thin axisymmetric mass distributions generate the same radial force with a somewhat smaller enclosed mass), the fact that we have used the projected rotational velocity as the galaxy circular velocity, probably results in an underestimation of the true enclosed mass. The velocity dispersion measurements obtained are rather small, with values that, once corrected for instrumental dispersion, are as low as 10-15\,km\,s$^{-1}$, even being compatible with zero in a few of the spaxels analyzed. We also find that the velocity dispersion values are slightly smaller in the E side of the nebula than in the W side of it, where the emission peak and the highest equivalent widths are found. 

\begin{figure*}
\includegraphics[width=0.48\textwidth,angle=0]{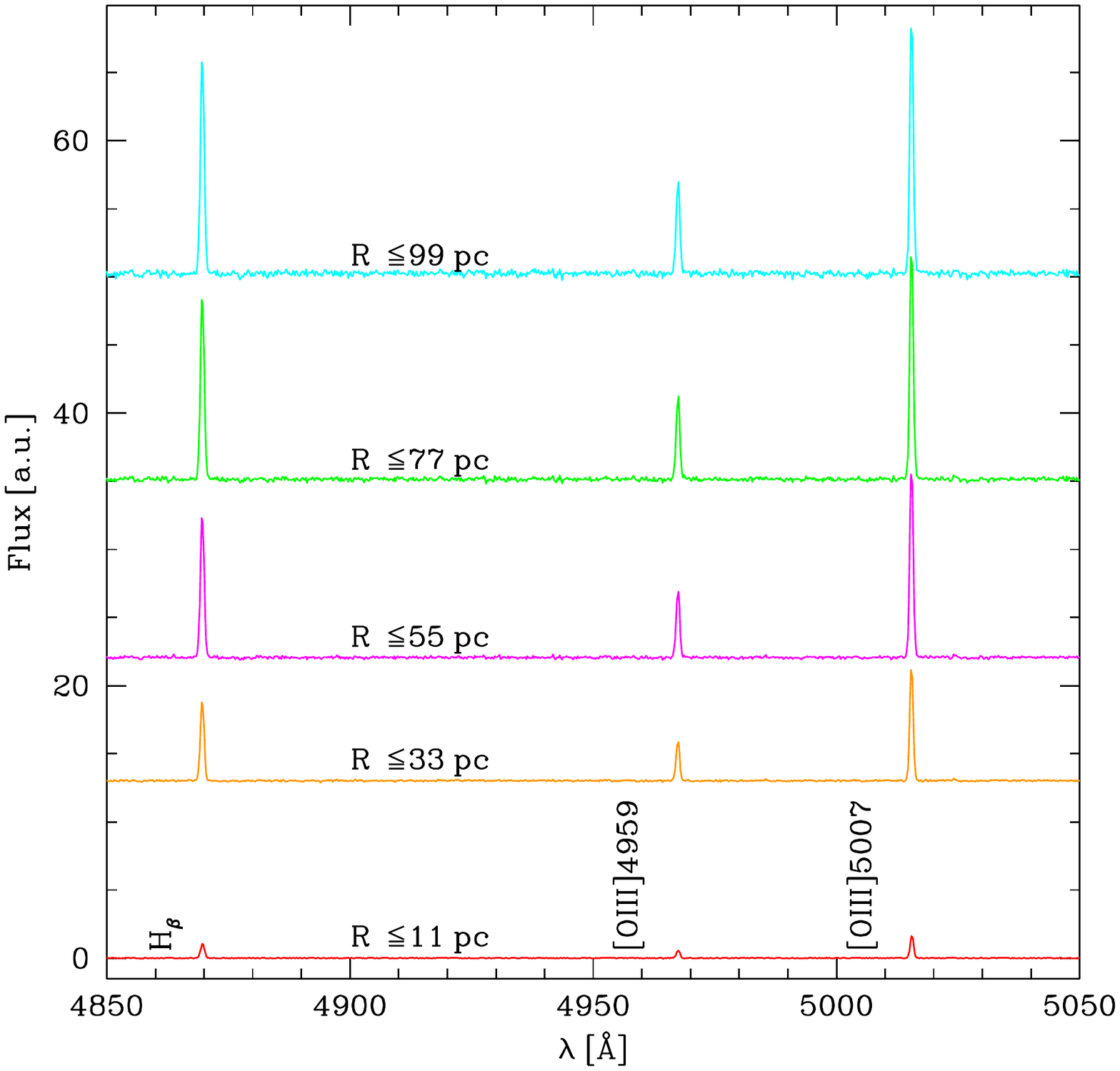}
\includegraphics[width=0.48\textwidth,angle=0]{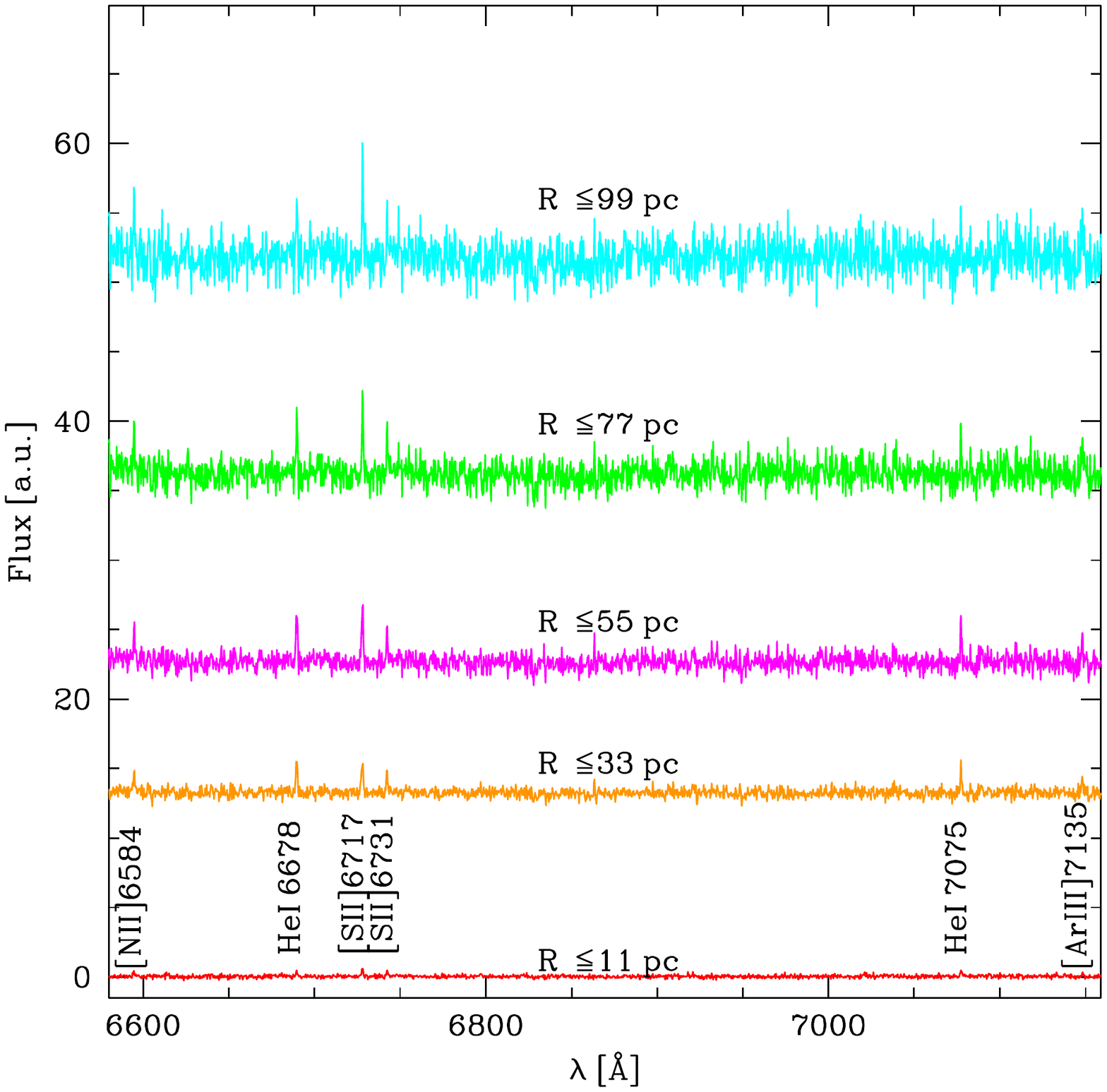}
\vspace{-1.6cm}
\caption{Accumulated spectra for LR-B (left) and LR-R (right) setups for increasing areas located at different distances of the center as labelled. Each spectrum is shifted an arbitrary distance for the  sake of clarity.}
\label{Spectra_ap}
\end{figure*}

\begin{figure*}
\includegraphics[width=0.47\textwidth,angle=0]{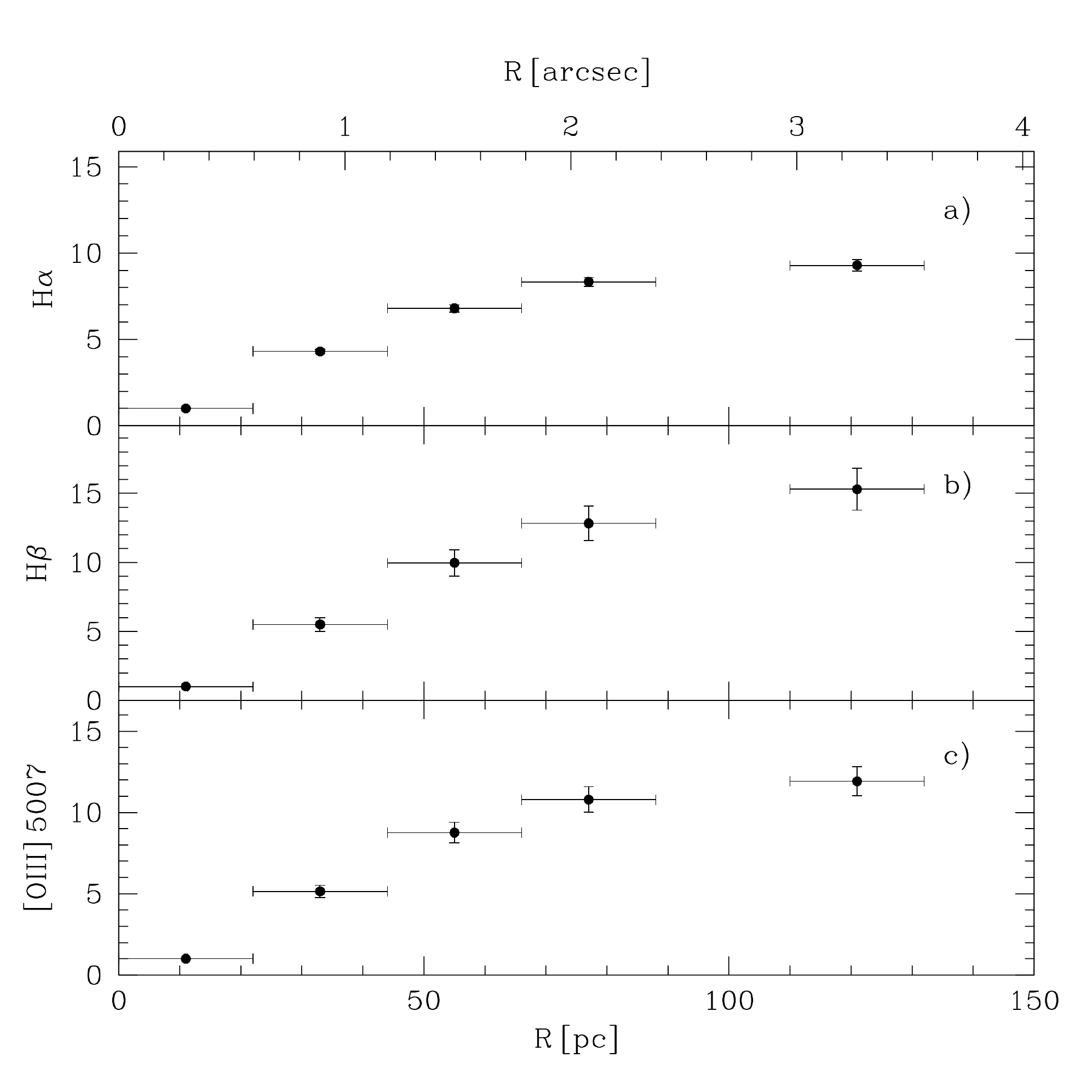}
\includegraphics[width=0.47\textwidth,angle=0]{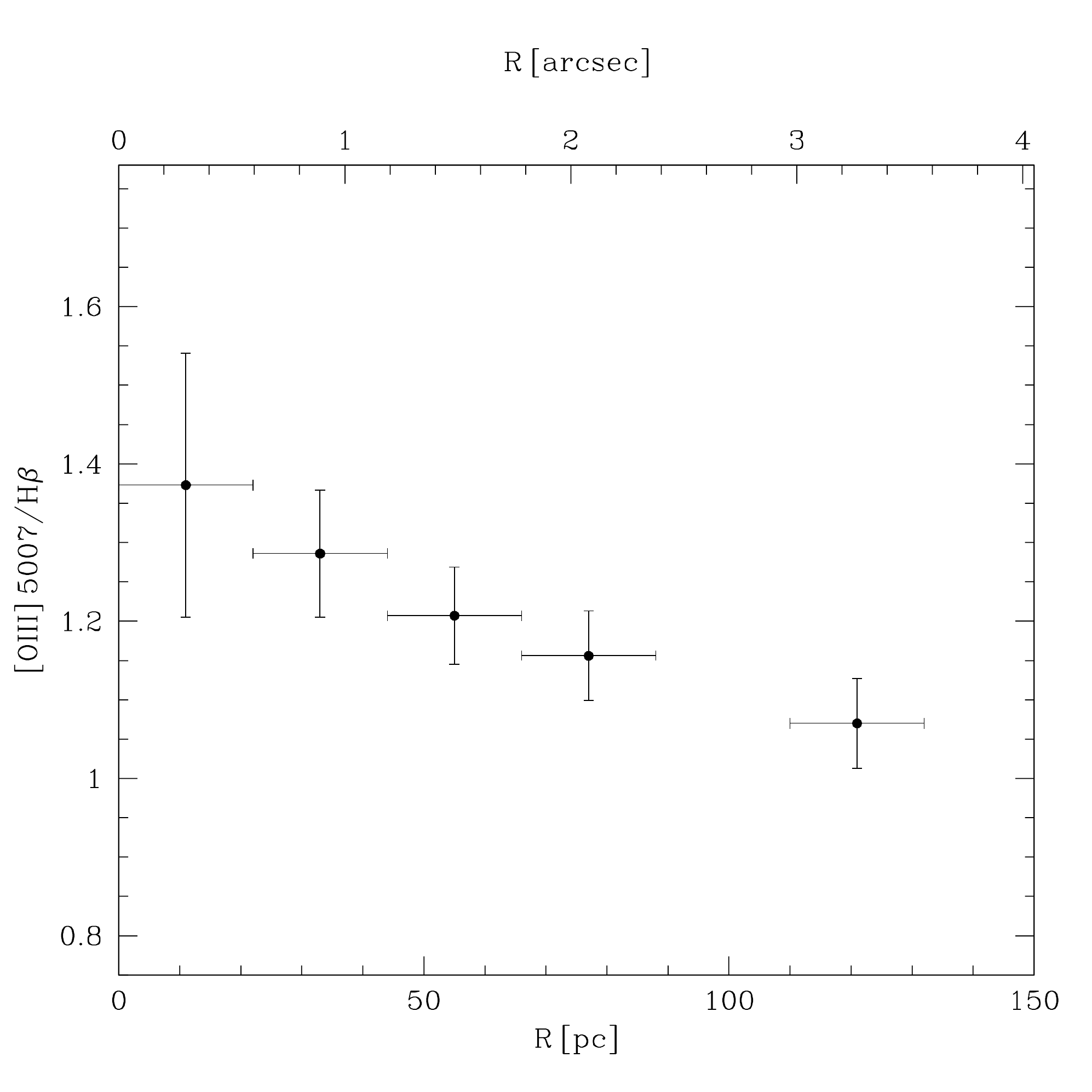}
\caption{Left: total flux accumulated in successive circular apertures around the most brightest spaxel, with respect to the flux in the central spaxel, for H${\alpha}$, H${\beta}$ and [\ion{O}{iii}]~5007. Each of the apertures includes the inner ones. Right: ratio [\ion{O}{iii}]~5007/H${\beta}$ as a function of the aperture.}
\label{LRBlinesaper}
\end{figure*}

\begin{figure*}
\includegraphics[width=0.23\textwidth,angle=-90]{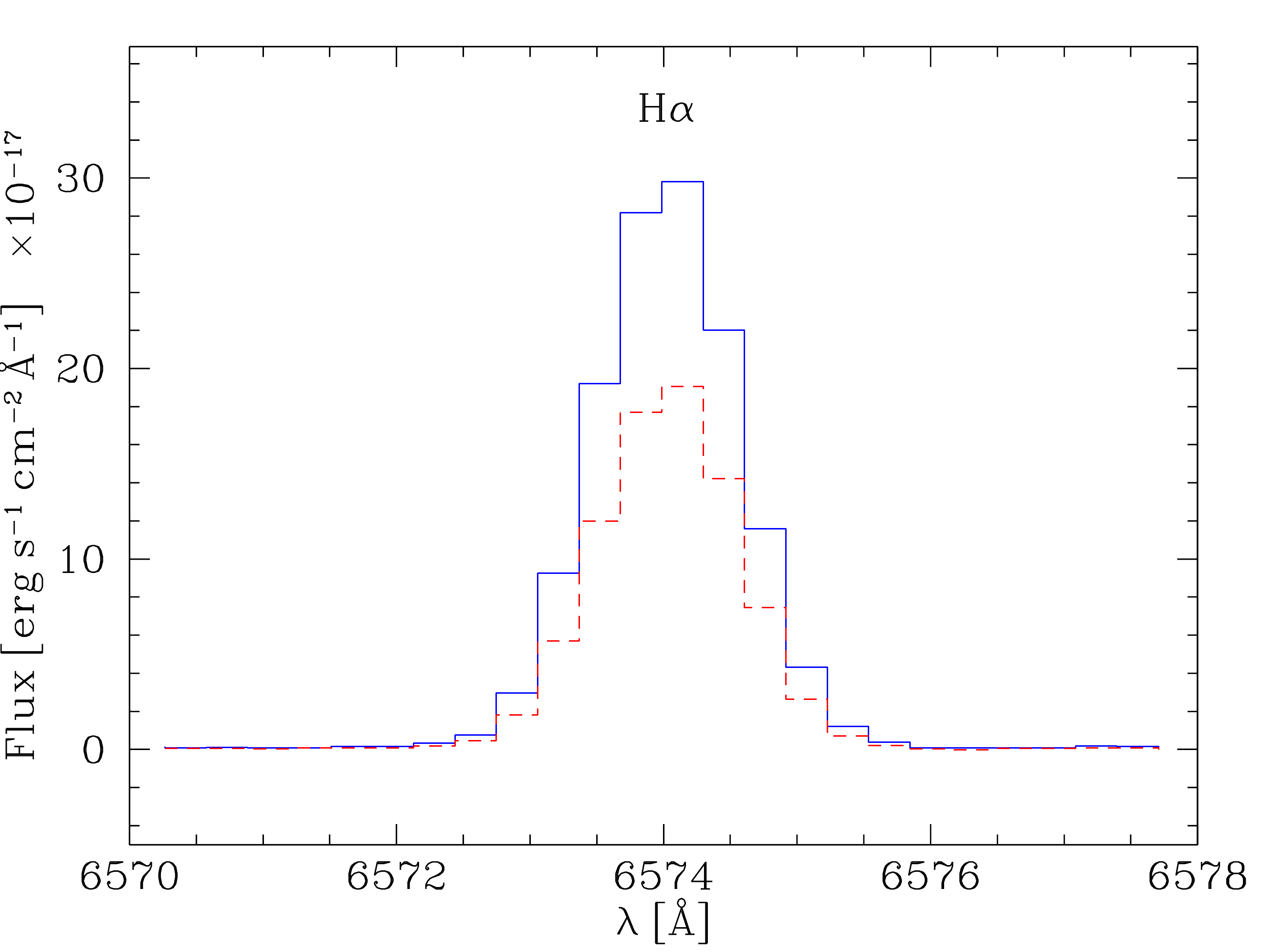}
\includegraphics[width=0.23\textwidth,angle=-90]{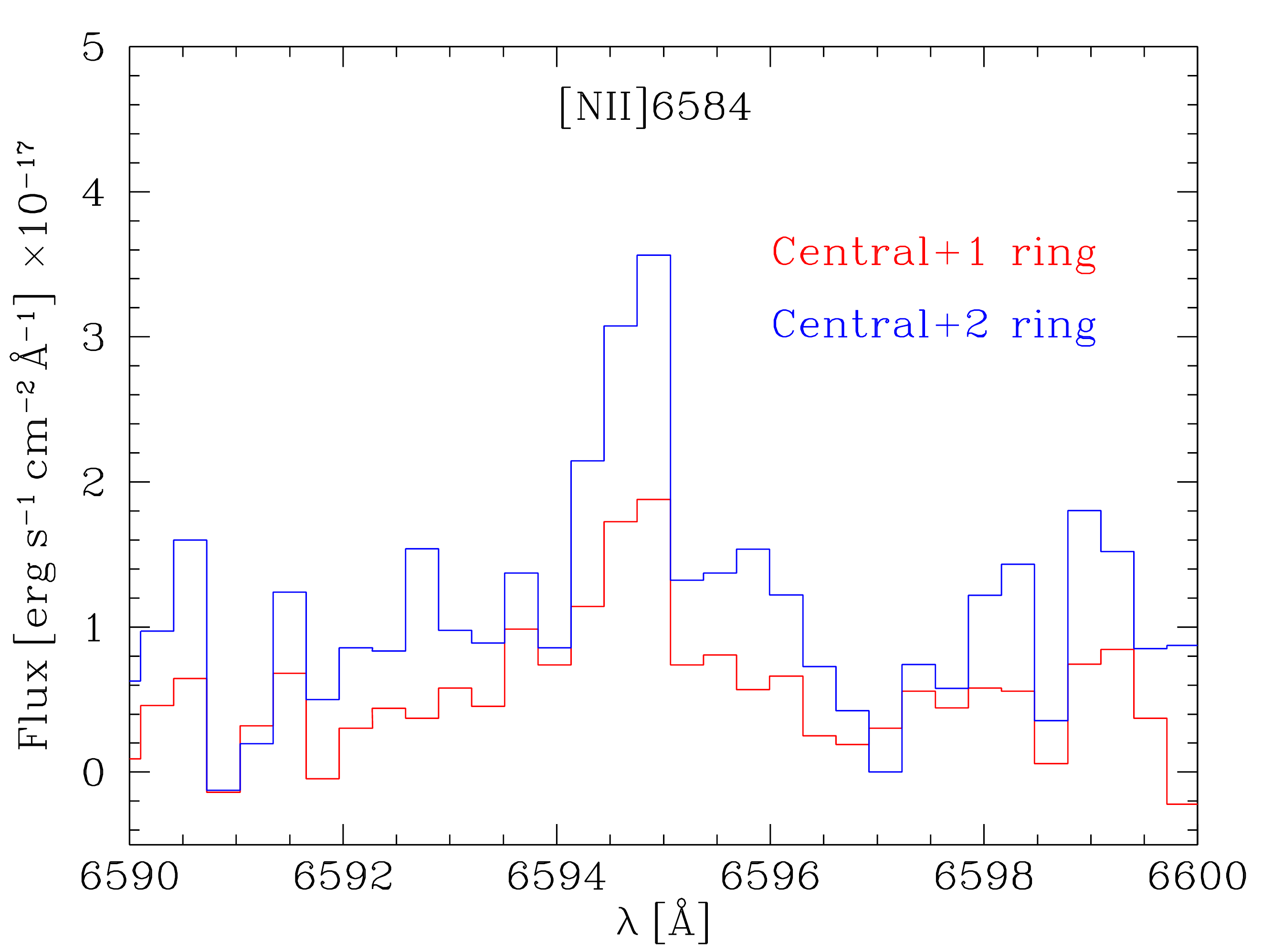}
\includegraphics[width=0.23\textwidth,angle=-90]{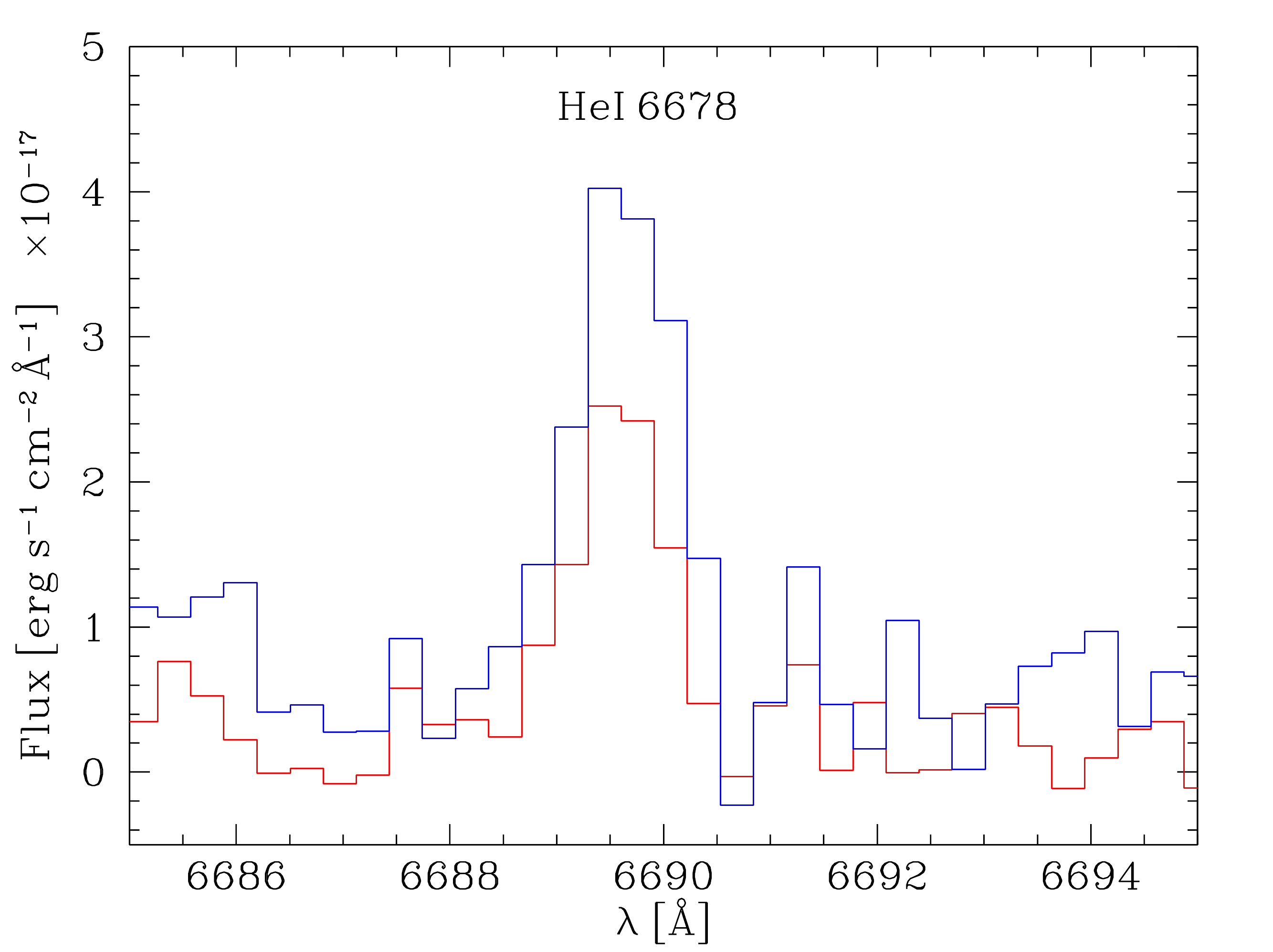}
\includegraphics[width=0.23\textwidth,angle=-90]{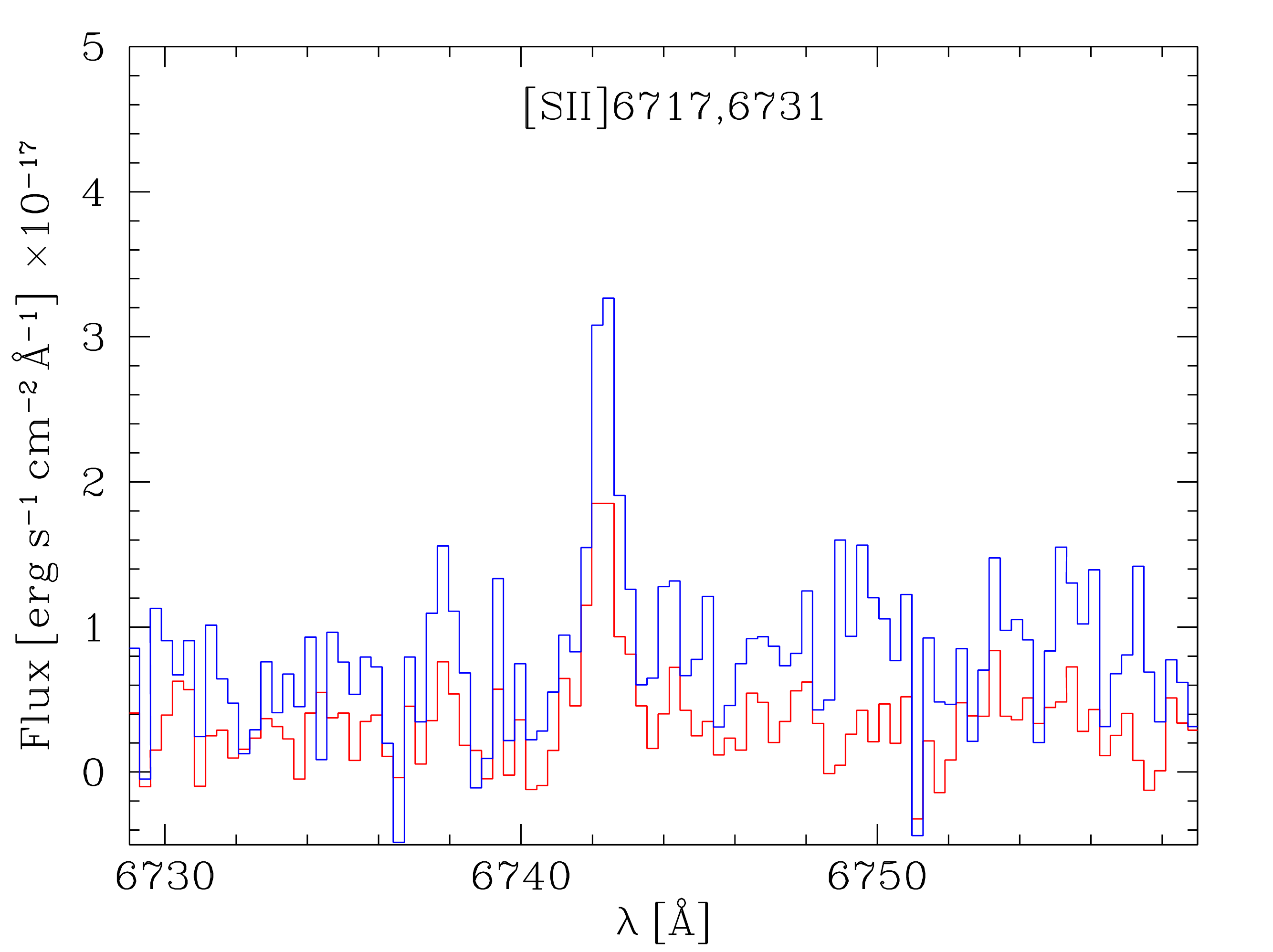}
\includegraphics[width=0.23\textwidth,angle=-90]{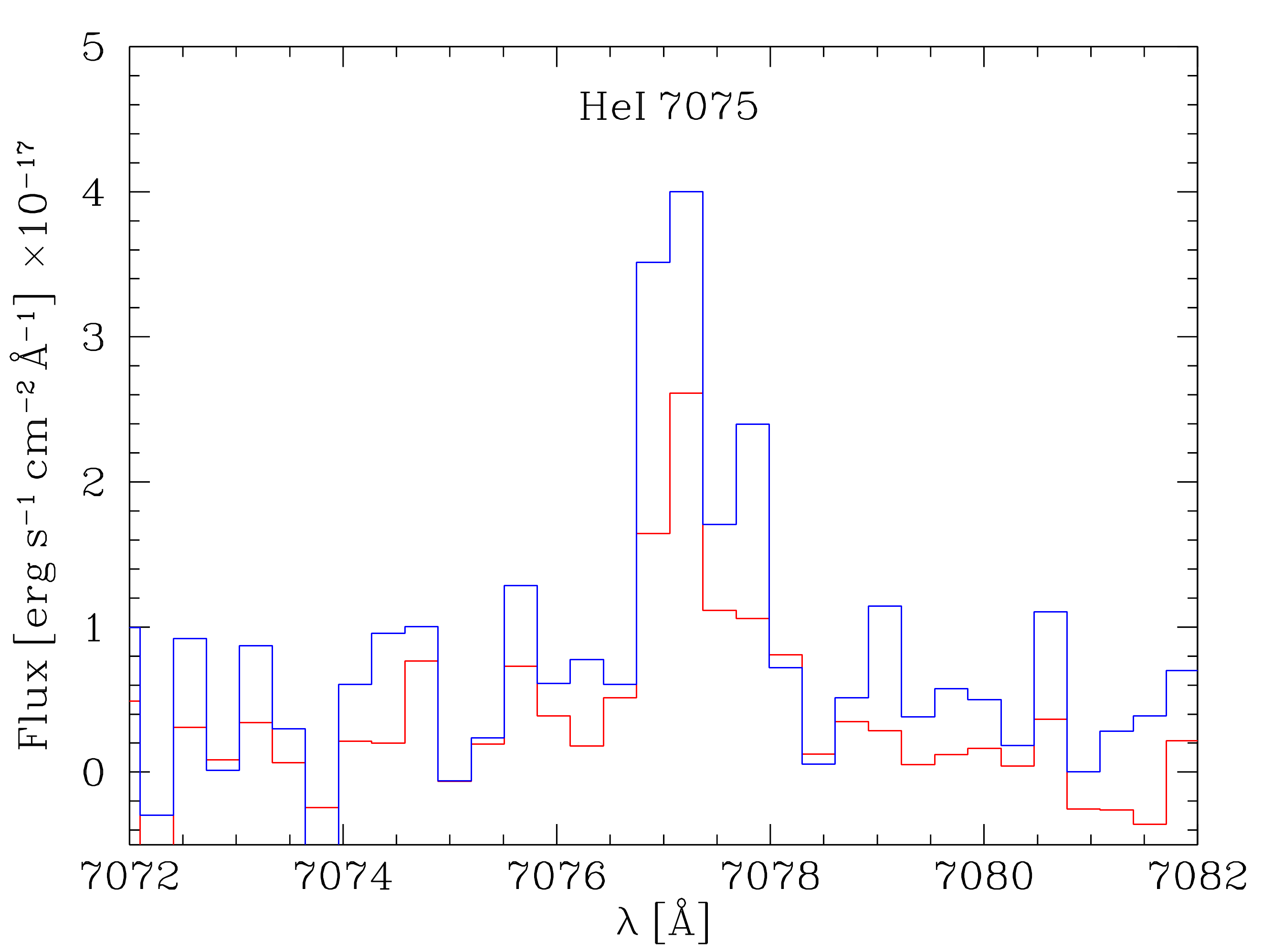}
\includegraphics[width=0.23\textwidth,angle=-90]{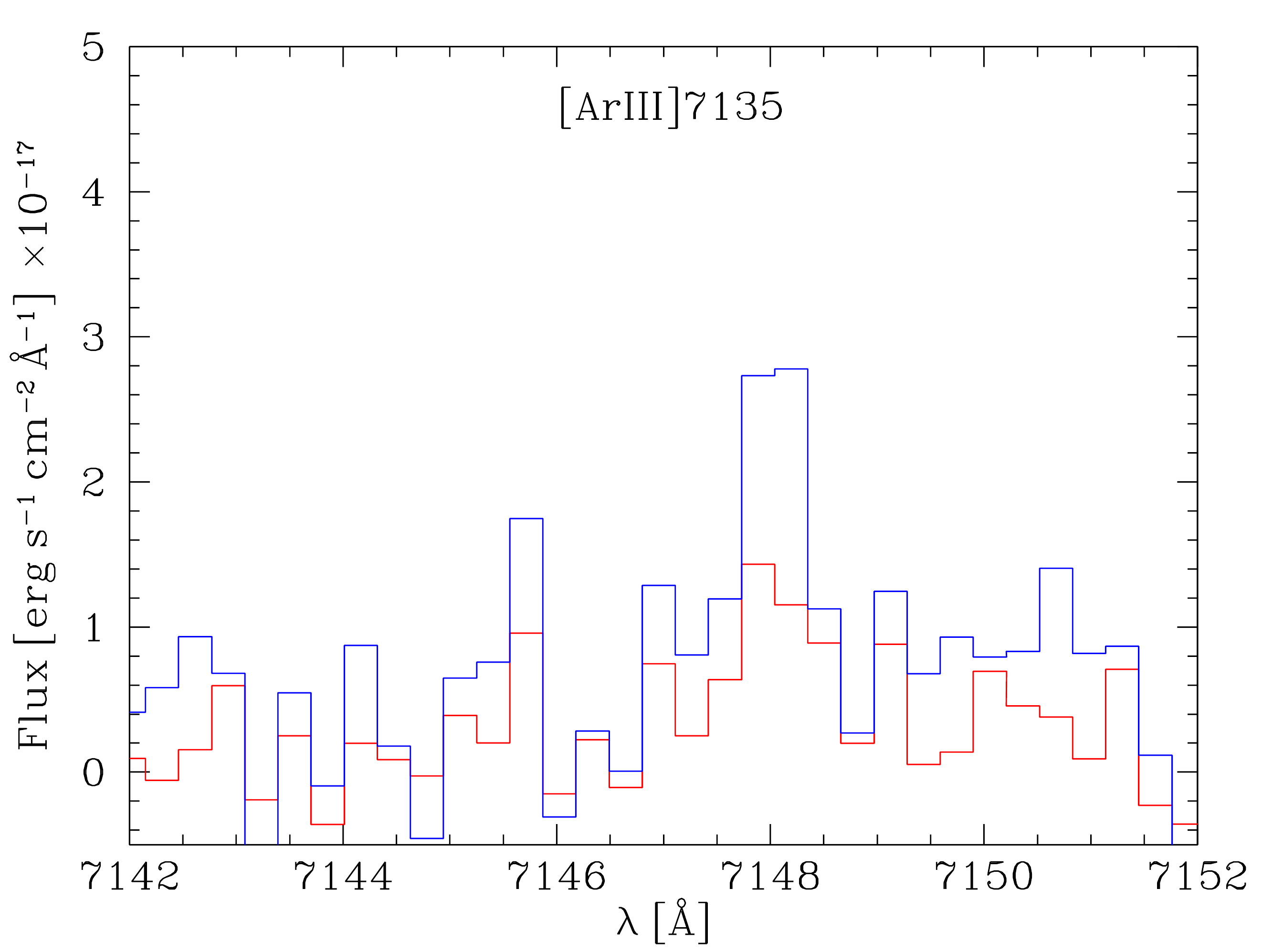}
\caption{Accumulated spectra for the LR-R  setup for two areas: one with the central ring and the first ring and the other adding contributions of the central ring and two rings, as labelled, for the spectral ranges where [\ion{N}{ii}]~6584, \ion{He}{i}~6678, [\ion{S}{ii}]~6717, [\ion{S}{ii}]~6731, \ion{He}{i}~7065 and  [\ion{Ar}{iii}]~7135 emission lines are located.}
\label{Spectra_acc}
\end{figure*}

\subsection{Aperture effect in the emission line regions and the ionizing photon budget}
\label{aperture-effect}

We measured the emission lines in the spectra extracted using different circular apertures that would be equivalent to different slit widths. The corresponding spectra are shown in Fig.~\ref{Spectra_ap}. In this case each spectrum includes the ones from the inner apertures, thus, the last one correspond to the integrated spectra of the global region of 140\,pc.
 The measurements of the emission lines are shown in Table~\ref{TabLRBaper} and Table~\ref{TabLRRaper}, for the LR-B and LR-R spectra, respectively. The  first four  columns in both are: (1) the distance from the central brightest spaxel to the limit of the extraction aperture, (2) the corresponding distance in arcsec, (3) the most external ionized ring included in the aperture  being 0 the central spaxel and considering 22\,pc between two consecutive rings and (4) the number of spaxels added in the spectrum for that aperture.  Columns\, 5, 7, 9, 10, 11, 12 and 13 of Table\,\ref{TabLRBaper} are the observed absolute fluxes, in units of $\times$ 10$^{-17}$\,erg\,cm$^{-2}$\,s$^{-1}$, of H${\beta}$, H${\gamma}$, [\ion{O}{iii}]\,4363, [\ion{O}{iii}]\,4959, [\ion{O}{iii}]\,5007, \ion{He}{i}\,4471 and \ion{He}{i}\,5015, respectively. Columns 6 and 8  give the EW of H${\beta}$ and H${\gamma}$. Table\,\ref{TabLRRaper} displays the observed flux and EW of H${\alpha}$ in columns\, 5 and 6, respectively, and columns\,7 to 13 show the flux, in units of $\times$ 10$^{-17}$\,erg\,cm$^{-2}$\,s$^{-1}$, of [\ion{N}{ii}]\,6584, \ion{He}{i}\,6678, [\ion{S}{ii}]\,6717, [\ion{S}{ii}]\,6731, \ion{He}{i}\,7065, [\ion{Ar}{iii}]\,7135 and \ion{He}{i}\,7281 emission lines, respectively. 

\begin{figure*}
\includegraphics[width=0.47\textwidth,angle=0]{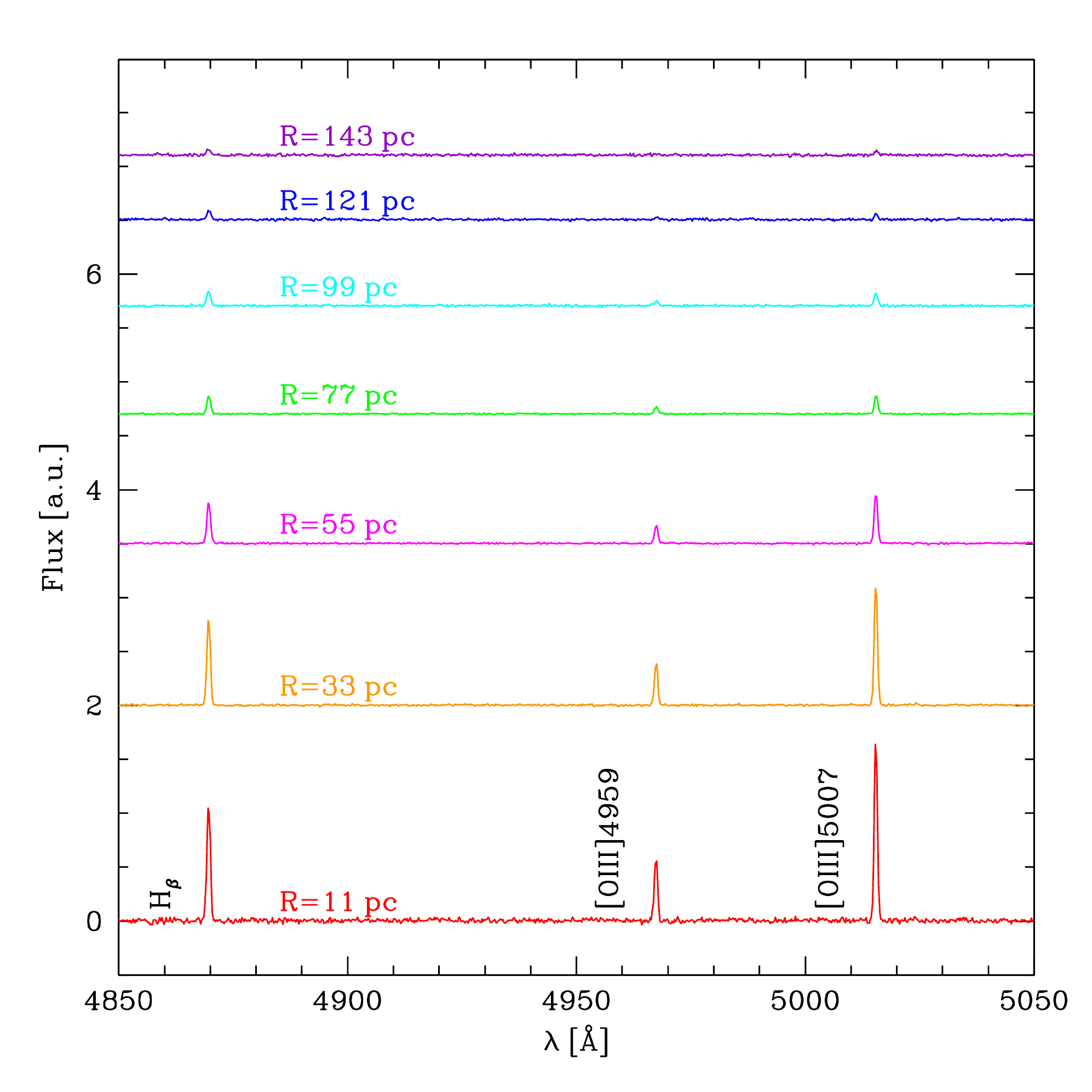}
\includegraphics[width=0.47\textwidth,angle=0]{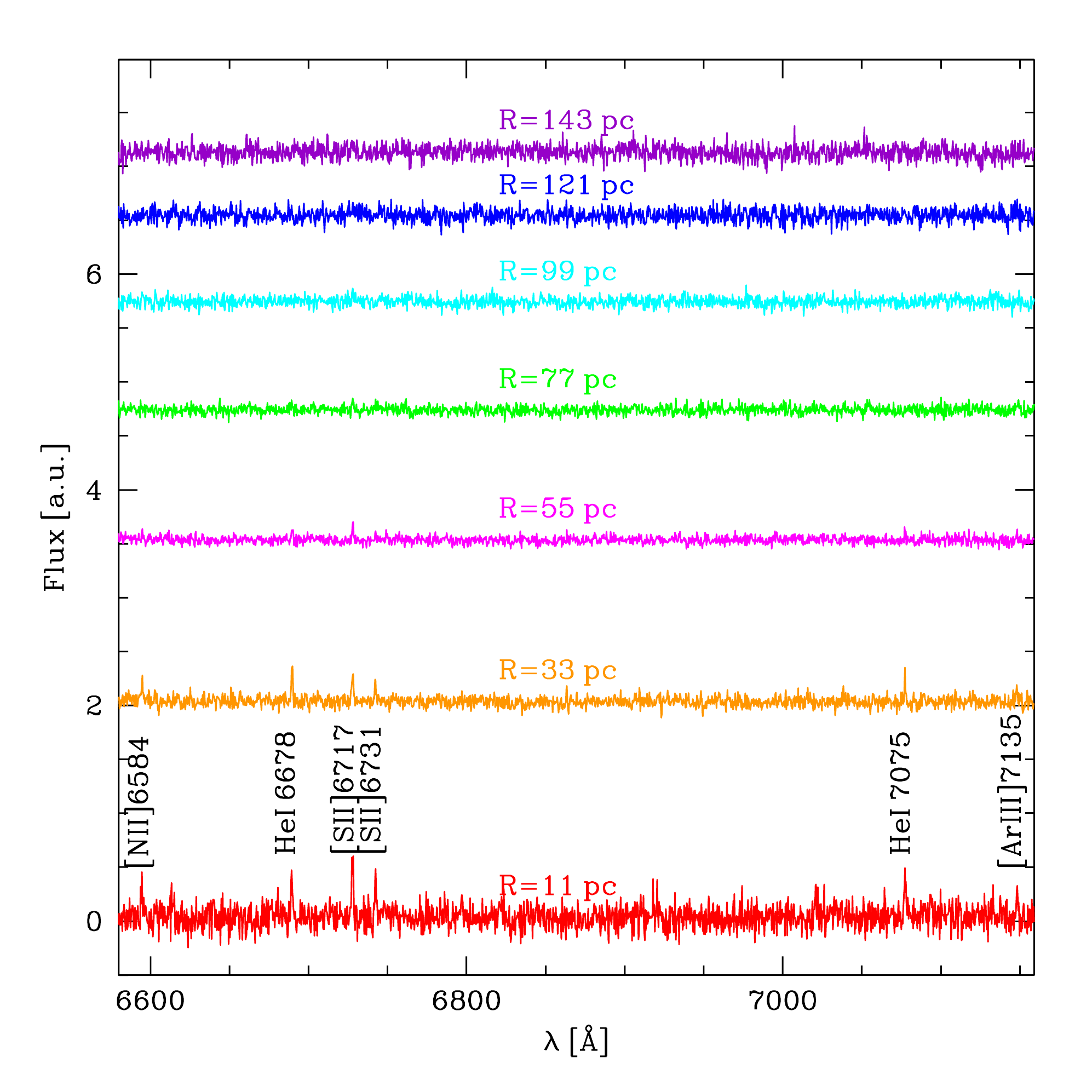}
\caption{Left: zoom view between 4850 and 5050\,\AA\ of the 1-spaxel equivalent LR-B spectra of the ionization rings.  Each spaxel  spectrum is  shifted in the Y-axis  for the sake of clarity. The spectrum of each ring is plotted in a different colour, whose label indicates the distance from the central brightest spaxel. Right: equivalent figure for LR-R  but we deliberately omit the  H${\alpha}$ 
region to show  other  very faint lines   between  6575 and  7150\,\AA.}
\label{Figspectrarings}
\end{figure*}

The fluxes  of  H${\alpha}$, H${\beta}$ and  [\ion{O}{iii}]\,5007, divided by the corresponding flux in the brightest spaxel, are shown in Fig.\,\ref{LRBlinesaper}. As expected, the flux of each line increases when the aperture is larger. In contrast, the ratio of [OIII]\,5007/H$\beta$, shown in the right panel, decreases as the size of the region increases.  The dilution of this ratio with the galactocentric distance, if confirmed in other XMD galaxies, might point to a selection effect  as the integrated detected [OIII]\,5007/H$\beta$, extensively used to extract the XMD galaxies samples,  could be linked not only to the metal content itself but also to the distance to the galaxy, in other words, to the spatial resolution. 

For the  LR-R setup, the addition of the spectra from the two first rings to the central spectrum, was particularly useful to increase the flux as the lines are not strong. This is illustrated in  Fig.\,\ref{Spectra_acc}, where we present, besides H$\alpha$, the regions of the spectra for [NII]\,6584, HeI\,6678, [SII]\,6717,\,6731, HeI\,7075 and [ArIII]\,7135. This  method allow us to see and to  measure these lines even  with  the uncertainties  shown in Table\,\ref{TabLRRaper}. However, adding more rings dilutes the weak lines increasing the noise (see Table\,\ref{TabLRRaper}) so that, given the uncertainties, we can assume that for these weak lines the   central plus the two first rings values  correspond to the global/integrated spectrum for the complete region of 140\,pc.

In our observations the spectral coverage is provided by the wavelength  range of LR-B and LR-R, as we do not have data in between. In addition, the data might have a slightly  different absolute flux calibration, despite clear observing conditions were reported by the observatory staff for the observations in both setups. To verify the general calibration in LR-B, we have compared the H${\beta}$ flux measured in our image with the reported fluxes by H16. On the one hand, in their observations at KPNO, for a  slit of 3.21\,arcsec length and 1.5\,arcsec width, these authors give a value of (18.70\,$\pm$\,0.30$)\,\times$10$^{-16}$\,erg\,cm$^{-2}$\,s$^{-1}$. This aperture is slightly smaller than our total 63 spaxels aperture, for which we measure a value for of H${\beta}$ flux of (16.07\,$\pm$\,0.06)\,$\times$\,10$^{-16}$\,erg\,cm$^{-2}$\,s$^{-1}$. Assuming an absolute flux calibration error of 10 percent  these measurements are coherent. On the other hand, with the MMT and a slit of 2.70\,arcsec $\times$ 1\,arcsec and seeing of 1.00\,arcsec, they report a value of (8.31\,$\pm$\,0.17)\,$\times$\,10$^{-16}$\,erg\,cm$^{-2}$\,s$^{-1}$. We  simulated the MMT slit in the {\sc qla tool} and  extracted the corresponding spectrum, measuring a flux of 7.56\,$\times$\,10$^{-16}$\,erg\,cm$^{-2}$\,s$^{-1}$, that is also coherent with the MMT value within 10 percent. This comparison allows us to validate the LR-B calibration. 

To verify the calibration in the LR-R setup we followed the same approach, i.e. we  simulated the MMT slit in the LR-R image, extracted the spectrum and derived the value of H${\alpha}$ flux. H16 reported for their slit a value of the emission line fluxes ratio between H${\alpha}$ and H${\beta}$ of 2.7663 while we obtain 3.6402. It is hard to say if these numbers are consistent within 10 percent or if we have a real increase of the flux in LR-R by a factor of 1.3 (3.6402\,/\,2.7663). For this reason we decided to compute the ionization budget from the luminosity of H${\beta}$ (see Section~\ref{Sec:Discussion}).

\begin{table*}
\footnotesize
\caption{LR-B emission lines measurements in different ionization rings from the central brightest spaxel. Fluxes are given in units of $\times$ 10$^{-17}$ erg cm$^{-2}$ s$^{-1}$.}
\label{TabLRBrings}
\begin{tabular}{rccrrrrrccccc}
\hline
(1) & (2) & (3) & (4) & \parbox{0.95cm}{(5)} & \parbox{0.65cm} {(6)} & \parbox{0.95cm}{(7)}  & \parbox{0.55cm}{(8)} & (9) & (10) & (11) & \parbox{0.6cm}{(12)} & \parbox{0.6cm}{(13)}\\
d & d & nr & Ns & \parbox{1cm}{H${\beta}$} & \parbox{0.7cm}{H${\beta}$} & \parbox{1cm}{H${\gamma}$}  & \parbox{0.55cm}{H${\gamma}$} & [\ion{O}{iii}]4363 & [\ion{O}{iii}]4959 & [\ion{O}{iii}]5007 &  \parbox{1.2cm}{\ion{He}{i}4471} & \parbox{1.2cm}{\ion{He}{i}5015} \\
 pc & '' &  &  & \parbox{1cm}{Flux} & EW (\AA) & \parbox{1cm}{Flux} & EW (\AA) & Flux & Flux & Flux & Flux & Flux \\
11	&	0.31	&	0 	&	1  &	10.50$\pm$1.03 & 213$\pm$168	&	5.09$\pm$1.45	& 44$\pm$43 &	0.38$\pm$0.44	& 5.18$\pm$0.73 & 14.42$\pm$1.06	& 0.26$\pm$0.26 &	0.20$\pm$0.17 \\
33	&	0.93	&	1	&	6  &	7.87$\pm$0.44	 &	339$\pm$193	&	3.91$\pm$0.58	& 21$\pm$~8 &	0.24$\pm$0.17	& 3.55$\pm$0.37	&  9.95$\pm$0.44	& 0.41$\pm$0.13 &	0.14$\pm$0.07 \\
55	&	1.55	&	2	&	12 &	3.91$\pm$0.24	 &	116$\pm$~43	&	2.21$\pm$0.39	& 15$\pm$~5 &	0.17$\pm$0.13	& 1.53$\pm$0.12	&  4.34$\pm$0.20	& 0.11$\pm$0.08 &	0.05$\pm$0.04 \\
77	&	2.17	&	3	&	18 &	1.67$\pm$0.11	 &	~41$\pm$~~9	&	1.24$\pm$0.21	&  ~8$\pm$~2 &	--	 	        & 0.64$\pm$0.07	&  1.63$\pm$0.10	&	--	          &	--              \\
99	&	2.79	&	4 	&	8  &	1.42$\pm$0.13	 &	~19$\pm$~~4	&	0.94$\pm$0.28	&  ~4$\pm$~1 &	--	            & 0.48$\pm$0.07	&  1.15$\pm$0.12	&	-- 	          &	-- 	            \\
121	&	3.41	&	5	&	6  &	1.00$\pm$0.11	 &	~17$\pm$~~4	&	0.46$\pm$0.28	&  ~3$\pm$~2 &	--	            & 0.22$\pm$0.09	&  0.43$\pm$0.09	&	--	          &	--              \\
143	&	4.03	&	6 	&	4  &	0.66$\pm$0.16	 &	~~16$\pm$~~9	&	0.42$\pm$0.34	&  ~2$\pm$~2 &	--              & 0.09$\pm$0.08	&  0.37$\pm$0.11	&	-- 	          &	--	            \\
\hline
\end{tabular}
\end{table*}
\normalsize

\begin{figure*}
\includegraphics[width=0.90\textwidth,angle=0]{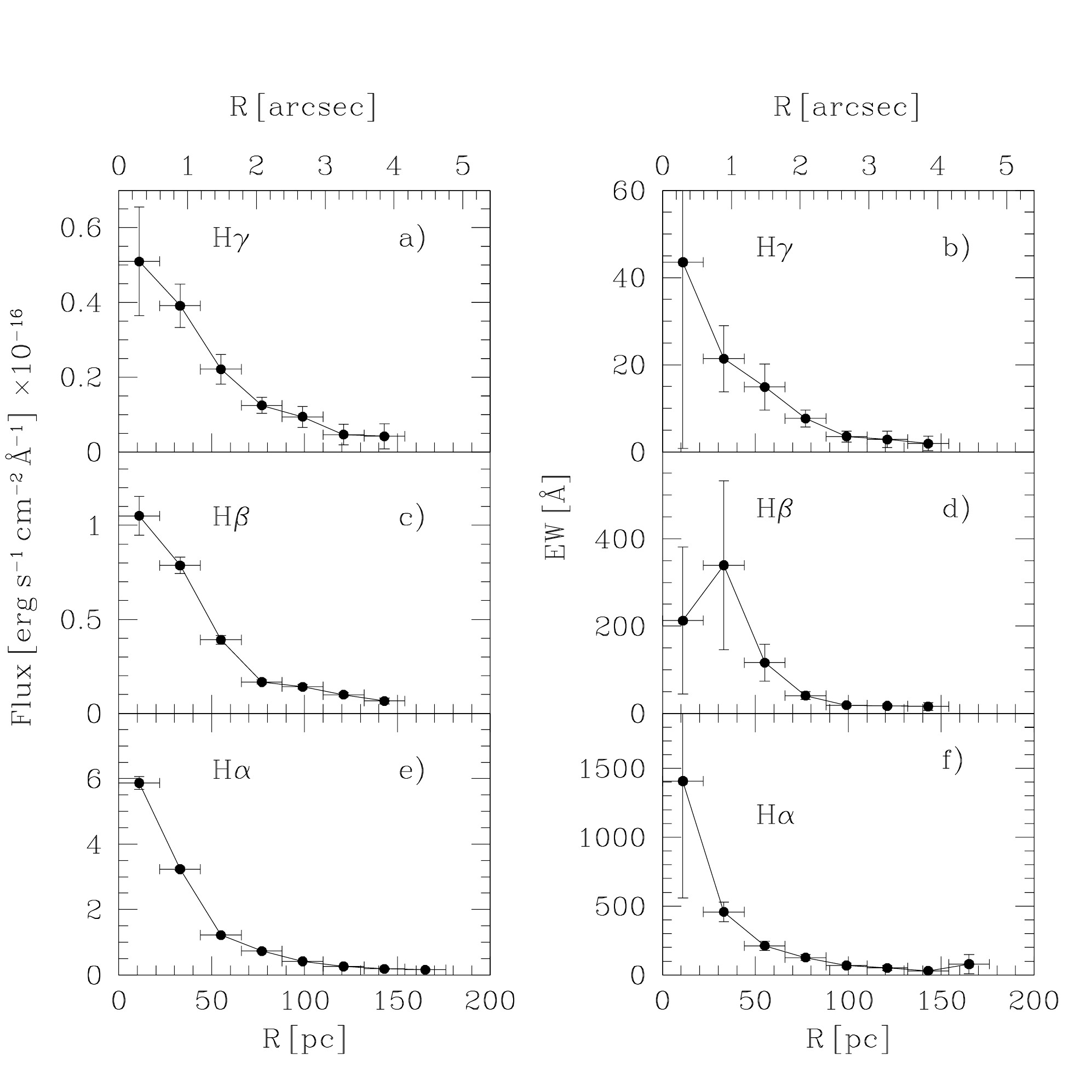}
\caption{Radial dependence of the observed flux (left) and EW (right) of the three Balmer lines present in the LR-B and LR-R spectral ranges.}
\label{FigBalmerradial}
\end{figure*}

Here, we briefly discuss the gas properties derived from the emission line spectrum. The low metallicity of this galaxy produces a large value of $T_{\rm e}$ as derived from the [\ion{O}{iii}]\,4363,\,4959,\,5007 emission lines. Due to the  low S/N of the [\ion{O}{iii}]\,4363 line in our data, we have derived $T_{\rm e}$ only in a central aperture adding 7 spaxels (the brightest one\,+\,6 around) equivalent to a radius of 33\,pc. We used the $R_{\rm O3}$ ratio, (I\,[\ion{O}{iii}]\,4959\,+\,I\,[\ion{O}{iii}]\,5007)\,/ I\,[\ion{O}{iii}]\,4363, with a value of 52.18\,$\pm$\,32.24, and derived $T_{\rm e}$ according to equation 5 from \citet{pm14}, obtaining a value of 17\,284\,$\pm$\,5\,789\,K. The intrinsic weakness of [\ion{O}{iii}]\,4363 and the associated error bars as large as 50 percent make difficult to have a more accurate determination. This value of $T_{\rm e}$ is consistent within the errors with the value of 19\,130\,$\pm$\,800\,K from \citet{cannon11}. We tried to  calculate the value of $n_{\rm e}$ in the same central aperture from the ratio $R_{S2}$ = I\,[\ion{S}{ii}]~6717\,/\,I\,[\ion{S}{ii}]\,6731, following equation 2 in \citet{pm14}, obtaining  1.59\,$\pm$\,0.49 compatible with a low density regime, $n_{\rm e}$ < 40\,cm$^{-3}$, for the derived $T_{\rm e}$. However, the error in $n_{\rm e}$ due to the low S/N of [\ion{S}{ii}]\,6717, 6731 lines makes impossible to obtain a better determination from our data. H16 did not present a more accurate estimate, giving a value in an equivalent aperture of $270 \pm 200$\,cm$^{-3}$. 

\subsection{Emission lines throughout the ionized region}
\label{ionization-rings}

We also  identified the spaxels with emission in the different ionization rings around the central brightest spaxel, and   averaged the spectra of all these spaxels, to make the measurement equivalent to one spaxel. The resulting spectrum for each ring is displayed in
Fig.\,\ref{Figspectrarings} where a zoom  view into regions of the spectrum in LR-B (left) and LR-R (right), are plotted in a different colour with a label, indicating the distance from the central brightest spaxel, using the scale factor of 22\,pc\,spaxel$^{-1}$.

\begin{figure}
\includegraphics[width=0.47\textwidth,angle=0]{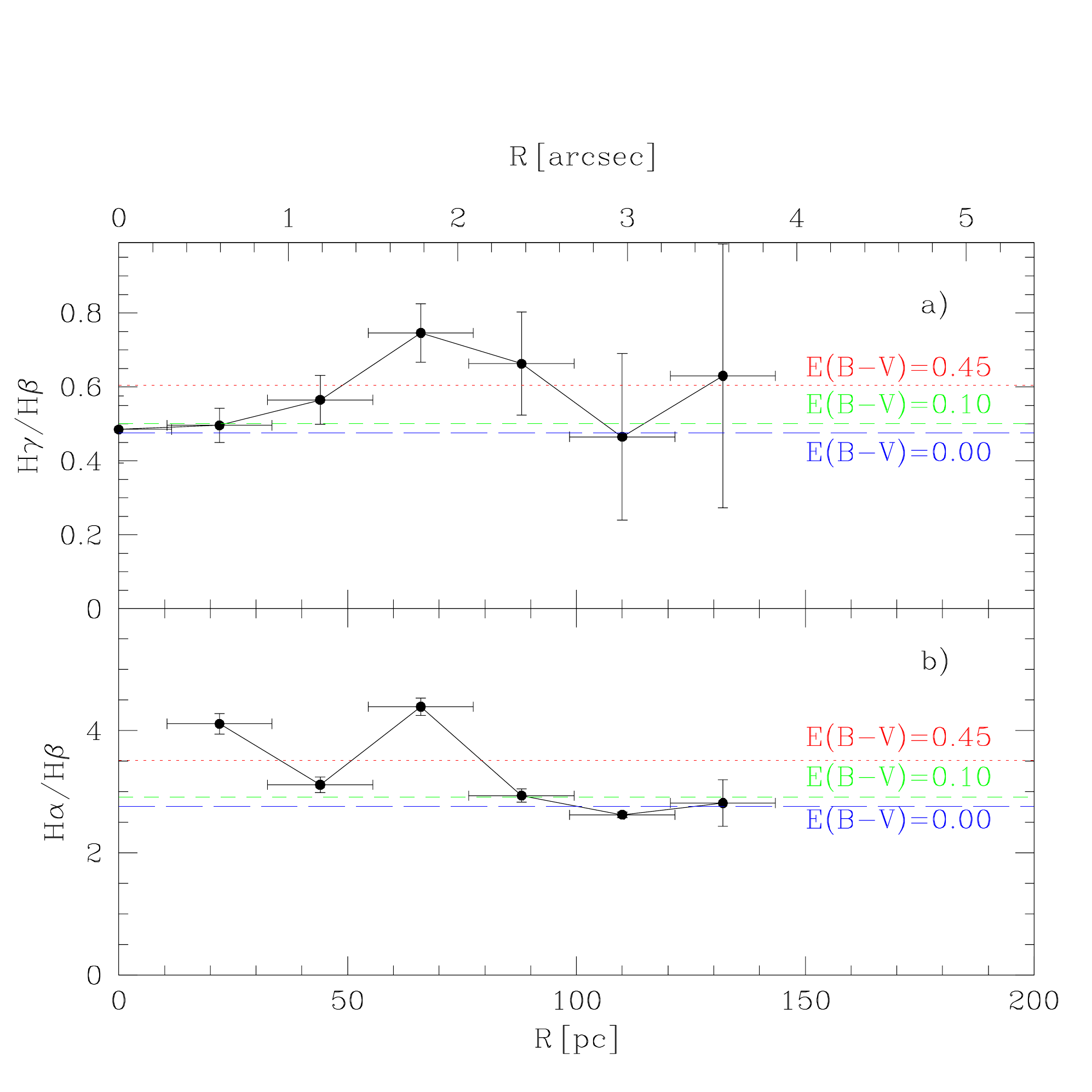}
\caption{Radial dependence of the Balmer lines ratios. The plot shows three lines with the theoretical ratios after applying a certain value of reddening, $E(B-V)$, as labelled in the plot. }
\label{FigBalmerratios}
\end{figure}

\begin{figure}
\includegraphics[width=0.47\textwidth,angle=0]{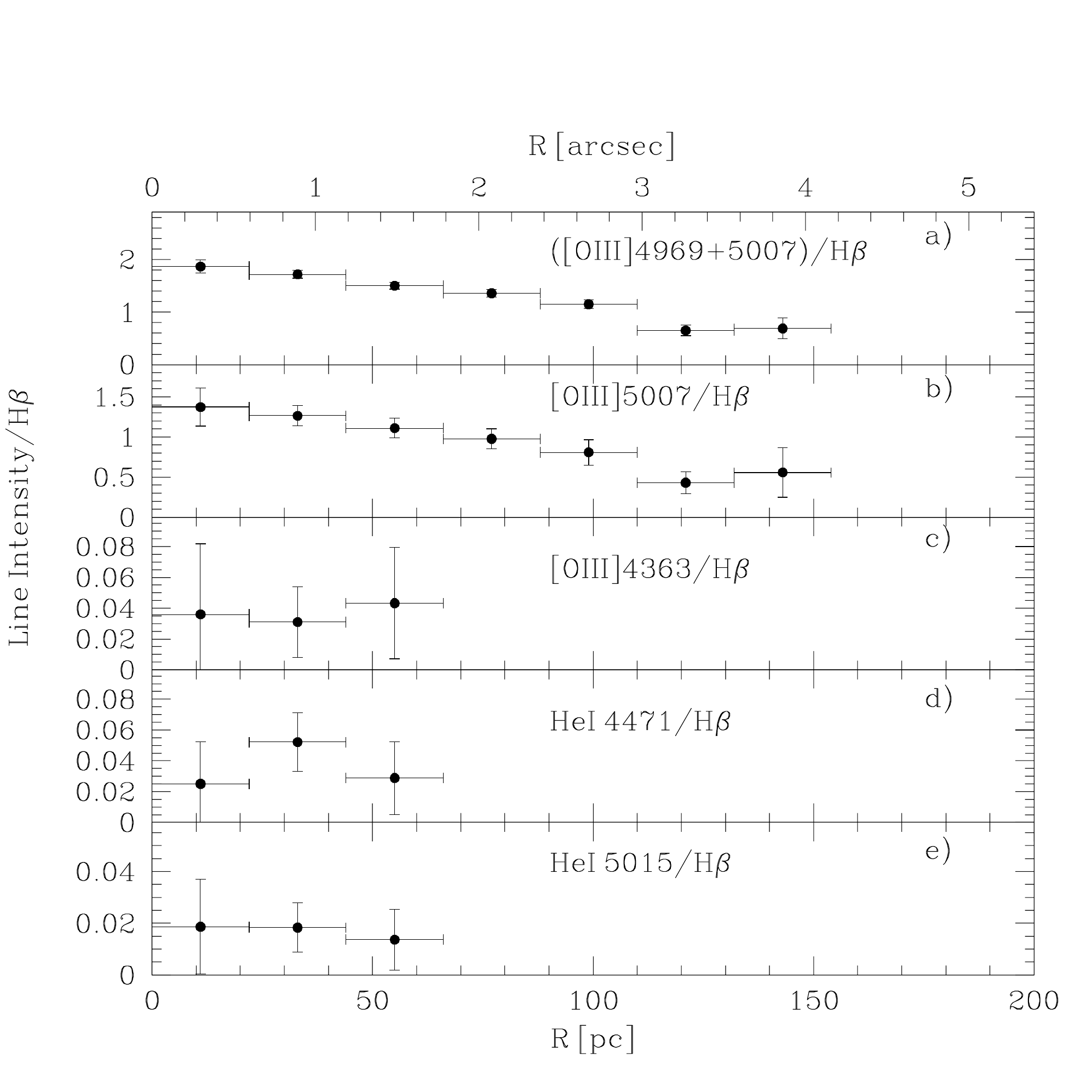}
\caption{Radial dependence of some emission lines from LR-B spectra. a) [OIII]\,4969+5007/H$\beta$; b) [OIII]\,5007/H$\beta$; c) [OIII]\,4363/H$\beta$; d) HeI\,4471/H$\beta$; and e) HeI\,5015/H$\beta$.}
\label{LRBlinesrings}
\end{figure}

Table\,\ref{TabLRBrings} presents the results for the emission lines in the LR-B spectra. The  first four columns  are: (1) the distance from the central brightest spaxel to the ionized ring, (2) the same distance in arcsec, (3) the identification of the ionized ring, being 0 the central spaxel and considering 22\,pc between two consecutive rings and (4) the number of spaxels in the averaged spectrum. Values in columns 5, 7, 9, 10, 11, 12 and 13 are the observed fluxes, with their corresponding errors, in units of $\times$ 10$^{-17}$\,erg\,cm$^{-2}$\,s$^{-1}$ for H${\beta}$, H${\gamma}$, [\ion{O}{iii}]\,4363, [\ion{O}{iii}]\,4959, [\ion{O}{iii}]\,5007, \ion{He}{i}\,4471 and \ion{He}{i}\,5015. The EW of H${\beta}$ and H${\gamma}$ with their errors are displayed in columns 6 and 8, respectively. 
In the case of the LR-R setup, with the exception of H${\alpha}$, the results for the measurements of emission lines per spaxel as a function of the distance from the brightest one, are not included in this section due to the low S/N, as illustrated in the right panel of Fig.\,\ref{Figspectrarings}.  Nevertheless, as we have  shown in the previous  section,  by accumulating data of different apertures we were able to measure some of these lines. 

Fig.\,\ref{FigBalmerradial} shows the Balmer lines fluxes (left), and their corresponding EW (right), as a function of the distance from the brightest spaxel. A clear decrease appears in both panels, although the EW of H$\beta$ shows a maximum in the second ring.
In Fig.\,\ref{FigBalmerratios} we plot the measured Balmer ratios H${\gamma}$/H${\beta}$ (top) and H${\alpha}/$H${\beta}$ (bottom). We  overlaid the theoretical values for   $T_{\rm e}$\,=\,20000\,K and three values of the reddening, $E(B-V)$: 0.00, 0.10 and 0.45\,mag with different colors as labelled. The extinction seems closer to the red line $E(B-V)$\,=\,0.45 for some rings, while is near $E(B-V)$\,=\,0 for others, so the averaged value is $<$E(B-V)$>$ $\sim 0.23$\,mag, in agreement with the one given by MQ20 of 0.29\,mag.  Fig.\,\ref{LRBlinesrings} presents the radial dependence of the ratio of some LR-B emission lines over H${\beta}$ as a function of the distance from the brightest spaxel. We  measured  only for the three central rings, except for [OIII]\,5007,\,4969  obtained, as the Balmer lines, up to 140\,pc of distance from  the central spaxel.

\section{DISCUSSION}
\label{Sec:Discussion}

\subsection{Constraining the parameters of the ionizing population}\label{Sec:constrainingparameters}
\begin{figure*}
\centering
\includegraphics[width=0.35\textwidth,angle=-90]{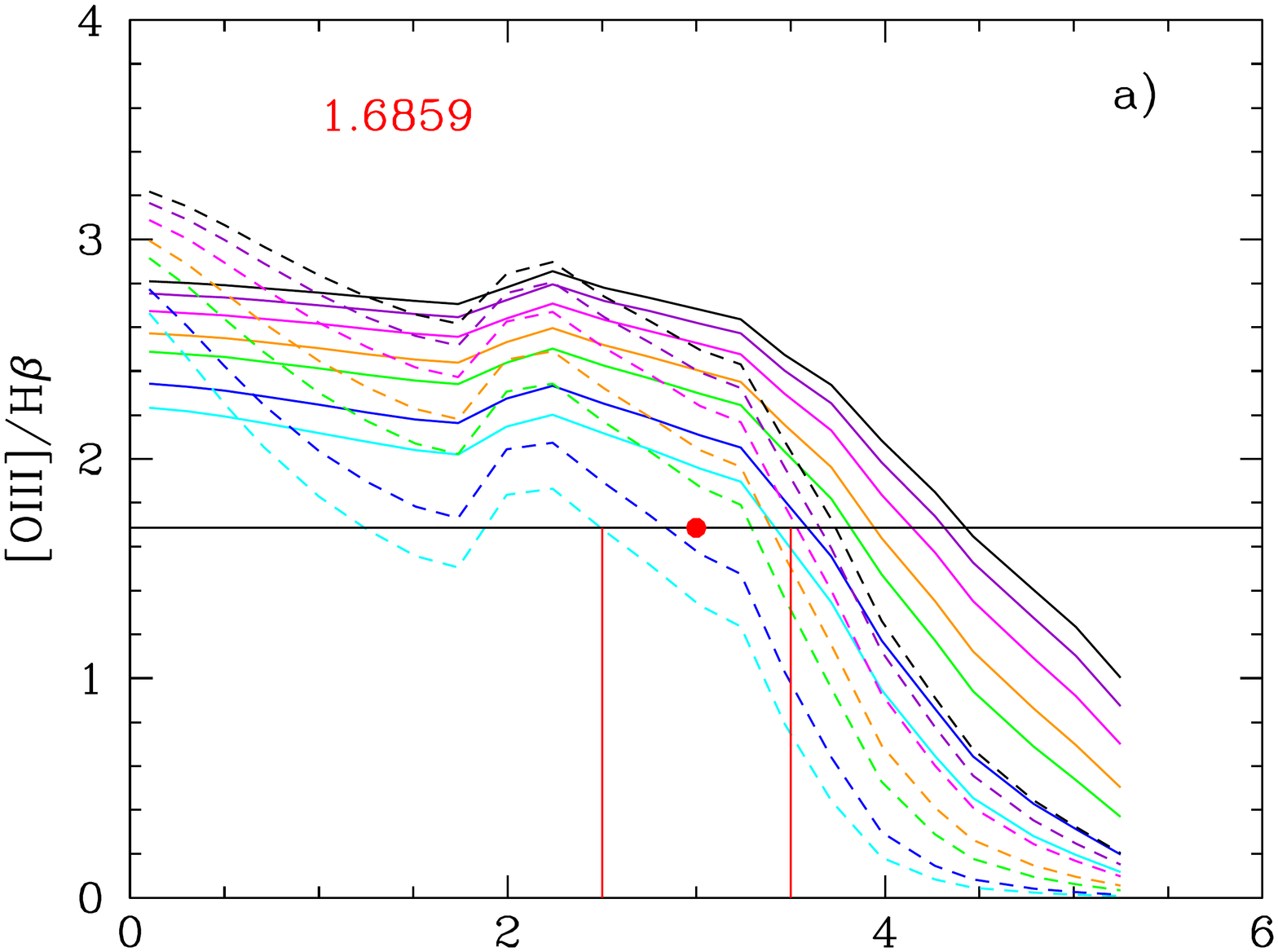}
\includegraphics[width=0.35\textwidth,angle=-90]{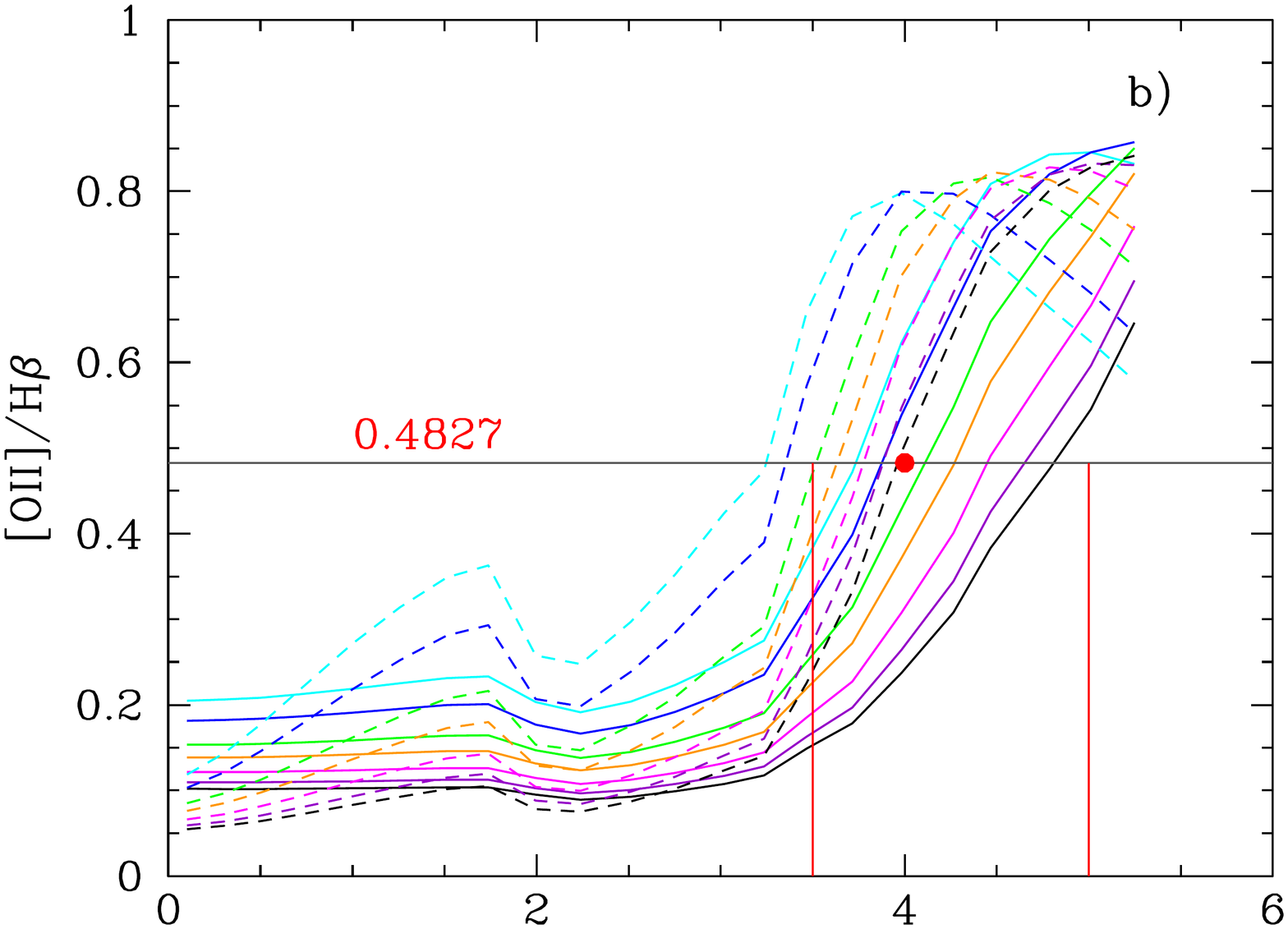}
\includegraphics[width=0.35\textwidth,angle=-90]{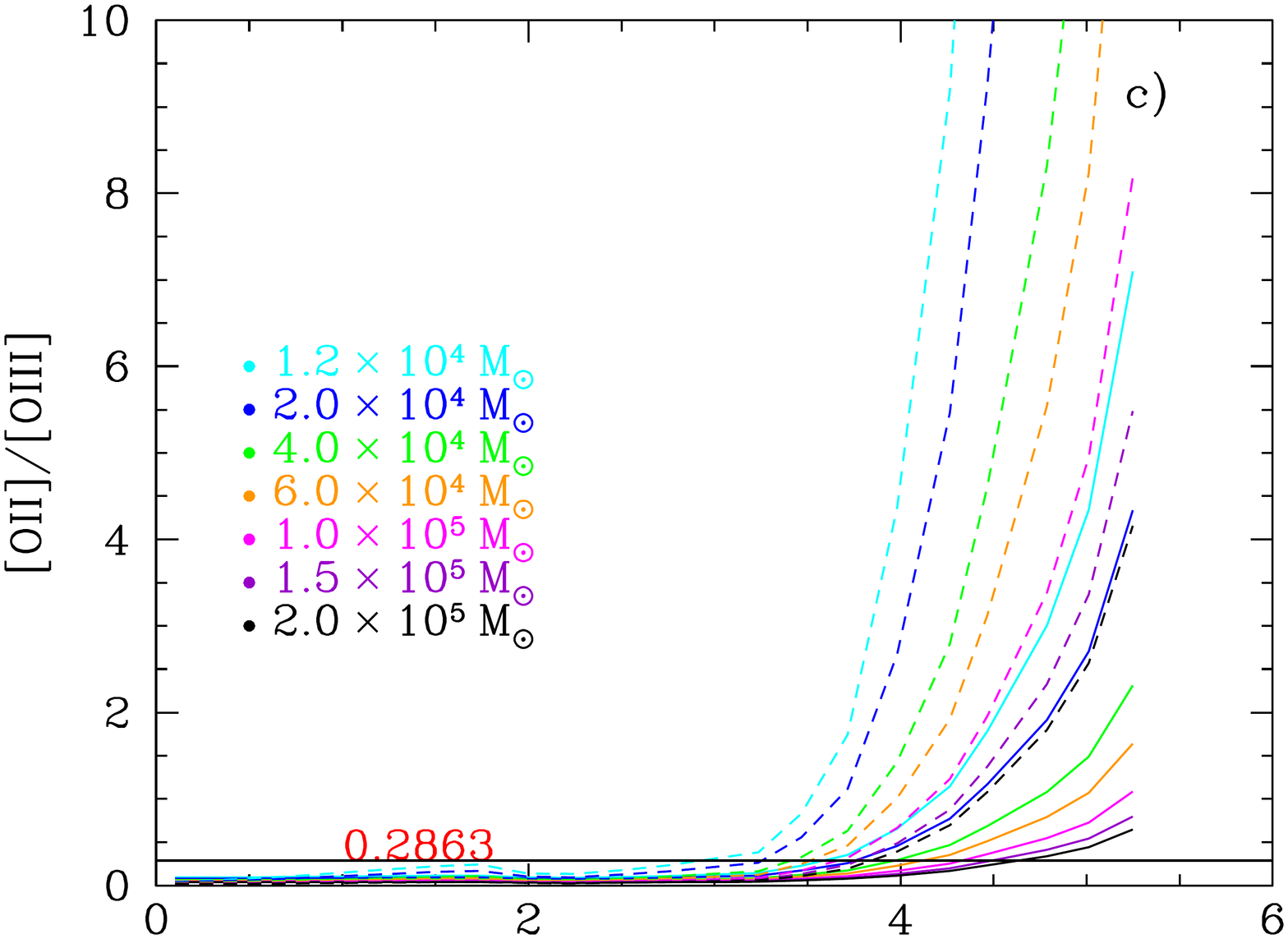}
\includegraphics[width=0.35\textwidth,angle=-90]{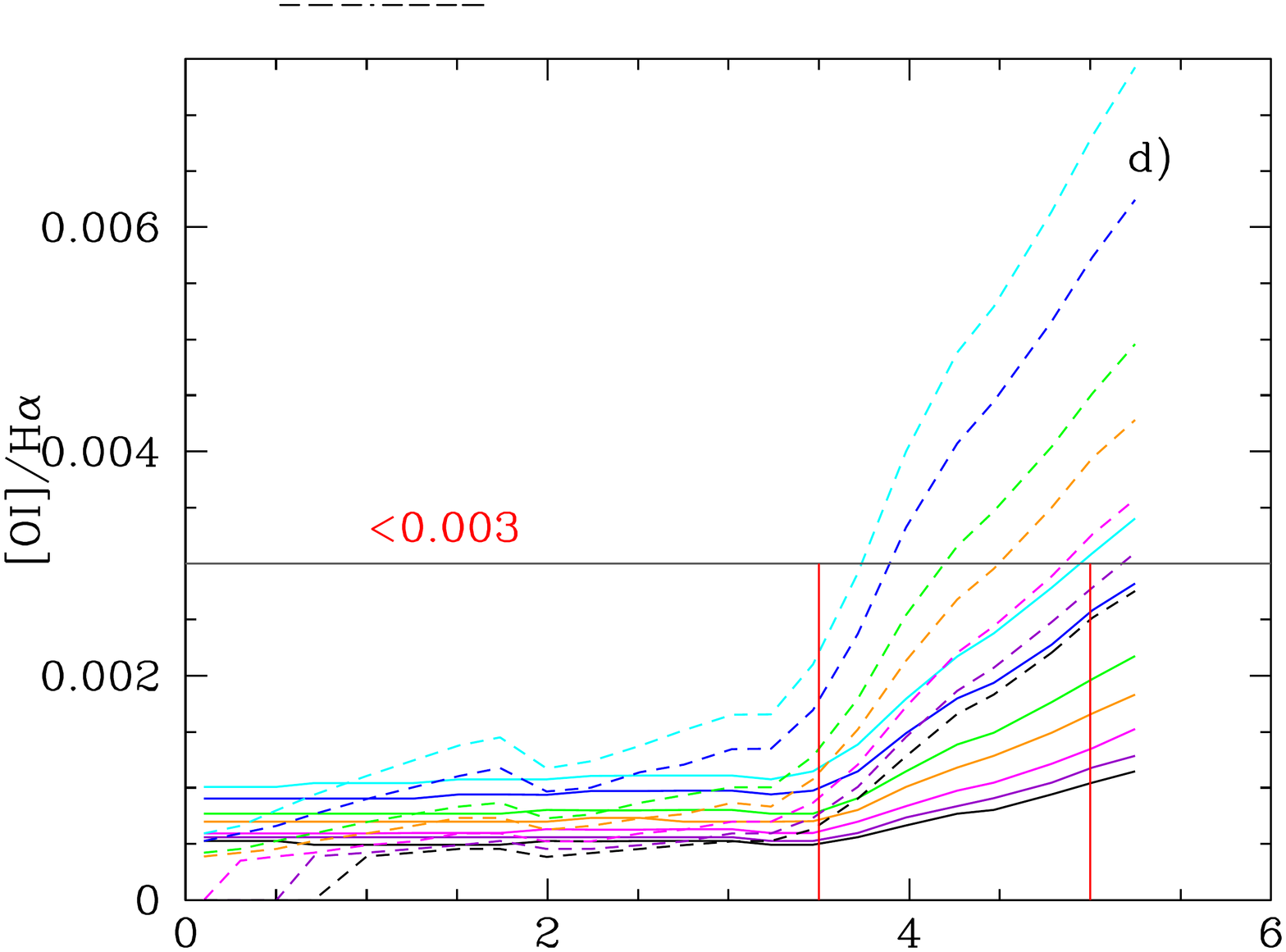}
\includegraphics[width=0.35\textwidth,angle=-90]{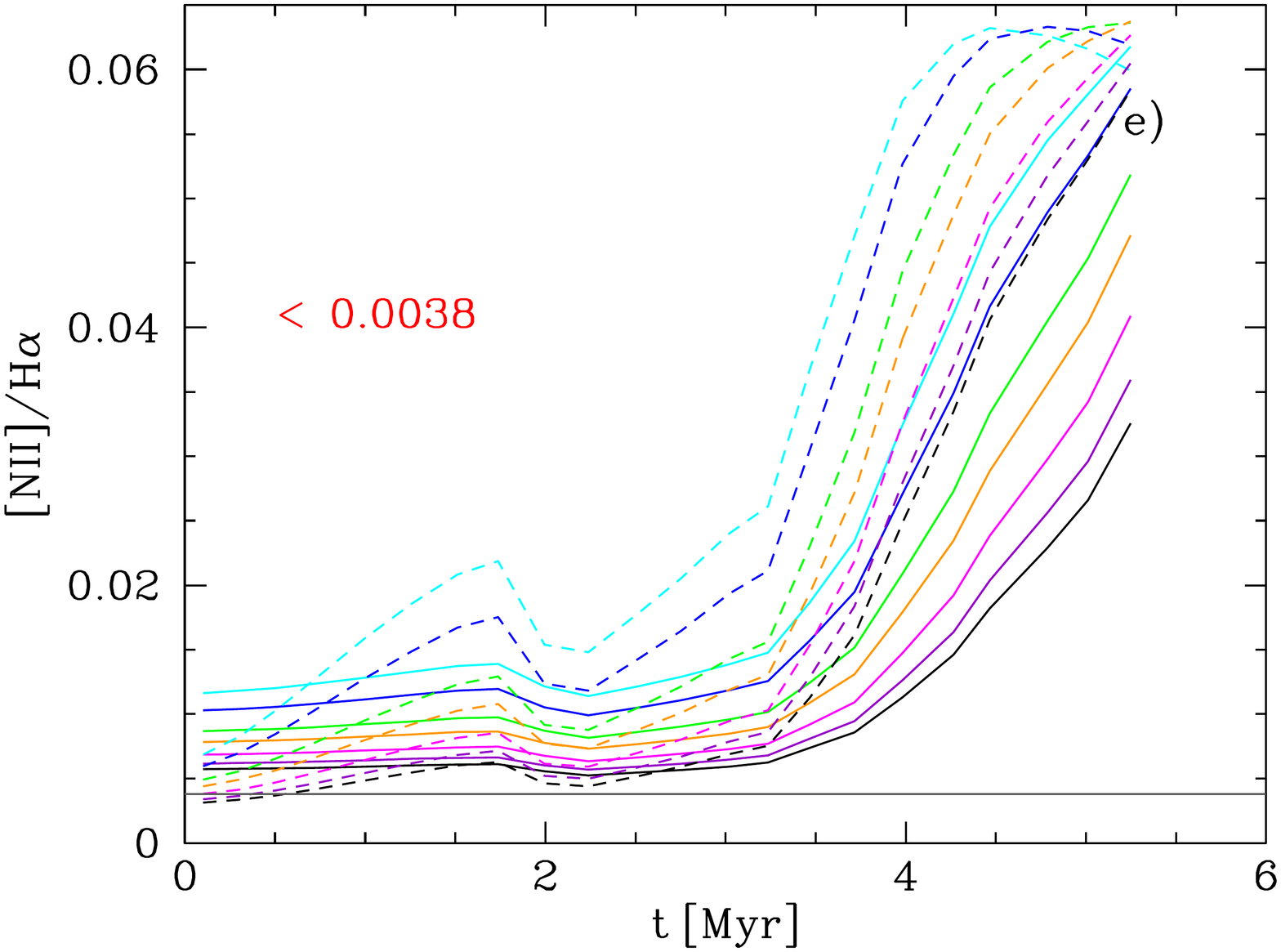}
\includegraphics[width=0.35\textwidth,angle=-90]{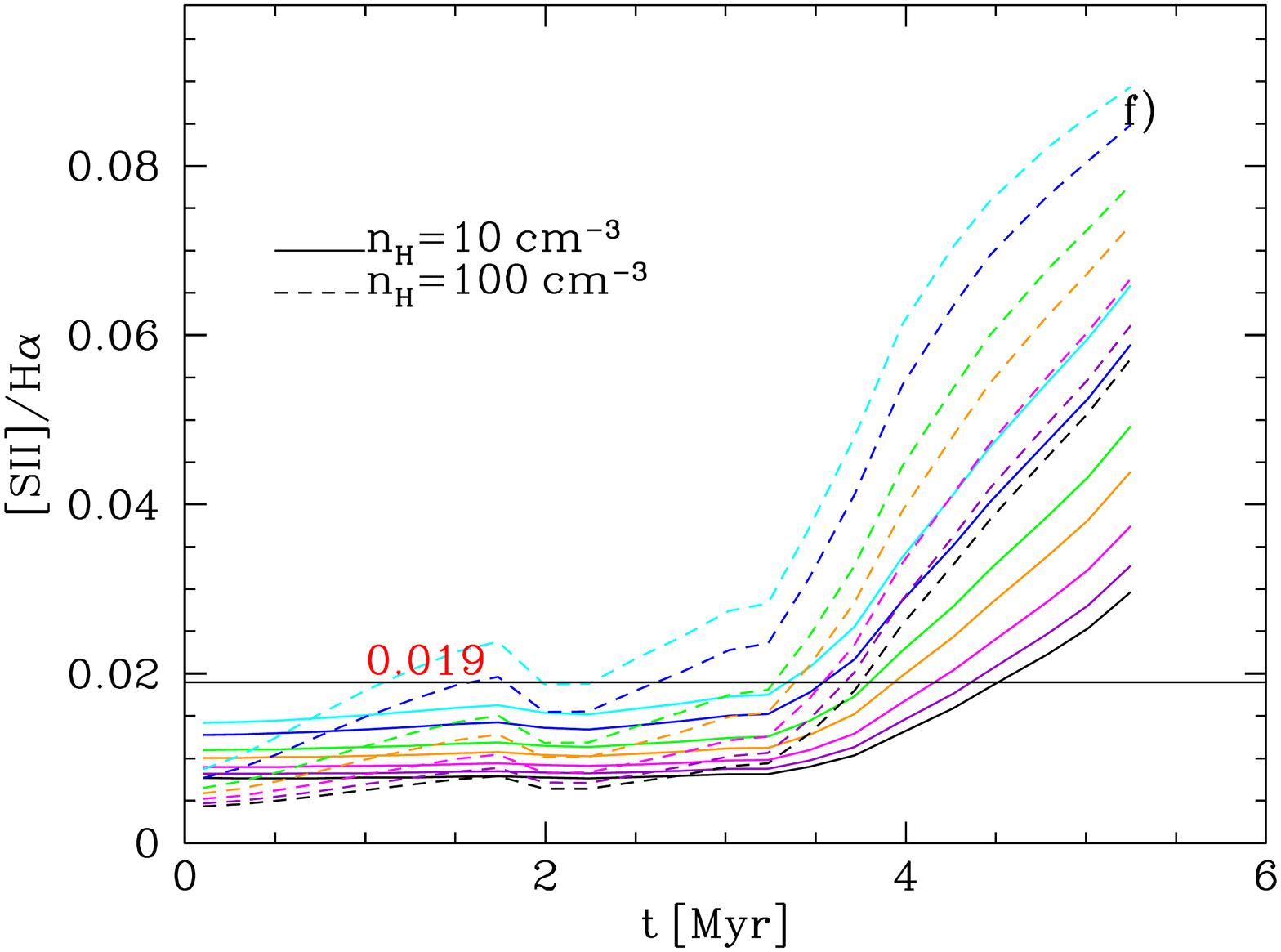}
\caption{Evolution of the emission line ratios from {\sc popstar} models (M10) of Z$ = 0.0004$ with Salpeter IMF (taken 0.15 and 100 M\textsubscript{\(\sun\)} as lower and upper mass values). The  models for different values of the cluster mass are plotted in  colors as labelled in panel c. Solid and dashed lines correspond to models with n$_{e}$ of 10 and 100\,cm$^{-3}$, respectively, as labelled in panel f. We have included as red dots, and labelling the numerical values, the measurements for the central region taken from H16.}
\label{Figpopstar1}
\end{figure*}

We dedicate this section to interpret the results of the detailed spatially-resolved study of the gas emission line spectra and kinematics presented in section~\ref{Sec:Results}. The first task is to constrain the metallicity, mass and age of the dominant young stellar cluster responsible for the ionization. To this aim we take as input: (1) the ionized gas properties obtained from the emission line spectrum, such as $T_{\rm e}$, $n_{\rm e}$ and abundance, derived in the central region, (2) the emission line ratios diagnostics in the same region, (3) the number of ionization photons, $Q(H)$, in the nebula and (4) the geometry of the region as seen in the MEGARA maps together with the information of the HST images from MQ20. 

We assume the same metallicity Z, and in particular the same oxygen abundance, for both gas and stars of the ionizing cluster. Taken the value of  $T_{\rm e}$ from the [\ion{O}{iii}] emission lines, the value of 12\,$+$\,log(O/H) is between 7.02 (H16) and 7.12 (MQ20). Considering 8.69 as the solar value, the metallicity, Z, would be in the range 0.021~Z\textsubscript{\(\sun\)}~$\leq$~Z~$\leq$~0.030~Z\textsubscript{\(\sun\)}. As the solar metallicity is Z\,$=$\,0.017, the value of the metallicity for Leoncino would be 0.0004\,$\pm$\,0.00008, so that for all comparisons in this paper with {\sc popstar} models, we will use those corresponding to Z\,$=$\,0.0004.

We  calculate the number of ionizing photons from the H$\beta$ luminosity, as:
\begin{equation}
\label{Eq:ionizingphotons}
{Q(H)} = 2.10 \times 10^{48}\, \mathrm{L(H{\beta}),}
\end{equation}

where L(H$\beta$) is the luminosity of H$\beta$, in units of 10$^{36}$\,ergs\,s$^{-1}$ calculated from the flux assuming a distance to the galaxy. We take the flux of H$\beta$ measured in our LR-B image for all the 63 spaxels, which is 16.07\,$\times$\,10$^{-16}$\,erg\,cm$^{-2}$\,s$^{-1}$. 
 Considering  the scale of 22\,pc\,spaxel$^{-1}$ (corresponding to a distance of 7.7 Mpc) we measure a L(H${\beta}$) total value of $1.140\times 10^{37}$\,erg\,s$^{-1}$ ($\log$ L(H${\beta}$) = 37.06) and a derived number of ionizing photons of 2.39\,$\times$\,10$^{49}$\,s$^{-1}$ ($\log\,Q(H)$\,=\,49.38). 
 
 We assume that the gas is ionized by the most massive stars of a young  SSP, whose properties should be derived from the gas emission line spectrum.  {\sc popstar} evolutionary synthesis models (M09) generated, for a given IMF and a set of the ionizing cluster parameters,  age and metallicity, the SEDs with the shape of the ionizing continuum, the number of ionizing photons and the inner radius of the H{\sc ii} region, $R_{\rm{in}}$, set by the action of the cluster mechanical energy.  {\sc popstar} models are the only ones computed for very low metallicity young populations, dominated by the massive stars ionizing spectrum. These models are given for a cluster mass of\,1\,M\textsubscript{\(\sun\)}. Hence, the luminosity and $Q(H)$ have to be scaled with the actual cluster mass (in M\textsubscript{\(\sun\)}) while the radius of the ionizing shell with M\textsubscript{\(\sun\)}$^{0.2}$ (see M09).

We used the estimated value of $Q(H)$ and the results from those {\sc popstar} models to infer the mass of the dominant ionizing cluster. M09 models predict values of 46.75, 46.58, 46.19 and 45.88 s$^{-1}$ M\textsubscript{\(\sun\)}$^{-1}$ when using the IMF from Chabrier (2003), Salpeter (1955), Kroupa (2001) and Ferrini, Penco and Palla (1990) respectively, at 3.2 Myr, and 46.71, 46.52, 46.16 and 45.85 at 3.5 Myr. The total value of log\,$Q(H)$, when adding all spaxels (63) with detected emission in H$\beta$, is
49.38\,s$^{-1}$. Assuming that all these photons are produced in a single ionizing cluster, the mass derived when using different IMFs ranges from 0.4 to 3.4\,$\times$\,10$^{3}$\,M\textsubscript{\(\sun\)}. We therefore conclude that, on the basis of the models and assuming an ionization bounded region, and no photon escaping from the nebula, the ionizing cluster mass is $\approx$1.7\,$\times$\,10$^{3}$\,M\textsubscript{\(\sun\)}, with an uncertainty of a factor of 2 including all hypotheses (IMF, reddening, and age, among others). We have to take into account that the
uncertainty in $Q(H)$ derived from stellar population synthesis models is large, since for a given cluster mass, $Q(H)$ depends not only on age and metallicity but overall on the selected IMF, and in the case of small clusters on possible stochastic effects, too. Furthermore, there is an uncertainty in the distance, e.g. assuming a distance of 13.8\,Mpc, the upper limit estimated by MQ20,  the spaxel size and the L(H${\beta}$) would increase up to a factor of 1.9 and 3.2, respectively.

The number of ionizing photons is derived not only from the value of L(H${\beta}$), which depends on the assumed distance to the galaxy, but also from the predictions of the synthesis evolutionary models. These models have general uncertainties common to any metallicity, e.g. the IMF --number of massive stars formed-- and the assumed isochrones and atmosphere models. These assumptions can change the number of ionizing photons up to a factor of 3. Moreover, the poor understanding of the evolution and atmospheric properties of low metallicity stars, add more uncertainty to the predictions for stellar populations of XMD galaxies. 

We have not considered any reddening when deriving $Q(H)$ as we have shown that the Balmer ratios are consistent with very low values of $E(B-V)$, see Fig.~\ref{FigBalmerratios}.

M10 used {\sc popstar} models and the photoionization code {\sc cloudy} \citep{ferland98} to predict the emission line spectrum produced by an ionizing SSP under certain hypotheses on the chemical composition of the gas (abundance and electron density) and an ionization bounded geometry. We  used the published results from M10 for models at Z\,$=$\,0.0004 (0.023 Z\textsubscript{\(\sun\)}), at all ionizing ages and the two values of $n_{\rm e}$ (10 and 100\,cm$^{-3}$) they used in the published models, to constrain the cluster parameters. As a reminder of the models reliability, M10 showed that the abundances derived from the computed emission line spectra through the [\ion{O}{iii}] electron temperature differ from the input metallicity in the model less than 0.05 dex in the range 0.0001\,$\leq$\,Z\,$\leq$\,0.008. Fig.\,\ref{Figpopstar1} shows the evolution up to 6\,Myr of six emission line ratios from M10 for  Z\,$=$\,0.0004 with a Salpeter IMF (taken 0.15 and 100 M\textsubscript{\(\sun\)} as lower and upper mass values). The models have cluster masses of 0.12, 0.20, 0.40, 0.60, 1.00, 1.50 and 2.00\, $\times$\,10$^{5}$ M\textsubscript{\(\sun\)}, plotted in different colors as labelled in panel c. Solid and dashed lines correspond to models with $n_{\rm e}$  of 10 and 100\,cm$^{-3}$, respectively. The emission line ratios are [\ion{O}{iii}]\,4959+5007/H$\beta$, [\ion{O}{ii}]\,3727+3729/H$\beta$, [\ion{O}{ii}]/[\ion{O}{iii}], [\ion{O}{i}]\,6300/H$\alpha$, [\ion{N}{ii}]\,6584/H$\alpha$ and [\ion{S}{ii}]\,6717, 6731/H$\alpha$. For comparison purposes, we  included the measurements for the central region taken from H16 (1.6859, 0.4827, 0.2863, <~0.003, <~0.0038, and 0.019, plotted as red dots in panels a, b, c, d, e and f, respectively). We used  H16 data because on  the one hand,  they are equivalent to our central 7-spaxel aperture, and for all lines present in both, H16 and this work, the intensity values are consistent within the errors, see section \ref{aperture-effect}. 
On the other hand, H16 measured [\ion{O}{ii}]\,3727,\,3729,  very useful to constrain  the gas density and the hardness of the ionizing radiation, but we could not measure it given the LR-B wavelength coverage.    

We compared the observed line ratios with the models for the smallest cluster (represented by the cyan lines), which is about one order of magnitude larger than the one we derive from the luminosity of H$\beta$. The values of [$\ion{O}{iii}$]/H$\beta$ (panel a) and [$\ion{O}{ii}$]/H$\beta$ (panel b) points to an age of 3\,--\,4\,Myr, although younger ages would be still compatible for higher values of electron density or less massive clusters (remember that in this case the electron density is $< 40$\,cm$^{-3}$). The observed ratio [$\ion{O}{ii}]/[\ion{O}{iii}$] (panel c) is compatible with the low density curves for clusters older than 3\,Myr. The values of both [\ion{O}{i}]\,6300/H$\alpha$ (panel d) and [$\ion{N}{ii}$]/H$\alpha$ (panel e)  fit the models with $n_{\rm e}$\,=10\,{cm}$^{-3}$ only. Finally, the value of [\ion{S}{ii}]\,6717, 6731/H$\alpha$ (panel f) discards a low-mass ionizing cluster in a high density regime (dashed cyan line) and predicts an age between 4 and 5.5\,Myr.

The EW(H$\beta$) can also constrain the ionizing cluster age in this central region. {\sc popstar} models predict a value of 250\,$\pm$\,50\,\AA\ for ages between 3 and 5\,Myr at Z\,$=$\,0.0004 and a Salpeter IMF. If we consider any other IMF used in M09 models, we obtain a similar cluster age within $\pm$ 1\,Myr. Despite  our measurement of EW(H$\beta$) in the central aperture  has a large error, 306\,$\pm$\,136\,\AA\,, see Table \ref{TabLRBaper}, this  result is compatible  with the predicted value by  M09 for the derived cluster age. 
In summary, the placement of the observed values of the central region of Leoncino in the emission-line evolutionary diagrams and the measurement of the EW(H$\beta$) are all coherent with a metallicity of Z\,$=$\,0.0004, an age of 4.0\,$\pm$\,1.0\,Myr for the ionizing cluster and low-density ($\approx$\,10\,cm$^{-3}$).

The equilibrium time scale increases as the metallicity decreases due to the low content of metals that can cool the gas. In addition, the effect is strengthened with decreasing  hydrogen density.  For this reason, we need to outline the importance of considering the equilibrium time of the photoionized nebula that, in low-metallicity galaxies, like Leoncino, is of the same order of magnitude as the age of the ionizing cluster. The time-step in {\sc popstar} models corresponds to the stellar cluster age step given by the isochrone time resolution. However, the output of the photoionization models shows the picture of the ionized region once the equilibrium state is reached. This implies that when we observe an ionized region, what we really see is the effect of an ionizing cluster of a given age, regardless the actual age of the ionized nebula. The consequence is that the age resolution with which we can date the ionizing cluster from the information of the gas ionizing spectrum depends on the speed at which the nebula reaches the equilibrium. In practice, this means that the ionizing stellar cluster would be older when we observe the region than the age we infer from the emission-line spectrum, since the cluster will have evolved during the nebula equilibrium time, controlled by its metallicity.

\begin{table*}
\caption{{\sc popstar\,$+$\,cloudy} models at Z\,$=$\,0.0004 assuming $n_{\rm e}$\,$=$\,10\,cm$^{-3}$ for Leoncino.The table shows the first rows of the full table presented in the Appendix.}
\label{Leomodels}
\begin{tabular}{rcccccccccccccc}
\hline
(1) & (2) & (3) & (4)  & (5) & (6) & (7) & (8) & (9)  & (10)	  & (11) & (12) & (13) & (14)\\
EW(H$\beta$) & $\log$\,Q(H)$^{*}$ & log\,age & $\log$\,Q(H) & $\log$\,R$_{\mathrm{in}}$ & [\ion{O}{ii}] & [\ion{O}{ii}] & H$\gamma$ & [\ion{O}{iii}] & [\ion{O}{iii}] & [\ion{O}{iii}] & H$\alpha$ & [\ion{S}{ii}] & [\ion{S}{ii}] \\
\AA & s$^{-1}$M\textsubscript{\(\sun\)}$^{-1}$ & yr & s$^{-1}$ & cm & 3726 & 3729 & 4340 &	4363 &	4959 & 5007 & 6563 & 6717 & 6731 \\

415	&	46.66	&	6.30	&	50.00	&	19.55	&	0.1608	&	0.1088	&	0.4759	&	0.0391	&	0.5021	&	1.4980	&	2.8252	&	0.0270	&	0.0190	\\
415	&	46.66	&	6.30	&	49.95	&	19.55	&	0.1654	&	0.1119	&	0.4759	&	0.0382	&	0.4939	&	1.4736	&	2.8258	&	0.0277	&	0.0195	\\
415	&	46.66	&	6.30	&	49.90	&	19.55	&	0.1701	&	0.1151	&	0.4758	&	0.0372	&	0.4855	&	1.4486	&	2.8266	&	0.0284	&	0.0200	\\
415	&	46.66	&	6.30	&	49.85	&	19.55	&	0.1749	&	0.1184	&	0.4758	&	0.0363	&	0.4770	&	1.4230	&	2.8272	&	0.0292	&	0.0205	\\
415	&	46.66	&	6.30	&	49.80	&	19.55	&	0.1800	&	0.1218	&	0.4758	&	0.0353	&	0.4683	&	1.3971	&	2.8280	&	0.0300	&	0.0211	\\
...	&	...	& ...	&	...	&	...	&	...	&	...	&	... & ...	&	...	&	...	&	...	&	... 	&	... \\
\hline
\end{tabular}
\end{table*}

M10 tested this effect by obtaining the equilibrium time for each {\sc cloudy} model. They concluded that neither the total number of ionizing photons (cluster mass) nor the ionizing spectrum shape (cluster age) have a significant role in setting the equilibrium time (actual age) of the ionized regions, which is mainly controlled by the electron temperature (given by the metallicity). According to those results, the average equilibrium age, t$_{\mathrm{eq}}$ is reached around 13.6 and 3.1\,Myr for metallicities Z\,$=$\,0.0001 and 0.0004 in {\sc popstar} + {\sc cloudy} models. This implies that, at such low metallicity, there is a delay between the actual time when we observe the ionized region and the age of the stellar cluster responsible for this ionization. At Z\,$=$\,0.0004, this delay is of the order of 3\,Myr, and therefore,  we could observe low-metallicity regions apparently and in practice ionized by a young cluster, but without finding the expected young and hot stars when we observe directly inside the nebula, just because they have evolved. Moreover, we could find more evolved stars radiating in the IR as RSGs (with an age between 6.5 to 7.5\,Myr at Z\,=\,0.004) as the resulting products of the evolution of the ionizing stars that can produce, among other effects, a decrease of the values of the EW of Balmer emission lines. All these effects complicate considerably the determination of the stellar populations in low metallicity and even more at low-density environments. 

Summarizing all the previous constraints, from the comparison between the observations and {\sc popstar} models and assuming an ionization bounded region and a low electron density ($\approx$10\,cm$^{-3}$), we obtain that a cluster with mass of young stars of $\approx$\,1.7\,$\times$\,10$^{3}$ M\textsubscript{\(\sun\)}, with an uncertainty of a factor of 3, a metallicity around or slightly lower than Z\,=\,0.0004, and an age of 4\,$\pm$\,1\,Myr, is responsible for the ionization. At this metallicity, the time for the nebula to reach the equilibrium is $\approx$\,3\,Myr,  implying that the ionizing cluster would have an age of 6\,$\pm$\,1\,Myr at the time of observing the ionized nebula. At this age, {\sc popstar} models (Padova's isochrones) predict the appearance of RSG and their absorption Balmer lines would produce a decrease of the effective emission EW measured in the region.

\subsection{Possible scenarios for the observed ionization structure}\label{possiblescenarios}

The strength and novelty of MEGARA is the capability of mapping the actual gas distribution with spatial resolution. This has allowed us to measure, for the first time, the total flux of the Balmer lines in emission and to produce maps in different emission lines ratios, unveiling the ionization structure of the galaxy and  the region kinematics, as discussed in sections \ref{maps} and   \ref{kinematics}, respectively.  From both, the published data and our MEGARA observations, we know that a young ionizing cluster is required to reproduce the observed emission line spectrum of the gas at least in the central two rings around the brightest spaxel. In this section we explore  different scenarios that could explain simultaneously the central emission line spectrum and  the spatial distribution of  [\ion{O}{iii}]/H$\beta$ and the EW(H$\beta$), whose values are summarized in Table \ref{TabLRBrings}. 

\subsubsection{{\sc popstar} + {\sc cloudy} models}\label{popstarcloudy}
We  carried out this task by comparing our data with the output of the photoionization code {\sc cloudy}, assuming Leoncino metallicity and an ionization bounded geometry, and considering that the ionizing source is a {\sc popstar} SSP. These models generate the SED of an ionizing cluster, of a certain mass, age and metallicity, which would have been formed in a very short time scale, comparing with the current evolutionary age, and would be physically concentrated in the center of the region. The winds and supernova produced by the most massive stars would push the neutral gas, placing it at a certain distance, $R_{\mathrm{in}}$, where the photons would start to ionize that gas. The resulting geometry would be a plane-parallel cell of ionized gas from $R_{\mathrm{in}}$ to a distance $R_{\mathrm{out}}$, with a thickness, $\Delta$R. The photoionization models assume that the region is not limited by the amount of neutral gas to be ionized but by the number of ionizing photons, i.e. the ionization bounded scenario. The metallicity controls the ionized region evolution until reaching the equilibrium in the nebula, and {\sc cloudy} stops the computation as soon as the nebula has been cooled. 

Under these assumptions \citet[][hereafter GV13]{gvargas13} published the results from a grid of {\sc popstar} + {\sc cloudy} simulations for an age range between 1 and 5.4 Myr, two values of the electron density, 10 and 100 cm$^{-3}$, and six values of the metallicity, considering the same abundance for gas and stars. GV13 also concluded that the emission line spectrum of elements heavier than hydrogen is not produced  under the same geometry  by older clusters, up to 20 Myr, which however still produce H{\sc ii} regions, with large envelopes of Balmer lines in emission, as the gas is pushed further by the mechanical energy of winds and SNe. The SEDs of these clusters can ionize hydrogen, but the high-energy photons are missing. The combination of the two effects, lack of hard photons and large distance between the cluster and the pushed gas, produces very low values of the ionization parameter, preventing the emission line spectrum of elements heavier than hydrogen.

For this paper we produce a customized {\sc popstar}\,+\,{\sc cloudy} grid of models with a global metallicity for gas and stars  of Z\,$=$\,0.0004  and considering that the abundances of the different elements are those derived from the emission line spectrum of the central region of Leoncino. All models assume an $n_{\rm e}$\,$=$\,10\,cm$^{-3}$. The  results are given in Table \ref{Leomodels} where we only show the first  rows   while giving the full table in the appendix. We run 631 different models resulting of the ionization by clusters of ages (in logarithm) of 6.30, 6.35, 6.40, 6.44, 6.48, 6.51, 6.54, 6.57, 6.60, 6.63, 6.65, 6.68, 6.70, 6.72, 6.74, 6.90, 7.15 and 8.35, corresponding to ionizing clusters with ages between 2 and 5.5\,Myr, plus some non-ionizing clusters at 8, 12.5 and 223\,Myr as an example of older populations. The grid covers a range in the logarithm of $Q(H)$, corresponding to a cluster mass range, that produces models with different value of the ionization parameter. Table \ref{Leomodels} shows, for each model, the value of the EW of H$\beta$ (in \AA\ column 1), the logarithm of the number of ionizing photons (s$^{-1}$) per solar mass (column 2) corresponding to the cluster age (column 3, logarithm, in yr), the total number of ionizing photons in s$^{-1}$ (column 4) and the value of the inner radius, $\log$\,$R_{\mathrm{in}}$ (columns 5). These five first columns come from the {\sc popstar} SSPs. Columns 6 to 14 display the emission line ratios normalized to H$\beta$  intensity  of [\ion{O}{ii}]\,3726,\,3729,  H$\gamma$\,4340, [\ion{O}{iii}]\,4363,\,4959,\,5007, H$\alpha$ and [\ion{S}{ii}]\,6717,\,6731.  

Given the lack of firm evidence for the presence of escaping gas through a disrupted nebula discussed in section \ref{kinematics},  we cannot determine whether ionizing photons are escaping from the nebula. Thus, for the sake of simplicity, we assumed an ionization bounded nebula with no ionizing photons being able to escape from it.  In the following sections we discuss different scenarios to explain the observed ionization structure.  The ionization parameter does not depend on the assumed distance to the galaxy, so  neither the main conclusions reached in the next subsections.

\subsubsection{Scenario 1: a single young cluster is responsible for observations}\label{singleyoung}

The first scenario is based on having a single young cluster with the values of mass, age and metallicity derived in section \ref{Sec:constrainingparameters} as responsible for all the ionizing photons. Therefore, we should reproduce the observed size of the ionized region, ranging from 22\,pc (at the the center) to   $\approx$\,150\,pc. We explore this scenario under different hypotheses on mass cluster and gas distribution.

\begin{figure}
\centering
\includegraphics[width=0.38\textwidth,angle=-90]{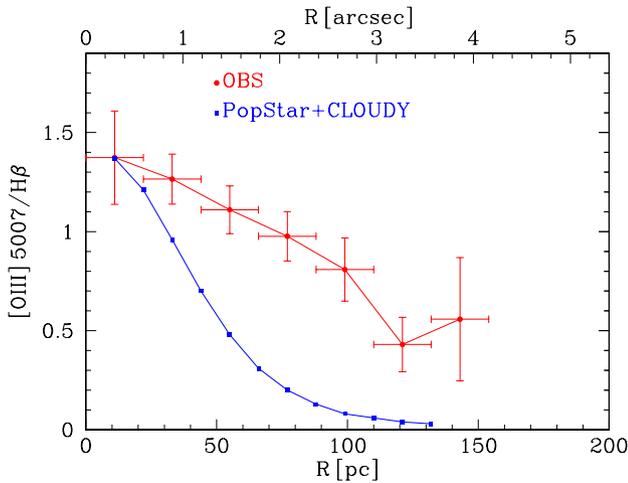}
    \caption{Values of [OIII]\,5007/H$\beta$ as a function of the distance from the center of the ionized region. The results obtained from the {\sc popstar}+{\sc cloudy} models,  assuming a single SSP as ionizing source and an uniform gas distribution throughout the nebula are shown in blue. The values and error bars derived from  MEGARA observations are shown in red. }
    \label{fig:oiii_rpc}
\end{figure}

As a  first attempt, we assume that a small ionizing cluster of mass $\approx$ 1.7\,$\times$\,10$^{3}$ M\textsubscript{\(\sun\)}, estimated from the ionization budget from L(H$\beta$),  is responsible for the ionization.
The mechanical energy from winds and supernova would push the gas placing at $R_{\mathrm{in}}$ of 29\,pc, the size of the central spaxel at this scale. To test this hypothesis, we use the results from GV13 models. We find that for such a low value of the mass cluster, these models predict $R_{\mathrm{out}}$ smaller than 40\,pc at all ionizing ages, in disagreement with the observed size of the ionization region of  $\approx$\,150\,pc. The first conclusion is that we cannot reproduce a region with a very small $R_{\mathrm{in}}$ and a relatively large $R_{\mathrm{out}}$ for  a low mass SSP, assuming that the distance to $R_{\mathrm{out}}$ is a consequence of the mechanical energy of this single small ionizing cluster exclusively. 

A second approach within this same scenario, a single cluster responsible for the ionization structure, is to assume that we might somehow have underestimated the number of ionizing photons, so that we would have a more massive cluster. Published models for a young cluster between 3 and 5\,Myr  at Z\,$=$\,0.0004 (see Table\,1 of GV13) give values of R$_{\mathrm{in}}$, $\Delta$R and R$_{\mathrm{out}}$ of 42, 46 and 88\,pc (at 3\,Myr), and 89, 3 and 92\,pc (at 5\,Myr) for a cluster of 1.2 $\times$\,10$^{4}$\,M\textsubscript{\(\sun\)}, the lowest mass of these published models. These values increase to 73, 247 and 320\,pc (at 3\,Myr) and 156, 16, 173 \,pc (at 5\,Myr) for a cluster of 2 $\times$\,10$^{5}$\,M\textsubscript{\(\sun\)}, the upper mass edge for the models presented in that paper. Scaling the models to more massive clusters, we obtain that a 3.5\,$\pm$\,0.5\,Myr cluster with a mass of 5\,$\pm$\,1\,$\times$\,10$^{4}$\,M\textsubscript{\(\sun\)} reproduces the size of the ionized region, with an inner radius of 30\,$\pm$\,5\,pc and an outer radius of 140\,$\pm$\,10\,pc. However, for such as massive cluster, the logarithm of $Q(H)$ is 51.1\,$\pm$\,0.1,  much larger than our estimated value of 49.38, for the total ionized nebula.  This mass difference is too high to be explained by distance estimate errors, a different IMF and/or intrinsic evolutionary model uncertainties. To explain that difference we would need a very large escaping fraction of photons of the region (98 per cent), leaving only 2 per cent of the produced photons to ionize the gas. From the kinematics study presented  in section \ref{kinematics}, there is no evidence at all of an enormous photon loss. 

We also explore a different geometry: instead of having the gas placed at a certain distance as resulting of the mechanical energy effect of the ionizing cluster, we consider that there is gas at all distances from the cluster, still located at the center of the region. This would be equivalent to having a superposition of ionized layers, each one corresponding to a different value of the ionization parameter, which decreases with the square of the distance between the ionizing source (the cluster) and the boundary of the ionized region. For this simulation  we  considered a SSP with $\log$\,t\,$=$\,6.54 (3.5\,Myr) and $\log$\,$Q(H)$\,$=$\,49.99. We  run ionization bounded models by changing the value of the inner radius from 11 to 132\,pc in steps of 11\,pc.  Fig.\ref{fig:oiii_rpc} presents  the ratio [\ion{O}{iii}]/H$\beta$ observed (in red) and  the models (in blue). We see that the observations cannot be reproduced either. A similar result would be obtained with a density gradient model, with the density decreasing outwards, since this would also decrease the ionization parameter. However, we do not have reliable measurements of density sensitive emission lines to be compared to the models.

Therefore, the conclusion is that a single ionizing young cluster cannot reproduce simultaneously $Q(H)$ derived from L(H$\beta$) and the size of the ionized region in an ionization bounded geometry and no photons escaping from the nebula. 

\subsubsection{Scenario 2: coexistence of two ionizing clusters}\label{twoyoung}

\begin{table}
\caption{Mass, $R_{\rm in}$, $\Delta R$, $R_{\rm out}$ and ionization parameters from {\sc popstar} model for an intermediate age cluster (6-20 Myr) at Z\,=\,0.0004 assuming $n_{\rm e}$\,=\,10\,cm$^{-3}$. }
\label{OldPop}
\begin{tabular}{rcrrcrc}
\hline
Age & $Q(H)$/M\textsubscript{\(\sun\)} & Mass & $R_{\mathrm{in}}$ & $\Delta R$ &  $R_{\mathrm{out}}$ & log~u \\
Myr & s$^{-1}$/M\textsubscript{\(\sun\)}$^{-1}$ & 10$^{3}$\,M\textsubscript{\(\sun\)} & pc & pc &  pc &  \\
    &           &           &         &       &         &          \\
6   &   45.80   &     1.2   &  67.1   & 0.8   &  67.9   &   -4.33  \\
7   &   45.61   &     1.9   &  83.5   & 0.4   &  83.9   &   -4.52  \\
8   &   45.43   &     2.8   &  97.9   & 0.2   &  98.1   &   -4.66  \\
9   &   45.29   &     3.9   & 113.1   & 0.2   & 113.3   &   -4.78  \\
10  &   45.01   &     7.4   & 139.7   & 0.1   & 139.8   &   -4.97  \\
15  &   44.16   &    52.5   & 269.9   & 0.0   & 269.9   &   -5.54  \\
20  &   43.74   &   138.0   & 386.1   & 0.0   & 386.1   &   -5.85  \\
\hline
\end{tabular}
\end{table}

 The second scenario assumes that the observed structure is the result of the combination of the actions of a very young cluster, responsible for the ionization of the two central rings, plus a few Myr older cluster, still capable to reproduce the ionization structure that we detect in [\ion{O}{iii}], H$\beta$ and H$\alpha$. Both clusters would be located at the center of the ionized region. In this case, the youngest cluster would be responsible for the ionization detected in the first two rings only, which contain 68 percent of the total H$\beta$ flux as shown in Table~\ref{TabLRBrings}, while the rest of the H$\beta$ flux (32 percent) would come from the ionization of the older cluster. This would give us values of $\log$\, $Q(H)$ of 49.21 and 48.88 for the young and the old cluster, respectively. 

The mass of the young ionizing cluster, with an age of 3.5\,$\pm$\,0.5\,Myr, would be (0.75\,$\pm$\,0.3)\,$\times$\,10$^{3}$\,M\textsubscript{\(\sun\)}. This cluster mass would be coherent with the size of the central ionized region 50\,$\pm$\,20\,pc\, and {\sc cloudy} simulations reproduce the central emission line spectrum. 

For the older population, we summarize {\sc popstar} results in Table~\ref{OldPop} for metallicity Z\,$=$\,0.0004 and $n_{\rm e}$\,=\,10\,cm$^{-3}$. Column 1 gives the cluster age in Myr, column 2 is the logarithm of $Q(H)$ for a cluster with 1 M\textsubscript{\(\sun\)}, column 3 is the actual cluster mass corresponding to the value of $\log$\,$Q(H)$\,$=$\,48.88, columns 4, 5 and 6 give the inner radius, the region thickness and the outer radius (all in pc) of the ionized nebula. Lastly, column 7 gives the ionization parameter, calculated from $Q(H)$, $n_{\rm e}$ and $R_{\mathrm{in}}$. 

We find that there is a range in age  8\,$\pm$\,2\,Myr  and mass (4.5\,$\pm$\,3.0)\,$\times$\,10$^{3}$\,M\textsubscript{\(\sun\)}  for the clusters, which can simultaneously reproduce $Q(H)$ and the region size. We  run {\sc cloudy} models for such a cluster. The conclusion is that this cluster can explain the extended region in H$\beta$ and H$\alpha$ but  cannot reproduce the observed values of [\ion{O}{iii}] across the nebula, due to the low values of the ionization parameter,  neither consequently  the large [\ion{O}{iii}]/H$\beta$ measured at large distances from the ionizing cluster(s).

\subsubsection{Scenario 3: one or several young clusters evolving while reaching region equilibrium time}\label{singleevolving}

This scenario considers that we have either, a single ionizing young cluster evolving while the nebula reaches the equilibrium time  or several small clusters at different locations throughout the nebula resulting from mass segregation.  

In the first case and considering that the region t$_{\mathrm{eq}}$\,$\approx$\,3\,Myr  for Leoncino low metallicity, the ionization would be the result of the photons produced by the evolving cluster until reaching t$_{\mathrm{eq}}$  and being cooled so that approximately between  3.5 and 6.5\,Myr.

\begin{figure}
\centering
\includegraphics[width=0.45\textwidth,angle=0]{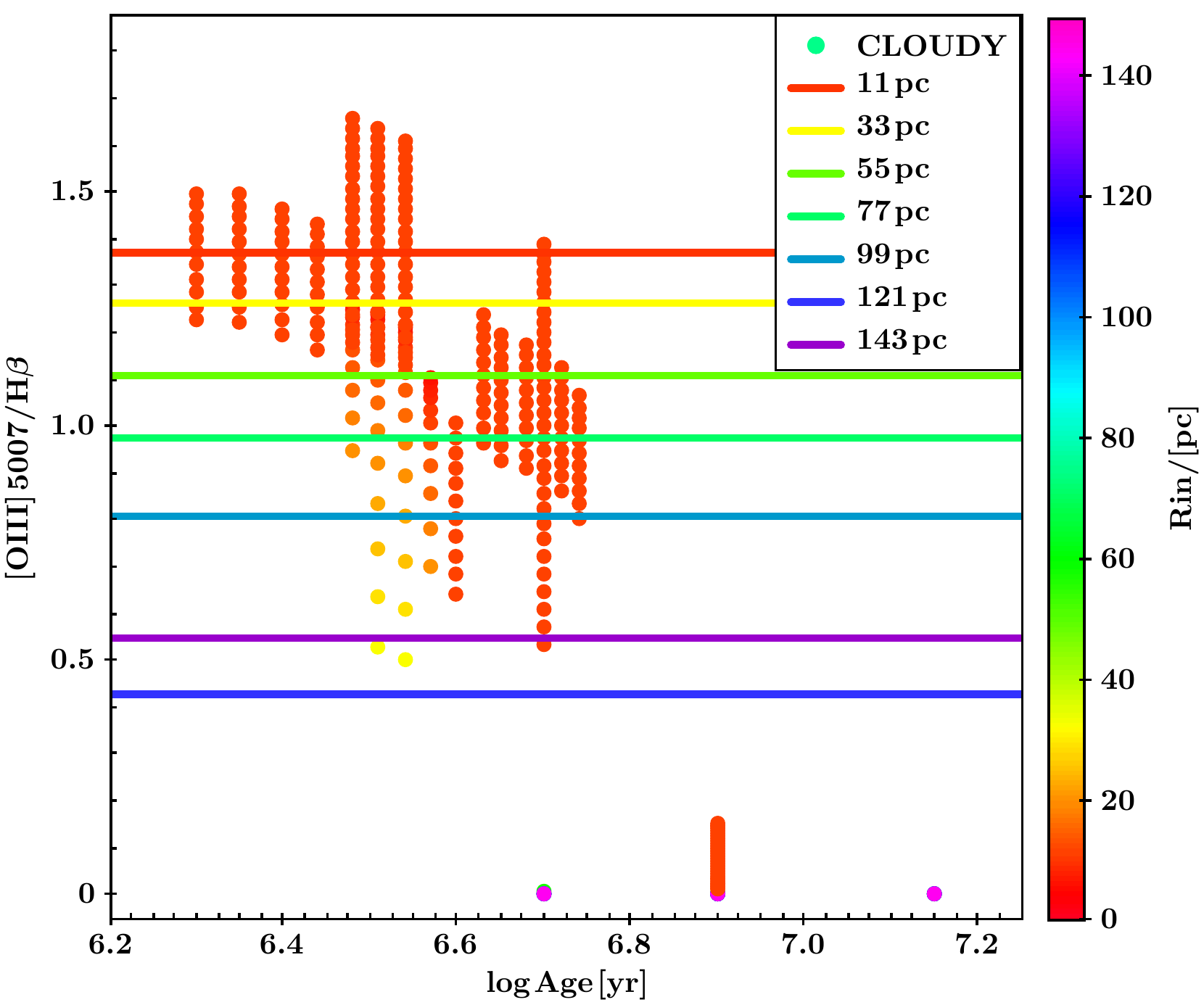}
\caption{The resulting [\ion{O}{iii}]\,5007/H$\beta$ as a function of the logarithm of the age as obtained from the set of {\sc cloudy} models shown as a scale in terms of R$_{\mathrm{in}}$ (in pc) given at right, with our observed data for this ratio of lines with different colors as labelled.}
\label{fig:oiii_logage}
\end{figure}

In the second case, we can expect some mass segregation throughout the region, producing smaller star associations along the nebula that could also explain the observed gradients. In fact, we could have both phenomena simultaneously.
The ionization structure would be equivalent to have several very young clusters of different ages, which are ionizing the gas in different locations of the nebula. The observed emission line spectrum would then be the result of the combination of the effects from the different clusters producing a series of ionized layers. We have explored this assumption by plotting in Fig.\,\ref{fig:oiii_logage} the results of the ratio [\ion{O}{iii}]\,5007/H$\beta$ as a function of the logarithm of the age of the SSP (in yr), obtained from the customized {\sc popstar}+{\sc cloudy} models described in subsection\,\ref{popstarcloudy}. Models are plotted as dots and we have overlaid the observed values of [\ion{O}{iii}]\,5007/H$\beta$ as solid lines, using different colours to distinguish the observations at different distances from the central spaxel (in red) to the outer zones (in dark blue and purple). The observations can be reproduced with SSPs of ages between 6.3 and 6.8 (in $\log$\,t). 

\begin{figure}
\centering
\includegraphics[width=0.45\textwidth,angle=0]{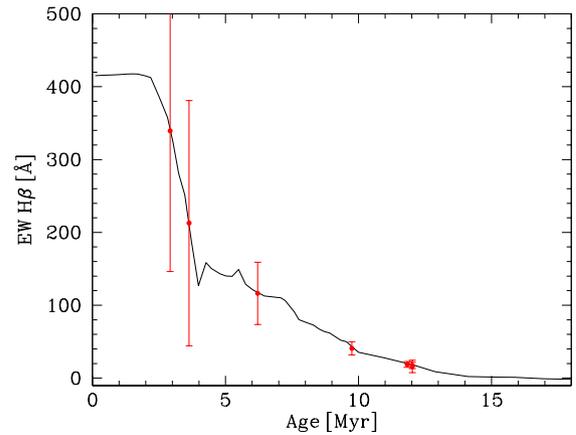}
    \caption{The evolution with time of EW(H$\beta$) as computed in popstar models compared with the results obtained in this work for different rings. The points give the possible age of each cluster in each ring.}
    \label{fig:ewhbeta}
\end{figure}

This scenario is also consistent with the measurements of the EW(H$\beta$), as it is shown in Fig.\,\ref{fig:ewhbeta}. The theoretical values (black curve) of EW(H$\beta$) from {\sc popstar} SSPs has been plotted as a function of the cluster age. The observations at different distances from the center (red points and error bars) have been overlapped. The distribution is compatible with the coexistence of several clusters of different ages. Nevertheless, the large error bars in the EW(H$\beta$) data, due to the low signal in the continuum, do not allow us to discriminate between having several young clusters with different ages, from having a single or multiple ionizing clusters and a much older stellar population in the bulge of the galaxy, which would produce the same dilution effect of the H$\beta$ emission line and the decreasing of its equivalent width.

\section{Summary and Conclusions}
\label{Sec:Conclusions}

We present for the first time spatially resolved spectroscopy of the low-metallicity galaxy AGC\,198691 (Leoncino Dwarf) obtained with the IFU of MEGARA at GTC, using the VPH gratings LR-B (4304 -- 5198\,\AA) and LR-R (6098 -- 7306\,\AA) with resolving power R\,$\approx$\,6\,000. The spatial resolution is 22\,pc\,spaxel$^{-1}$, assuming a distance to Leoncino of 7.7 Mpc. The emission line spectrum in the central region confirms the published data. 

We  obtained maps for the brightest emission lines and measured the first momenta (flux, velocity, sigma), detecting H$\alpha$, H$\beta$ and [\ion{O}{iii}] emission with S/N\,$>$\,5 as far as $\approx$\,150\,pc away from the peak emission. We find that the ionized region is slightly asymmetric, being more extended towards the NE than to the SW of the line emission peak. Our kinematics study discards density bounded conditions in the SW of the galaxy, which would indeed favor the escape of hot gas from inside the nebula. This is because we detect high excitation gas present at all velocities, i.e. both, at the core and the wings of the emission lines, and the overall spatial variation of the [\ion{O}{iii}]\,5007/H$\beta$ ratio is also rather similar, confirming that there is no high excitation gas flowing out of the region. The ionized region shows little velocity structure, indicating that either all the emission comes from a region with a common bulk motion or any rotational motion is mainly taking place in the plane of the sky. 
 We have not found any evidence of recent gas infall or loss of metals by means of outflows. This result is supported by the closed-box model predictions, which gives, for the gas fraction measured, a metallicity consistent with the oxygen abundance found by other authors and points towards Leoncino being a genuine XMD galaxy.
 
The IFU observations allow the measurement of all detected  H${\beta}$ in emission, giving a logarithm of the L(H${\beta}$)  $\approx$\,37.06 (erg\,s$^{-1}$), from which we derive 49.38 (s$^{-1}$) as logarithm of $Q(H)$ at the assumed distance to the galaxy. This implies a mass of ionized gas of $\approx$\,8\,$\times$\,10$^{3}$\,M\textsubscript{\(\sun\)}. The total mass of the ionizing clusters would be $\approx$\,2\,$\times$\,10$^{3}$\,M\textsubscript{\(\sun\)} assuming a Salpeter IMF (with 1 and 100 M\textsubscript{\(\sun\)} as mass limits) and the {\sc popstar} models. 

The distribution in equivalent width of the Balmer lines shows a negative gradient from the SW, where the line-emission peak is located, towards the NE of the galaxy. The highest values of the EW(H${\beta}$) is reached not at the position of the spaxel with the brightest line emission but in the two spaxels located immediately SW from it, suggesting a spatial segregation between the ionized gas and the ionizing stars in spatial scales of a few tens of parsecs. This is also the conclusion from the {\sc popstar}\,$+$\,{\sc cloudy} grid of models we have done to explain the ionization structure throughout the nebula. Although the emission line spectrum at the center of the ionized region (2 spaxels in radius) can be explained by a single ionizing cluster with an age of $\approx$\,3.5\,$\pm$\,0.5\,Myr, the extended observations of [\ion{O}{iii}] and Balmer lines throughout the ionized region demand photoionization by clusters of different ages, between 3.5 and 6.5\,Myr that might respond to the evolution of a single cluster evolving along the cooling time of the nebula ($\approx$\,3\,Myr at the metallicity of Leoncino, Z\,$=$\,0.0004), to mass segregation of the cluster in smaller associations or a combination of these two phenomena.

\section{Acknowledgements}
 The authors are grateful to the referee, Alec S. Hirschauer, for his constructive comments and suggestions. This work  is based on observations made with the Gran Telescopio Canarias (GTC), installed in the Spanish Observatorio del Roque de los Muchachos, on the island of La Palma. The work is based on data obtained with the MEGARA instrument, funded by European Regional Development Funds (ERDF), through the Programa Operativo Canarias FEDER 2014–2020. The authors thank the support given by Dr. Antonio Cabrera and Dr. Daniel Revert\'e, GTC Operations Group staff, during the preparation and execution of the observations at the GTC.
This work has been supported by DGICYT grant  RTI2018- 096188-B-I00, which is partly funded by the European Regional Development Fund (ERDF). 
Jorge Iglesias acknowledges finantial support from the following projects: Estallidos6 AYA2016-79724-C4 (Spanish Ministerio de Econom\'ia y Competitividad), Estallidos7 PID2019-107408GB-C44 (Spanish Ministerio de Ciencia e Innovaci\'on), grant P18-FR-2664 (Junta de Andaluc\'ia), and grant SEV-2017-0709 “Centro de Excelencia Severo Ochoa Program” (Spanish Science Ministry).

\section{Data Availability}
The fluxes of the  principal emission lines used in this work are available in the article. The reduced fits files on which these data are based will be shared on reasonable request to the first author.

\label{lastpage}


\begin{thebibliography}{}


\bibitem[\protect\citeauthoryear{Abazajian et al.} {2005}]{Aba05}Abazajian K. et al., 2005, AJ, 129, 1755
\bibitem[\protect\citeauthoryear{Ahn et al.}{2014}]{ahn14} Ahn C.P. et al., 2014, ApJS, 211, 17
\bibitem[\protect\citeauthoryear{Andrews and Martini}{2013}]{andrews13} Andrews B-H., Martini P., 2013, ApJ, 765, 140
\bibitem[\protect\citeauthoryear{Asplund et al.}{2009}]{asplund09} Asplund M., Grevesse N., Sauval A.J, Scott P., 2009, ARA\&A, 47, 481
\bibitem[\protect\citeauthoryear{Berg et al.}{2012}]{bergetal12} Berg D.A. et al., 2012, ApJ, 754, 98
\bibitem[\protect\citeauthoryear{Aver et al.}{2021}]{aver21} Aver E., Berg D. A.,  Hirschauer A.S., Olive K.A.,   Pogge R. W. Rogers N. S. J.,  Salzer J. J.,  Skillman E. D. 2021, submitted to MNRAS, arXiv:2109.00178v1 [astro-ph.GA]

\bibitem[\protect\citeauthoryear{Bouwens et al.}{2015}]{bouwens15} Bouwens R.J., Illingworth G.D., Oesch P.A., Caruana J., Holwerda B., Smit R., Wilkins S.M., 2015, ApJ, 811, 140
\bibitem[\protect\citeauthoryear{Borthakur et al.}{2014}]{borthakur14} Borthakur S., Heckman T.M., Leitherer C., Overzier R.A., 2014, Science 346, 216
\bibitem[\protect\citeauthoryear{Bressan et al.}{2012}]{bressan12} Bressan S., Marigo P., Girardi L. et al., 2012, MNRAS, 427, 127 
\bibitem[\protect\citeauthoryear{Bruzual \& Charlot}{2003}]{bc03} Bruzual G., Charlot S., 2003, MNRAS, 344, 1000 
\bibitem[\protect\citeauthoryear{Cannon et al.}{2011}]{cannon11} Cannon J.M., Giovanelli R., Haynes M.P. et al., 2011, ApJL, 739, L22 
\bibitem[\protect\citeauthoryear{Cardiel \& Pascual}{2018}]{carandpas18} Cardiel N., Pascual S.,  2018, DOI:10.5281/zenodo.2270518
\bibitem[\protect\citeauthoryear{Carrasco et al.}{2018}]{carspie18} Carrasco E. et al., 2018, Proceedings of the SPIE, Volume 10702, id. 1070216, DOI: 10.1117/12.2313040
\bibitem[\protect\citeauthoryear{Castillo-Morales, Pascual \& Gil de Paz}{2018}]{casetal18} Castillo-Morales \'A., Pascual S., Gil de Paz A., 2018, MEGARA Data Reduction Cookbook. DOI:10.5281/zenodo.3932063  
\bibitem[\protect\citeauthoryear{Chabrier}{2003}]{Chabrier2003} Chabrier G., 2003, ApJL, 586, L133
\bibitem[\protect\citeauthoryear{Chrisholm et al.}{2017}]{chrisholm17} Chrisholm J., Orlitová I., Schaerer D., Verhamme A., Worseck G., Izotov Y.I., Thuan T.X., Guseva N.G., 2017, A\&A, 605, A67
\bibitem[\protect\citeauthoryear{Cyburt et al.}{2016}]{cyburt16} Cyburt, R.H., Fields B.D., Olive K.A., Yeh T.H., 2016, RvMP, 88, 015004 

\bibitem[\protect\citeauthoryear{Dalcanton}{2007}]{dalcanton07} Dalcanton J. J., 2007, ApJ, 658, 941 
\bibitem[\protect\citeauthoryear{Dolphin}{2000}]{dolphin00} Dolphin A. E., 2000, PASP, 112, 1383 
\bibitem[\protect\citeauthoryear{Ekta \& Chengalur}{2010}]{ekta10} Ekta B., Chengalur J. N., 2010, MNRAS, 397, 963 
\bibitem[\protect\citeauthoryear{Ferland et al.}{1998}]{ferland98} Ferland G. J., Korista K. T., Verner D. A., Ferguson J. W. et al., 1998, PASP, 110, 761
\bibitem[\protect\citeauthoryear{Ferrini, Penco \& Palla}{Ferrini, Penco \& Palla}{1990}]{Ferrini_Penco_Palla_1990} Ferrini F., Penco U., Palla F., 1990, A\&A, 231, 391
\bibitem[\protect\citeauthoryear{Garc{\'{\i}}a-Vargas, Bressan, D{\'{\i}}az}{1995a}]{gvbd95a} Garc{\'{\i}}a-Vargas M.~L., Bressan A., D{\'{\i}}az A.~I., 1995a, A\&ASS, 112, 13
\bibitem[\protect\citeauthoryear{Garc{\'{\i}}a-Vargas, Bressan, D{\'{\i}}az}{1995b}]{gvbd95b} Garc{\'{\i}}a-Vargas M.~L., Bressan A., D{\'{\i}}az A.~I., 1995b, A\&ASS, 112, 35
\bibitem[\protect\citeauthoryear{Garc{\'{\i}}a-Vargas, Moll{\'a}, Mart{\'{\i}}n-Manj{\'o}n}{2013}]{gvargas13} Garc{\'{\i}}a-Vargas M.~L., Moll{\'a} M., Mart{\'{\i}}n-Manj{\'o}n M.~L., 2013, MNRAS, 432, 2746 
\bibitem[\protect\citeauthoryear{Gavil\'an et al.}{2013}]{gavilan13} Gavil\'an M.,   Ascasibar Y.,  Moll\'a M., D\'iaz A. I., MNRAS 434, 2491
\bibitem[\protect\citeauthoryear{Gil de Paz et al.}{2018}]{gilspie18} Gil de Paz A. et al., 2018, Proceedings of the SPIE, Volume 10702, id. 1070217, DOI: 10.1117/12.2313299
\bibitem[\protect\citeauthoryear{Gil de Paz, Pascual \& Chamorro-Cazorla}{2018}]{gilspie18b}{Gil de Paz A., Pascual S., Chamorro-Cazorla M.}, 2018,  guaix-ucm/megara-tools: Release v0.1.1, DOI:10.5281/zenodo.4264048 
\bibitem[\protect\citeauthoryear{Giovanelli et al.}{2005}]{giovanelli05} Giovanelli R., Haynes M. P., Kent B. R. et al., 2005, AJ. 130, 2598
\bibitem[\protect\citeauthoryear{Giovanelli et al.}{2013}]{giovanelli13} Giovanelli R., Haynes M. P., Adams E. A. K. et al., 2013, AJ. 146, 15
\bibitem[\protect\citeauthoryear{G{\'o}mez-Alvarez et al.}{2018}]{gomezalv18} G{\'o}mez-Alvarez et al., 2018, Proceedings of the SPIE, Volume 10707 id. 107071L, DOI:10.1117/12.2314220
\bibitem[\protect\citeauthoryear{Grimes et al.}{2009}]{grimes09} Grimes J. P., Heckman T., Aloisi A., Calzetti D., Leitherer C.; Martin C. L., Meurer G.,  Sembach K.,  Strickland, D., 2009, ApJS, 181, 272
\bibitem[\protect\citeauthoryear{Guseva et al.}{2015}]{gusevaetal15} Guseva N. G., Izotov Y. I., Fricke K. J., Henkel C., 2015, A\&A, 579, A11
\bibitem[\protect\citeauthoryear{Guseva et al.}{2017}]{gusevaetal17} Guseva N. G., Izotov Y. I., Fricke K. J., Henkel C., 2017, A\&A, 599, A65
\bibitem[\protect\citeauthoryear{Haynes et al.}{2011}]{haynes11} Haynes M. P., Giovanelli R., Martin A. M. et al., 2011, AJ. 142, 170
\bibitem[\protect\citeauthoryear{Haynes et al.}{2018}]{haynes18} Haynes M. P., Giovanelli R., Kent B. R. et al., 2018, ApJ. 861, 49
\bibitem[\protect\citeauthoryear{Hirschauer et al.}{2016}]{hirs16} Hirschauer A.S. et al., 2016, ApJ. 822, 108
\bibitem[\protect\citeauthoryear{Hirschauer et al.}{2018}]{hirs18} Hirschauer A.S., Salzer J. J., Janowiecki S., Wegner WG. A., 2018, AJ. 155, 82
\bibitem[\protect\citeauthoryear{Izotov \& Thuan}{2007}]{izotovandthuan07} Izotov Y. I., Thuan T. X., 2007, ApJ, 665, 1115
\bibitem[\protect\citeauthoryear{Izotov et al.}{2012}]{izotovetal12} Izotov Y. I., Thuan T. X., Guseva N., 2012, A\&A, 546, A122
\bibitem[\protect\citeauthoryear{Izotov et al.}{2015}]{izotovetal15} Izotov Y. I., Guseva N. G., Fricke K. J., Henkel C., 2015, MNRAS, 451, 2251
\bibitem[\protect\citeauthoryear{Izotov et al.}{2016a}]{izotovetal16a} Izotov Y. I., Orlitov\'a I., Schaerer D., Thuan T. X., Verhamme A. et al., 2016a, Nature 529, 178
\bibitem[\protect\citeauthoryear{Izotov et al.}{2016b}]{izotovetal16b} Izotov Y. I., Schaerer D., Thuan T. X., Worseck G., et al., 2016b, MNRAS 461, 3683
\bibitem[\protect\citeauthoryear{Izotov et al.}{2016c}]{izotovetal16c} Izotov Y. I., Guseva N. G., Fricke K. J., Henkel C., 2016c, MNRAS, 462, 4427 
\bibitem[\protect\citeauthoryear{Izotov, Thuan \& Guseva}{2017}]{izotovtg17} Izotov Y. I., Thuan, T.X., Guseva, N. G., 2017,MNRAS, 471, 1, 548

\bibitem[\protect\citeauthoryear{Izotov et al.}{2018a}]{izotovetal18a} Izotov Y. I., Thuan T. X., Guseva N. G., Liss S. E., 2018a, MNRAS 473, 1956 
\bibitem[\protect\citeauthoryear{Izotov et al.}{2018b}]{izotovetal18b} Izotov Y. I., Schaerer, D., Worseck G., Guseva, N. G., Thuan T. X., Verhamme A, Orlitov\'a I., Fricke K. J., 2018b, MNRAS 474, 4514 
\bibitem[\protect\citeauthoryear{Izotov et al.}{2018c}]{izotovetal18c} Izotov Y. I. et al., 2018c, MNRAS 478, 4851
\bibitem[\protect\citeauthoryear{Izotov et al.}{2019a}]{itg19} Izotov Y. I., Thuan T. X., Guseva N. G., 2019, MNRAS 483, 5491 
\bibitem[\protect\citeauthoryear{Izotov et al.}{2019b}]{izotovetal19} Izotov Y. I., Guseva N. G., Fricke K. J., Henkel C., 2019, A\&A, 623, A40 
\bibitem[\protect\citeauthoryear{Jaskot \& Oey}{2013}]{jaskot13} Jaskot A. E., Oey M.S., 2013, ApJ. 766, 91
\bibitem[\protect\citeauthoryear{Kewley \& Dopita}{2002}]{KD02}
Kewley L. J.,  Dopita M. A., 2002, ApJS, 142, 35
\bibitem[\protect\citeauthoryear{Khaire et al.}{2016}]{khaire16} Khaire V., Srianand R., Choudhury T. R., Gaikward P., 2016, MNRAS, 457, 4051
\bibitem[\protect\citeauthoryear{Kojima et al.}{2020}]{kojima20}  Kojima T.,    Ouchi M., Rauch M., Ono Y. et al., 2020, ApJ. 898, 142
\bibitem[\protect\citeauthoryear{Kroupa}{2001}]{Kroupa2001} Kroupa P., 2001, MNRAS, 322, 231
\bibitem[\protect\citeauthoryear{Kunth \& \"Ostlin}{2000}]{kunth00} Kunth D., \"Ostlin G., 2000, A\&A\ Rev, 10, 1
\bibitem[\protect\citeauthoryear{Leitherer et al.}{1999}]{lei99} Leitherer C., et al., 1999, ApJS, 123, 3 
\bibitem[\protect\citeauthoryear{Leitherer et al.}{2010}]{l10} Leitherer C., Ortiz Otálvaro P. A., Bresolin F., et al., 2010, ApJS, 189, 309
\bibitem[\protect\citeauthoryear{Leitherer et al.}{2014}]{l14} Leitherer C, Ekström S., Meynet G., Schaerer D., et al. 2014, ApJS. 212, 14
\bibitem[\protect\citeauthoryear{Leitherer et al.}{2016}]{l16} Leitherer C., Hernández S., Lee J. C., Oey M.S., 2016, ApJ, 823, L64
\bibitem[\protect\citeauthoryear{Marino et al.}{2013}]{MA13}
Marino R. A., Rosales-Ortega F. F., S{\'a}nchez S. F., et al. 2013, A\&A 559, A114 
\bibitem[\protect\citeauthoryear{Mart{\'{\i}}n-Manj{\'o}n et al.}{2010}]{mman10} Mart{\'{\i}}n-Manj{\'o}n M. L., Garc{\'{\i}}a-Vargas M. L., Moll{\'a} M., D{\'{\i}}az A. I., 2010, MNRAS, 403, 2012 
\bibitem[\protect\citeauthoryear{McGaugh \& Schombert}{2014}]{mcgaugh14} McGaugh S. S., Schombert J. M., 2014, AJ, 148, 77 
\bibitem[\protect\citeauthoryear{McQuinn et al.}{2015a}]{mcquinn15a} McQuinn K. B. W., Skillman E. D., Dolphin A. et al., 2015a, ApJ, 812, 158 
\bibitem[\protect\citeauthoryear{McQuinn et al.}{2015b}]{mcquinn15b} McQuinn K. B. W., Skillman E. D., Dolphin A. et al., 2015b, ApJL, 815, L17 
\bibitem[\protect\citeauthoryear{McQuinn et al.}{2020}]{mcquinn20} McQuinn K. B. W., Berg D. A., Skillman E. D. et al., 2020, ApJ, 891, 2, 181
\bibitem[\protect\citeauthoryear{Menacho et al.}{2019}]{menacho2019} Menacho V., \"Ostlin G., Bik A., Della Bruna L., Melinder J., Adamo A., Hayes M., Herenz  E. C., Bergvall N., 2019, MNRAS, 487, 3183
\bibitem[\protect\citeauthoryear{Moll{\'a}, Garc{\'{\i}}a-Vargas, Bressan}{2009}]{mol09} Moll{\'a} M., Garc{\'{\i}}a-Vargas M.~L., Bressan A., 2009, MNRAS, 398, 451 
\bibitem[\protect\citeauthoryear{Moll{\'a} et al.}{2015}]{mol15} Moll{\'a} M., Cavichia O., Gavil{\'a}n M., Gibson B. K., 2015, MNRAS, 451, 3693
\bibitem[\protect\citeauthoryear{Nakajima \& Ouchi}{2014}]{naka14} Nakajima K. \& Ouchi M., 2014, 442, 900
\bibitem[\protect\citeauthoryear{Osterbrock \& Ferland}{2006}]{osterbrockandferland} Osterbrock D. E., Ferland G.  J., 2006, \textit{Astrophysics of gaseous nebulae and active galactic nuclei}, Sausalito, CA,  University Science Books, 2nd ed.
\bibitem[\protect\citeauthoryear{Ouchi et al.}{2009}]{ouchi09} Ouchi M.. et al., 2009, ApJ. 706, 1136
\bibitem[\protect\citeauthoryear{Pascual et al.}{2018}]{pasetal18}Pascual S., Cardiel N., Picazo-S{\'a}nchez P., Castillo-Morales A., Gil de Paz, A., 2018, guaix-ucm/megaradrp: v0.8, DOI:10.5281/zenodo.2206856
\bibitem[\protect\citeauthoryear{P{\'e}rez-Montero}{2014}]{pm14} P{\'e}rez-Montero E., 2014, MNRAS, 441, 2663
\bibitem[\protect\citeauthoryear{Pustilnik, Tepliakova \& Makarov}{2019}]{pustilnik19} Pustilnik S. A., Tepliakova A. L., Makarov D. I., 2019, MNRAS, 482, 4329
\bibitem[\protect\citeauthoryear{Robertson et al.}{2013}]{robertsonetal13} Robertson B. E. et al., 2013, ApJ., 768, 71
\bibitem[\protect\citeauthoryear{Robertson et al.}{2015}]{robertsonetal15} Robertson B. E., Ellis R. S., Furlanetto S. R., Dunlop J. S., 2015, ApJL, 802, L19
\bibitem[\protect\citeauthoryear{Salpeter}{1955}]{Salpeter1955} Salpeter E. E., 1955, ApJ, 121, 161
\bibitem[\protect\citeauthoryear{S{\'a}nchez-Almeida et al.}{2016}]{sanchezalmeida16} S{\'a}nchez-Almeida J., et al., 2016, ApJ, 819, 110
\bibitem[\protect\citeauthoryear{Sz{\'e}csi}{2016}]{szecsi16} Sz{\'e}csi D., 2016, PhD Thesis, Mathematisch-Naturwissenschaftlichen Fakultät der Universität Bonn, DOI: 10.5281/zenodo.998070
\bibitem[\protect\citeauthoryear{Stasinska et al.}{2015}]{stasinskaetal15} Stasinska G., Izotov Y, Morisset C. and Guseva N., 2015, A\&A, 576, A83
\bibitem[\protect\citeauthoryear{Skillman, Kennicutt \& Hodge}{1989}]{skillman89} Skilmann E.D., Kennicutt R.C., Hodge P.W, 1989, ApJ, 347, 875 
\bibitem[\protect\citeauthoryear{Skillman et al.}{2013}]{skillman13} Skilmann E.D., Salzer J. J., Berg D. A. et al., 2013, AJ, 146, 3 
\bibitem[\protect\citeauthoryear{Tikhonov \& Galazutdinova}{2019}]{tg19} Tikhonov, N.A., Galazutdinova, O.A., 2019, Astronomy Letters, 45, 11
\bibitem[\protect\citeauthoryear{Tremonti et al.}{2004}]{tremontietal04} Tremonti C.A. et al., 2004, ApJ, 613, 898

\end{thebibliography}
\end{document}